\documentclass[preprint2,tighten]{aastex_reep}

\usepackage{epstopdf}
\usepackage{subfigure}  
\usepackage{amsmath}

\usepackage{longtable}

\shorttitle{NLTE Effects in Multithreaded Light Curves}
\shortauthors{Reep et al.}

\begin{document}

\title{Efficient Calculation of Non-Local Thermodynamic Equilibrium Effects in Multithreaded Hydrodynamic Simulations of Solar Flares}

\author[0000-0003-4739-1152]{Jeffrey W. Reep}
\affiliation{Space Science Division, Naval Research Laboratory, Washington, DC 20375, USA; \href{mailto:jeffrey.reep@nrl.navy.mil}{jeffrey.reep@nrl.navy.mil}}

\author{Stephen J. Bradshaw}
\affiliation{Department of Physics \& Astronomy, Rice University, Houston, TX 77005, USA}

\author{Nicholas A. Crump}
\affiliation{Space Science Division, Naval Research Laboratory, Washington, DC 20375, USA}

\author[0000-0001-6102-6851]{Harry P. Warren}
\affiliation{Space Science Division, Naval Research Laboratory, Washington, DC 20375, USA}

\begin{abstract}
Understanding the dynamics of the solar chromosphere is crucial to understanding energy transport across the solar atmosphere.  The chromosphere is optically thick at many wavelengths and described by non-local thermodynamic equilibrium (NLTE), making it difficult to interpret observations.  Furthermore, there is considerable evidence that the atmosphere is filamented, and that current instruments do not sufficiently resolve small scale features.  In flares, it is likely that multithreaded models are required to describe the heating.  The combination of NLTE effects and multithreaded modeling requires computationally demanding calculations, which has motivated the development of a model that can efficiently treat both.  We describe the implementation of a solver in a hydrodynamic code for the hydrogen level populations that approximates the NLTE solutions.  We derive an accurate electron density across the atmosphere, that includes the effects of non-equilibrium ionization for helium and metals.  We show the effects on hydrodynamic simulations, which are used to synthesize light curves using a post-processing radiative transfer code.  We demonstrate the utility of this model on IRIS observations of a small flare.  We show that the Doppler shifts in \ion{Mg}{2}, \ion{C}{2}, and \ion{O}{1} can be explained with a multithreaded model of loops subjected to electron beam heating, so long as NLTE effects are treated.  The intensities, however, do not match observed values very well, which is due to assumptions about the initial atmosphere.  We briefly show how altering the initial atmosphere can drastically alter line profiles, and therefore derived quantities, and suggest that it should be tuned to pre-flare observations.
\end{abstract}


\nopagebreak

\section{Introduction}
\label{sec:intro}

The solar chromosphere and corona are intimately connected through the transport of mass, momentum, and energy.  In order to understand the coronal response to solar flare heating events, a chromospheric model must be developed.  The chromosphere, however, is not in local thermodynamic equilibrium (LTE), and therefore radiative processes cannot be ignored in general \citep{carlsson2002}.  In order to calculate an accurate electron density, for example, one must determine the ionization fractions of the most abundant elements, which in turn requires knowledge of the radiation field.  It therefore is crucial to deal with non-LTE (NLTE) and radiative transfer effects.  The computational treatment of radiative transfer is a hugely demanding task, and the development of computational schemes remains an active field of research (\textit{e.g.} \citealt{judge2017}).  

To make matters worse, the solar atmosphere is generally filamented, meaning that there are many small-scale structures and features beneath the resolution of modern instrumentation.  Sub-structuring of active region loops has been found when comparing AIA images to higher resolution Hi-C data (\textit{e.g.} \citealt{brooks2013}), as well as in coronal rain when comparing AIA images to higher resolution H-$\alpha$ data \citep{antolin2015}, and in flare spectral data with IRIS \citep{warren2016}.  These facts point to the necessity of multithreaded models to accurately capture the details of the emission, and thus the underlying energy release and transport processes.  We therefore wish to have a model that can both capture multithreaded details of dynamic events, while simultaneously treating NLTE effects in the chromosphere, both of which are computational challenges in their own right.

In this work, we describe the implementation of such a model.  Using a hydrodynamic code, we have implemented an approximation to the radiative transfer equations in order to solve the hydrogen level populations in NLTE, and thereby determine a more accurate electron density.  We give full details of this implementation, and examine an example simulation in detail.  We then follow the methodology of our previous work to create a multithreaded model of a solar flare \citep{reep2016b}, which we then contrast against observations of light curves and Doppler shifts.  

Our comparison focuses on observations of a small flare by IRIS \citep{depontieu2014}, in which \citet{warren2016} found red-shifts in spectral lines that persisted for well over 30 minutes.  Specifically, \ion{Si}{4} 1402.77\,\AA\ and \ion{C}{2} 1334.54\,\AA\ were red-shifted for long periods of time near the flare ribbon, while \ion{Mg}{2} 2796.354\,\AA\ showed red-shifts that gradually decayed.  Similarly-behaved red-shifts are seen routinely in flares with IRIS, \textit{e.g.} \citet{graham2015,li2015,polito2016,brosius2017,tian2018}, less commonly with SOHO/CDS \citep{brosius2004}, and possibly in \ion{He}{2} 303.78\,\AA\ with Hinode/EIS (\citealt{lee2017}, though that line has multiple blends with coronal iron lines).  These observations are surprising in light of the results of \citet{fisher1989}, who showed that chromospheric condensation events only last for about a minute, regardless of the strength or duration of the heating, which might sufficiently explain the \ion{Mg}{2} emission, but not the other lines.  Although it was not explicitly shown in \citet{warren2016}, \ion{O}{1} 1355.60\,\AA\ was essentially stationary (shown in Section \ref{sec:forward} of this paper).  

In order to explain the persistent red-shifts, therefore, \citet{reep2016b} used a multi-threaded model where a large number of loops rooted within one IRIS pixel are successively heated, thereby causing successive condensation events and producing a long-lasting red-shift.  This model was consistent with the \ion{Si}{4} emission, though the \ion{C}{2} line had a much stronger stationary component than the model.  Furthermore, the model predicted that the \ion{O}{1} line should be red-shifted at around 10\,km\,s$^{-1}$, in contradiction with the observations of a stationary line.  One possible explanation is that the forward modeling erroneously assumed that the lines were optically thin, which may or may not hold true for \ion{C}{2} \citep{rathore2015} and \ion{O}{1} \citep{lin2015}, depending on the particular conditions of the event.  It is also possible that the line shapes may be non-Maxwellian, which can affect the opacity \citep{dudik2017}.  We seek to directly test the importance of NLTE effects by redoing the simulations with the improved model, and then synthesizing the emission with a radiative transfer solver.

In this paper, we first describe the method in Section \ref{sec:implementation} that we implement to give an approximation to the level populations of hydrogen in the chromosphere, thereby improving the resultant electron density and better treating the effects of NLTE.  We then examine the details of a loop subjected to heating by an electron beam in Section \ref{sec:modeling}.  Then, in Section \ref{sec:forward}, we use a multi-threaded model of a beam-heated flare to forward model line profiles, light curves, and Doppler shifts in order to compare with observations.  We show that many features are consistent with observations, but the assumed initial atmosphere can have a large impact on the results.
\section{Implementation}
\label{sec:implementation}

In this Section, we describe a method by which we solve the NLTE level populations and ionization state of hydrogen, which we have added to the field-aligned HYDrodynamics and RADiation code (HYDRAD, \citealt{bradshaw2003,bradshaw2013}).  HYDRAD solves the equations describing the conservation of mass, momentum, and energy for a two-fluid plasma constrained to a magnetic flux tube.  The code solves the full loop length in an arbitrary geometry, in terms of loop shape, inclination, and expanding cross-section (equivalently, varying magnetic field strength).  It includes the effects of full non-equilibrium ionization for any desired element, returning the ion populations and an accurate calculation of the radiative losses.  HYDRAD also makes use of adaptive mesh refinement of arbitrary order, important for accurately resolving sharp gradients in density and temperature.  The code is light enough to run on a desktop (tested on Linux, Mac, and Windows), but general enough to run on high performance computing machines.  

In previous versions, the chromospheric ionization fraction was calculated with LTE assumptions, \textit{i.e.}, that the ionization fraction is determined by the local density and temperature and that it is collisionally dominated.  To improve upon this, we use the method for computing NLTE effects outlined in \citet{leenaarts2006} and \citet{leenaarts2007}, which is in turn based on the method derived by \citet{sollum1999}.  For completeness, we reiterate many of the details from those works, and note a few small points that differ from these other works (\textit{e.g.} atomic parameters).  

We wish to solve the level populations of the hydrogen atom with the equation
\begin{equation}
\frac{\partial n_{i}}{\partial t} + \frac{\partial}{\partial s} \Big(n_{i} v\Big) = \sum_{j \neq i} n_{j} P_{ji} - n_{i} \sum_{j \neq i} P_{ij}
\label{eqn:levels}
\end{equation}
\noindent where $n_{i}$ is the fractional level population of level $i$ and $v$ is the bulk flow velocity.  The rate coefficient $P_{ij}$ represents the rate (s$^{-1}$) at which atoms transition from level $i$ to level $j$ (and vice versa for $P_{ji}$).  We solve this equation for a six-level hydrogen atom, including the the first five levels (principle quantum number $n$ from 1 to 5) plus the ionized state.

\subsection{Rate Coefficients}
\label{subsec:rates}

The rate coefficients $P_{ij}$ are the sum of collisional and radiative rates:
\begin{equation}
P_{ij} = C_{ij} + R_{ij}
\end{equation}
\noindent The collisional rate coefficients $C_{ij}$ are taken directly from the tables bundled with the RH1.5D code \citep{uitenbroek2001,pereira2015}, which in turn are based on \citet{johnson1972}.  They are look-up tables as a function of density and temperature, which we interpolate to calculate the local coefficients.

The radiative rate coefficients $R_{ij}$ require a detailed calculation to accurately determine.  In general, a full solution of the radiative transfer equations is necessary to determine the radiation field, which in turn determines the rates.  In the RADYN code \citep{carlsson1992,allred2015}, the transfer equation is solved with the method from \citet{scharmer1981,scharmer1985}, while the Flarix code \citep{heinzel2016} uses an accelerated lambda-iteration method \citep{rybicki1991}.  In this work, we follow the prescription derived by \citet{sollum1999} that gives a method for approximating the radiation field in the chromosphere, which is somewhat less accurate than the methods used in RADYN or Flarix, but significantly less computationally demanding.  For similar reasons, the method is also implemented in the Bifrost MHD code \citep{gudiksen2011}.  

The prescription for the radiation field has a few key assumptions.  First, it is assumed that below a certain height in the chromosphere, the population is in LTE.  This can be due to either of two options: that the collisional rates dominate the radiative rates, or if the radiation field is well-described by a Planck function, both of which are true at the photosphere and in the lower chromosphere.  Below this critical height, therefore, the radiation field is assumed to be described by a Planck function.

Second, the Lyman transitions are assumed to be in detailed balance in the chromosphere \citep{carlsson2002}: $n_{1}R_{1j} = n_{j} R_{j1}$, which says that the number of transitions into the ground state from state $j$ equals the number from the ground state into state $j$.  This allows the simplification that the net rates for the Lyman transitions are collisionally dominated $P_{1j} \approx C_{1j}$.  \citet{sollum1999} tested this approximation in depth, finding that the errors are negligible in the chromosphere, and become more significant in the transition region.

Third, it is assumed that the radiation field for each transition at a given location can be characterized by a local brightness temperature $T_{b}$, defined by \citet{sollum1999} as the temperature where the Planck function $B_{\nu}(T_{b})$ equals the intensity at that wavelength.
\begin{equation}
J_{\nu} = \frac{2 h \nu^{3}}{c^{2}} \frac{1}{\exp{\Big(\frac{h \nu}{k_{B} T_{b}}\Big)} - 1}
\label{eqn:jnu1}
\end{equation}
\noindent where $\nu$ is the frequency of the transition, $h$ Planck's constant, $c$ the speed of light, and $k_{B}$ the Boltzmann constant.  The brightness temperature is not an actual temperature, but a convenient parameter that characterizes the radiation field (as commonly done in radio astronomy).  Note that \citet{sollum1999} and \citet{leenaarts2006} refer to this temperature as the ``radiation temperature'' rather than brightness temperature.

Finally, for each transition, at each location, and at each time, the brightness temperature must be determined.  At the top of the chromosphere, the brightness temperature $T_{b}^{\text{top}}$ is taken as input based on the \citet{sollum1999} study, which we list in the appendix.  Next, we determine a critical height $z_{\text{crit}}$ for each transition defined as the lowest point in the atmosphere where $J_{\nu}(T_{e}) = 2 J_{\nu}(T_{b}^{\text{top}})$, which for most transitions is near the temperature minimum region.  Below this height we assume $T_{b}(z < z_{\text{crit}}) = T_{e}(z)$, and above that height the brightness temperature is a function of the column mass (equivalently, the optical depth):
\begin{multline}
J_{\nu}(z) = B_{\nu}(T_{b}^{\text{top}}) + \\
\Big[B_{\nu}(T_{e}(z_{\text{crit}})) - B_{\nu}(T_{b}^{\text{top}})\Big]\Bigg(\frac{m_{c}(z)}{m_{c}(z_{\text{crit}})}\Bigg)^{H}
\label{eqn:jnu2}
\end{multline}
\normalsize
\noindent where $B_{\nu}(T)$ is the Planck function at temperature $T$, $m_{c}(z)$ is the column mass at height $z$ in the chromosphere, and $H$ is a constant prescribed by \citet{sollum1999} ($H = 2$ for bound-bound transitions and $H = 4$ for bound-free transitions).  This equation is identical to Equation (3) of \citet{leenaarts2006}, but it differs from Equation 5.3 of \citet{sollum1999}, where $B_{\nu}(T_{e}(z_{\text{crit}}))$ is replaced by $B_{\nu}(T_{e}(z))$.  The difference between the two is small, but the former is computationally simpler (Leenaarts 2017, private communication).

From Equation \ref{eqn:jnu1}, we can solve for the brightness temperature $T_{b}(z > z_{\text{crit}})$:
\begin{equation}
T_{b}(z > z_{\text{crit}}) = \frac{h \nu}{k_{B}}\Bigg(\ln{\Big[1 + \frac{2 h \nu^{3}}{J_{\nu}c^{2}}\Big]}\Bigg)^{-1}
\end{equation}
\noindent where $J_{\nu}$ is given by Equation \ref{eqn:jnu2}.

We can now calculate the radiative coefficients at a given location and time using this prescription for the radiation field.  Following \citet{sollum1999} as before, for radiative excitation, with $i < j$,
\begin{equation}
R_{ij} = \frac{8 \pi^{2} e^{2} f_{ij}}{m_{e}c^{3}} \frac{\nu_{0}^{2}}{\exp{\Big(\frac{h \nu_{0}}{k_{B}T_{b}}\Big)} - 1}
\end{equation}
\\
\noindent where $e$ is the elementary charge, $f_{ij}$ the oscillator strength, $m_{e}$ the electron mass, and $\nu_{0}$ the rest frequency of the line.  We take the atomic parameters like oscillator strengths, rest wavelengths, and statistical weights from \citet{wiese2009}.  The radiative de-excitation coefficient can be found similarly
\begin{equation}
R_{ji} = \frac{g_{i}}{g_{j}} \exp{\Big(\frac{h \nu_{0}}{k_{B}T_{b}}\Big)} R_{ij}
\end{equation}
\noindent where $g_{i}$ is the statistical weight of level $i$.

The bound-free (radiative ionization) rate $R_{ic}$ is calculated with a Kramers cross-section:
\begin{equation}
R_{ic} = \frac{8\pi \alpha_{0} \nu_{0}^{3}}{c^{2}} \sum_{q = 1}^{\infty} E_{1} \Bigg(\frac{q h \nu_{0}}{k_{B} T_{b}}\Bigg)
\end{equation}
\noindent where $\nu_{0}$ is now the edge frequency, $\alpha_{0}$ is the cross-section at $\nu_{0}$, $E_{1}$ is the first exponential integral, and $q$ is just a summation index.  We have evaluated this summation over the first 10,000 terms, and found that the value of the sum has converged for all transitions in this study.

Finally, the free-bound (radiative recombination) rate $R_{ci}$ is similar:
\begin{multline}
R_{ci} = \frac{8\pi \alpha_{0} \nu_{0}^{3}}{c^{2}} \Big(\frac{n_{i}}{n_{c}}\Big)_{\text{LTE}} \times \\
\sum_{q=0}^{\infty} E_{1} \Bigg( \Big( \frac{q T_{e}}{T_{b}} + 1\Big) \frac{h \nu_{0}}{k_{B} T_{e}}\Bigg)
\end{multline}
\noindent We similarly sum over the first 10,000 terms here, again finding that all transitions in this work have converged.  In this expression, the population ratio $\frac{n_{i}}{n_{c}}$ is the LTE ratio, given by the Saha equation:
\begin{equation}
\Big(\frac{n_{i}}{n_{c}}\Big)_{\text{LTE}} = n_{e} \frac{g_{i}}{2g_{c}} \Bigg( \frac{2 \pi m_{e} k_{B} T_{e}}{h^{2}}\Bigg)^{-3/2} \exp{\Bigg(\frac{h \nu_{0}}{k_{B}T_{e}}\Bigg)}
\end{equation}

All of these coefficients are pre-calculated as look-up tables as functions of temperature(s) and density, as appropriate.  In simulations, we then interpolate the values using the local brightness temperature, electron temperature, and density.

\subsection{Brightness Temperature}
\label{subsec:brightness}

Since the thesis is not publicly available, in Table \ref{Tb}, we reproduce the list of the brightness temperatures at the top of the chromosphere $T_{b}^{\text{top}}$ for each transition ($i \rightarrow j$), derived for the VAL C model by \citet{sollum1999} from average quiet sun observations (M. Carlsson, private communication).  
\begin{table}
\centering
\caption{The brightness temperatures at the top of the chromosphere for each transition $i \rightarrow j$, derived by \citet{sollum1999}, Table B.1 in that work. \label{Tb}}
\begin{tabular}{ c | c }
$i \rightarrow j$ & $T_{b}^{\text{top}}$ [K] \\
\hline
$2 \rightarrow 3$ & 4500 \\
$2 \rightarrow 4$ & 4550 \\
$2 \rightarrow 5$ & 4500 \\
$3 \rightarrow 4$ & 4000 \\
$3 \rightarrow 5$ & 4300 \\
$4 \rightarrow 5$ & 3700 \\
$2 \rightarrow 6$ & 5493 \\
$3 \rightarrow 6$ & 4850 \\
$4 \rightarrow 6$ & 4750 \\
$5 \rightarrow 6$ & 4470 \\
\end{tabular}
\end{table}

Since these values are based on the quiet sun, however, we wish to scale them appropriately for flaring simulations.  In order to scale $T_{b}^{\text{top}}$ with time, we first note that its value scales linearly with the intensity of the given line.  Starting with Equation \ref{eqn:jnu1}, we can solve to find:
\begin{align}
T_{b} &= \frac{h \nu}{k_{B}\ln{\Big(1+\frac{2 h \nu^{3}}{J_{\nu} c^{2}}\Big)}} \\ \nonumber
	 &\approx \frac{c^{2} J_{\nu}}{2k_{B}\nu^{2}} \nonumber
\end{align}
where we have expanded the logarithmic term to first order in the last step: $\ln{(1+x)} \approx x$ for small values of $x$.  The intensity of the line, then, can be used to scale the brightness temperature with time.  

We have found empirically that the intensity of each line scales well with the footpoint density $n_{FP}$ at the base of the transition region (where hydrogen transitions from neutral to ionized).  We base this scaling on the more accurate treatment of radiative transfer from RADYN, using the publicly available simulations on the F-CHROMA website\footnote{\href{https://star.pst.qub.ac.uk/wiki/doku.php/public/solarmodels/start}{F-CHROMA model database}}. 

\begin{figure}
  \begin{minipage}[b]{\linewidth}
    \centering
    \includegraphics[width=\textwidth]{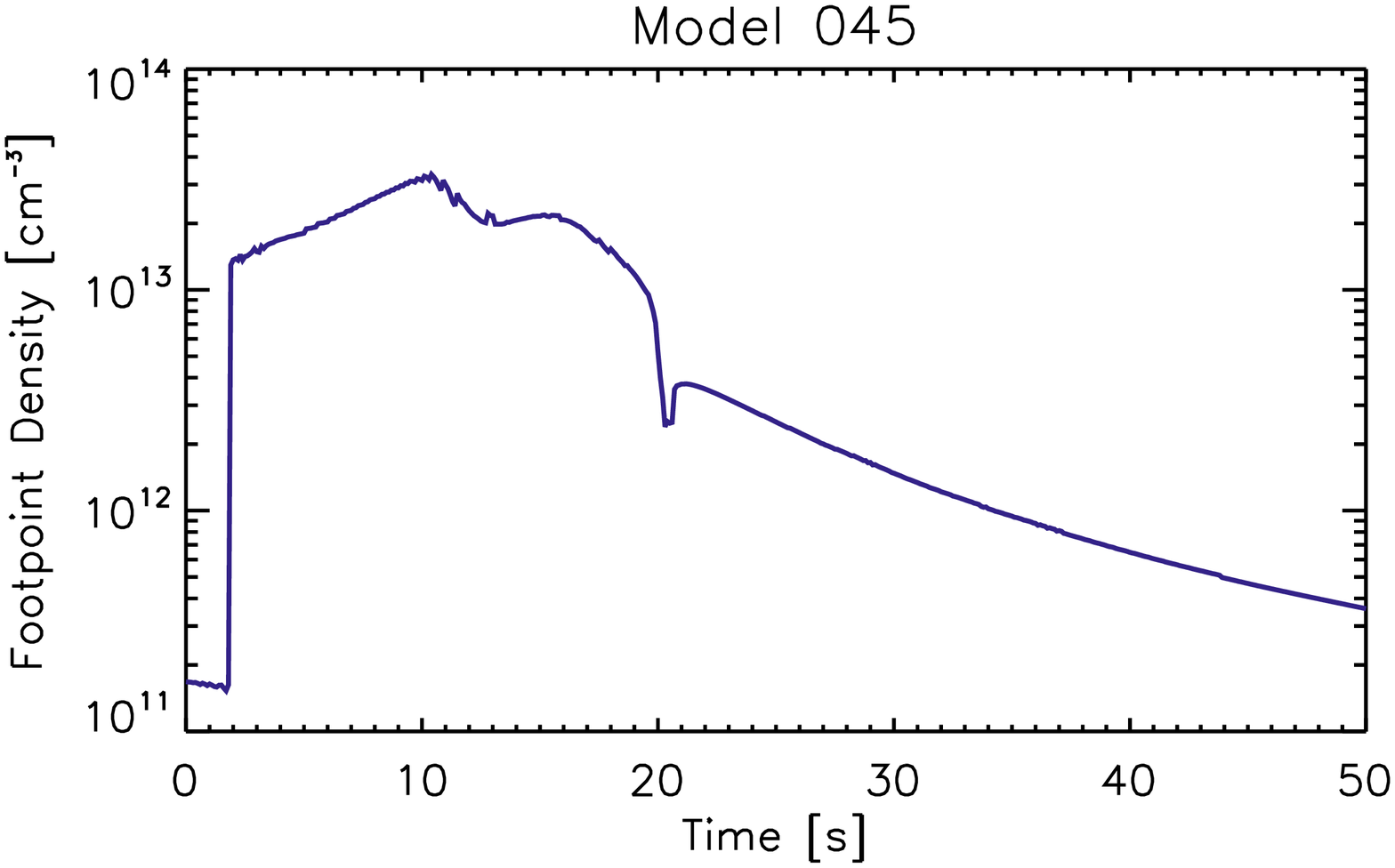}
  \end{minipage}
  \begin{minipage}[b]{\linewidth}
    \centering
    \includegraphics[width=\textwidth]{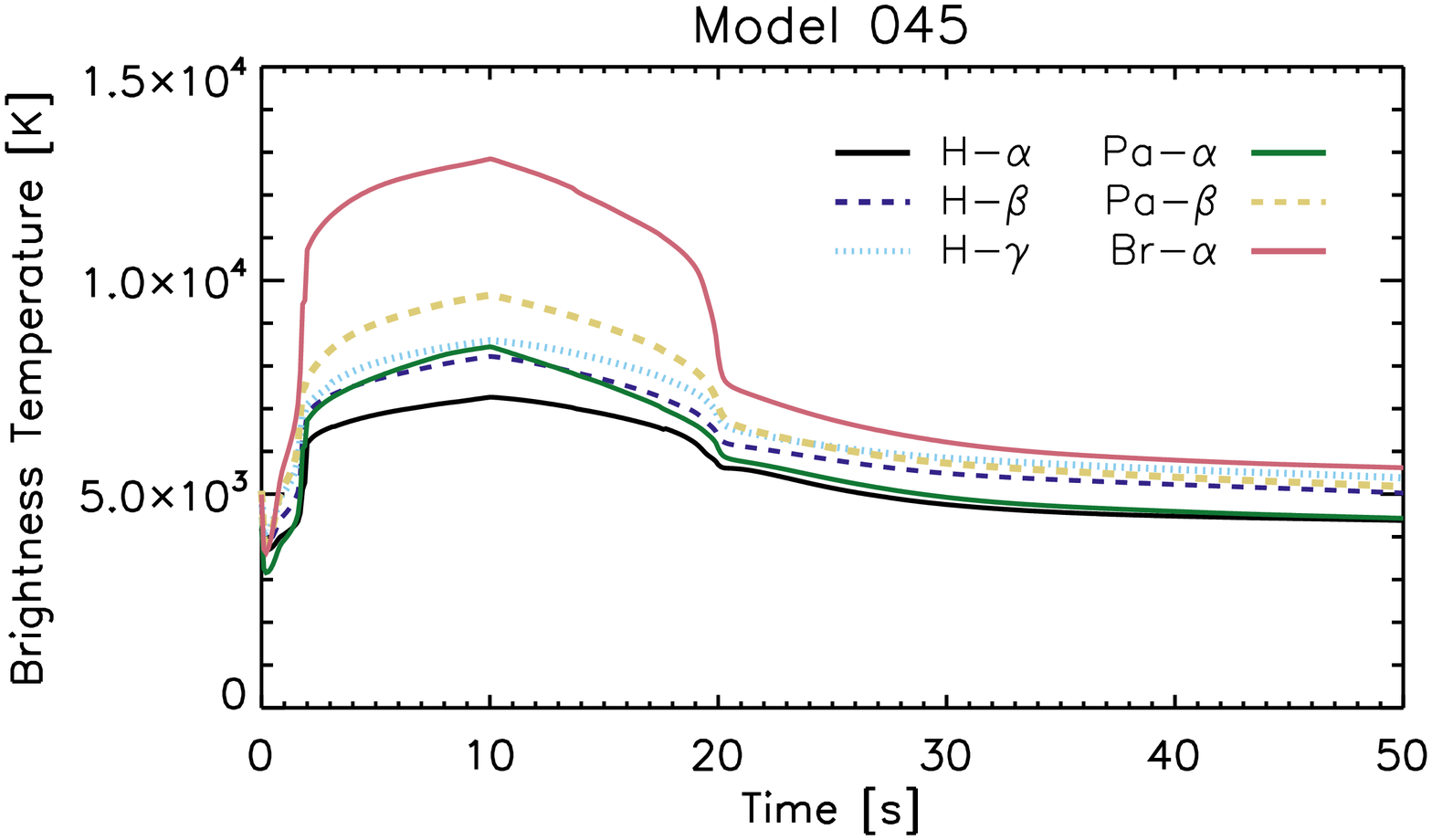}
  \end{minipage}
    \begin{minipage}[b]{\linewidth}
    \centering
    \includegraphics[width=\textwidth]{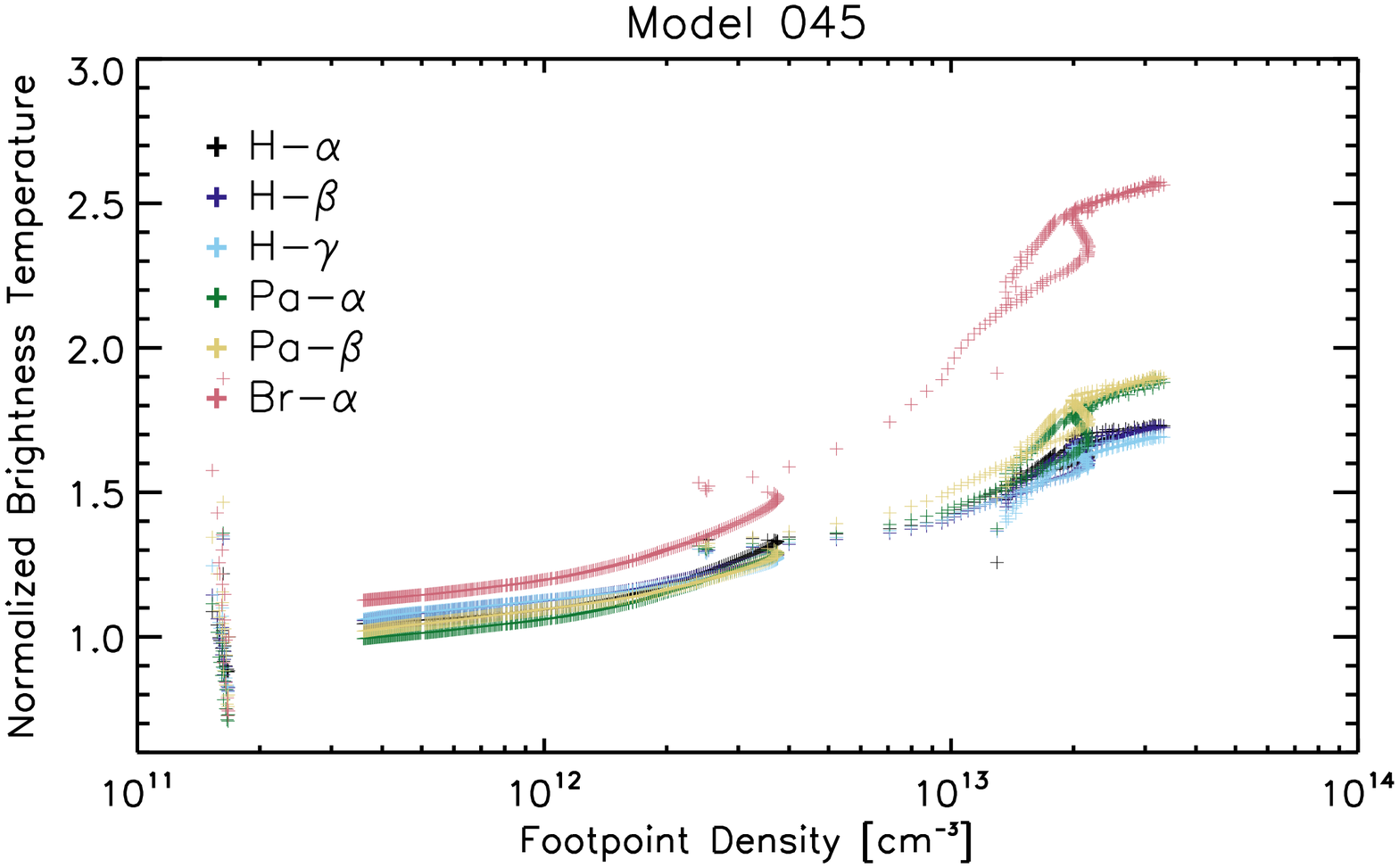}
  \end{minipage}
\caption{The scaling of brightness temperature $T_{b}^{top}$ with footpoint density, using Model 045 from the F-CHROMA database.  Top: the footpoint density as a function of time.  Center: the brightness temperatures with time, for the six bound-bound transitions under consideration.  Bottom: a scatter plot showing the relation between the two variables, with the brightness temperature normalized to its initial value.  The brightness temperatures can be scaled directly in relation to the footpoint density.}
\label{fig:scaling}
\end{figure}
For example, in Figure \ref{fig:scaling}, we show the footpoint density variation with time (top), normalized brightness temperature variation with time for the six bound-bound transitions considered (center), and a scatter plot showing the relation between the two variables (bottom) for Model 045 in the database.  In all six cases, there is a direct correlation between the footpoint density and the brightness temperature (with noticeable scatter).  We disregard bound-free transitions because the brightness temperatures of these do not vary with time in the RADYN simulation.  We fit a line in log-log space, and rewrite the relation: 
\begin{align}
\log{\Bigg(\frac{T_{b}^{top}(t)}{T_{b}^{top}(t = 0)}\Bigg)} &= m \log{n_{FP}} + C   \\
 T_{b}^{top}(t) &= C (n_{FP})^{m}\ T_{b}^{top}(t = 0) 
\end{align}
\noindent where $C$ is a constant, and $m$ is the slope.  This relation holds particularly well in simulations of intermediate-strength heating, where both the footpoint densities and brightness temperatures are sampled across a wide range of values.  From Model 045, we find the following values of $m$, which we use in our approximation:
\begin{itemize}
\item H-$\alpha$: $m = 0.1188 \pm 0.0009$
\item H-$\beta$: $m = 0.1116 \pm 0.0010$
\item H-$\gamma$: $m = 0.1061 \pm 0.0011$
\item Paschen-$\alpha$: $m = 0.1460 \pm 0.0014$
\item Paschen-$\beta$: $m = 0.1402 \pm 0.0017$
\item Brackett-$\alpha$: $m = 0.1979 \pm 0.0026$
\end{itemize}
\noindent The values of $C$ can be chosen such that the initial value of density $n_{0}$ gives the initial brightness temperature:
\begin{align}
 T_{b}^{top}(t) &= \Big(\frac{n_{FP}}{n_{0}}\Big)^{m}\ T_{b}^{top}(t = 0)
\end{align}
\noindent We do not scale the temperatures to densities lower than the initial value.

\begin{figure*}
  \begin{minipage}[b]{0.5\linewidth}
    \centering
    \includegraphics[width=\textwidth]{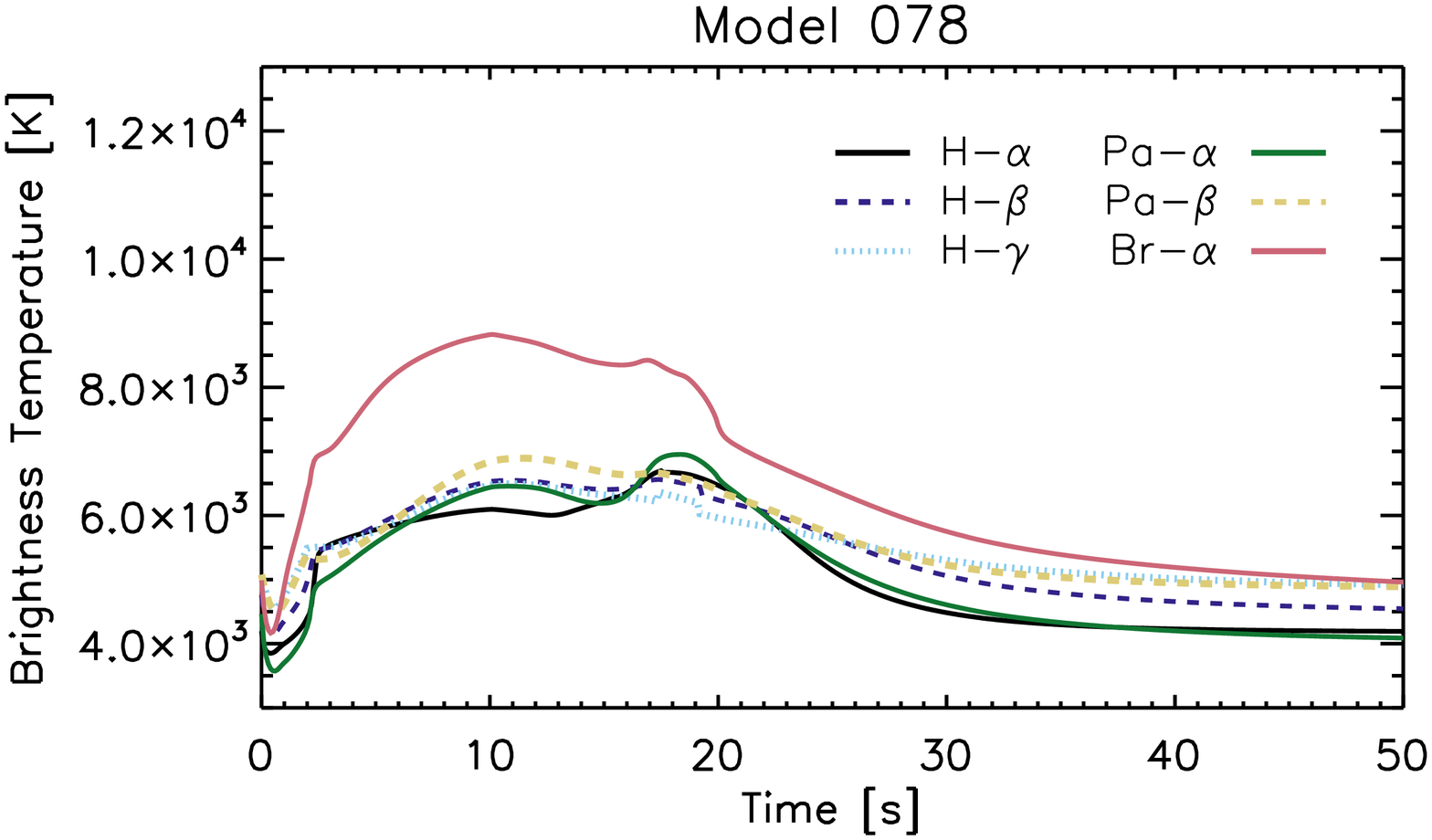}
  \end{minipage}
  \begin{minipage}[b]{0.5\linewidth}
    \centering
    \includegraphics[width=\textwidth]{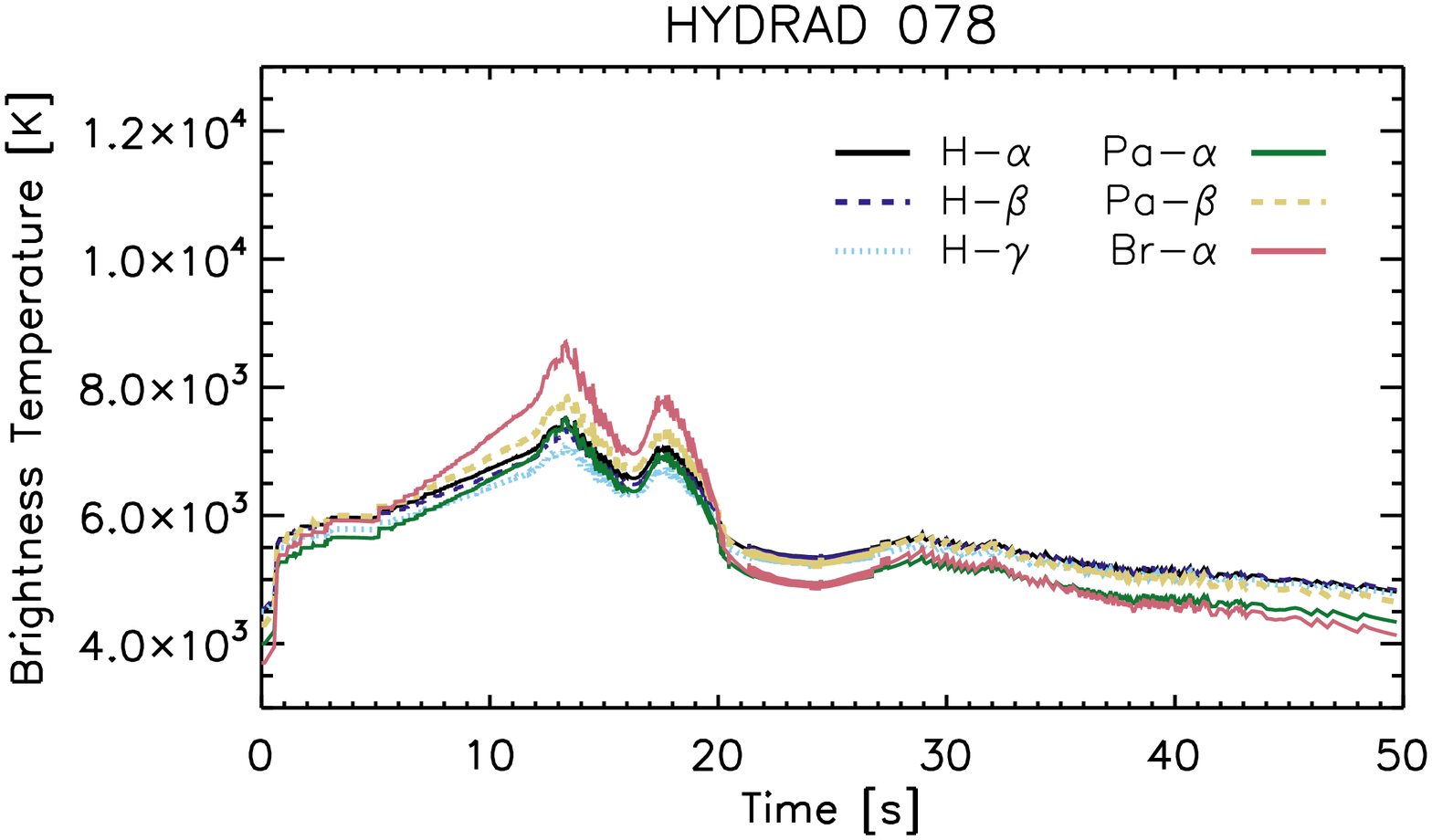}
  \end{minipage}
    \begin{minipage}[b]{0.5\linewidth}
    \centering
    \includegraphics[width=\textwidth]{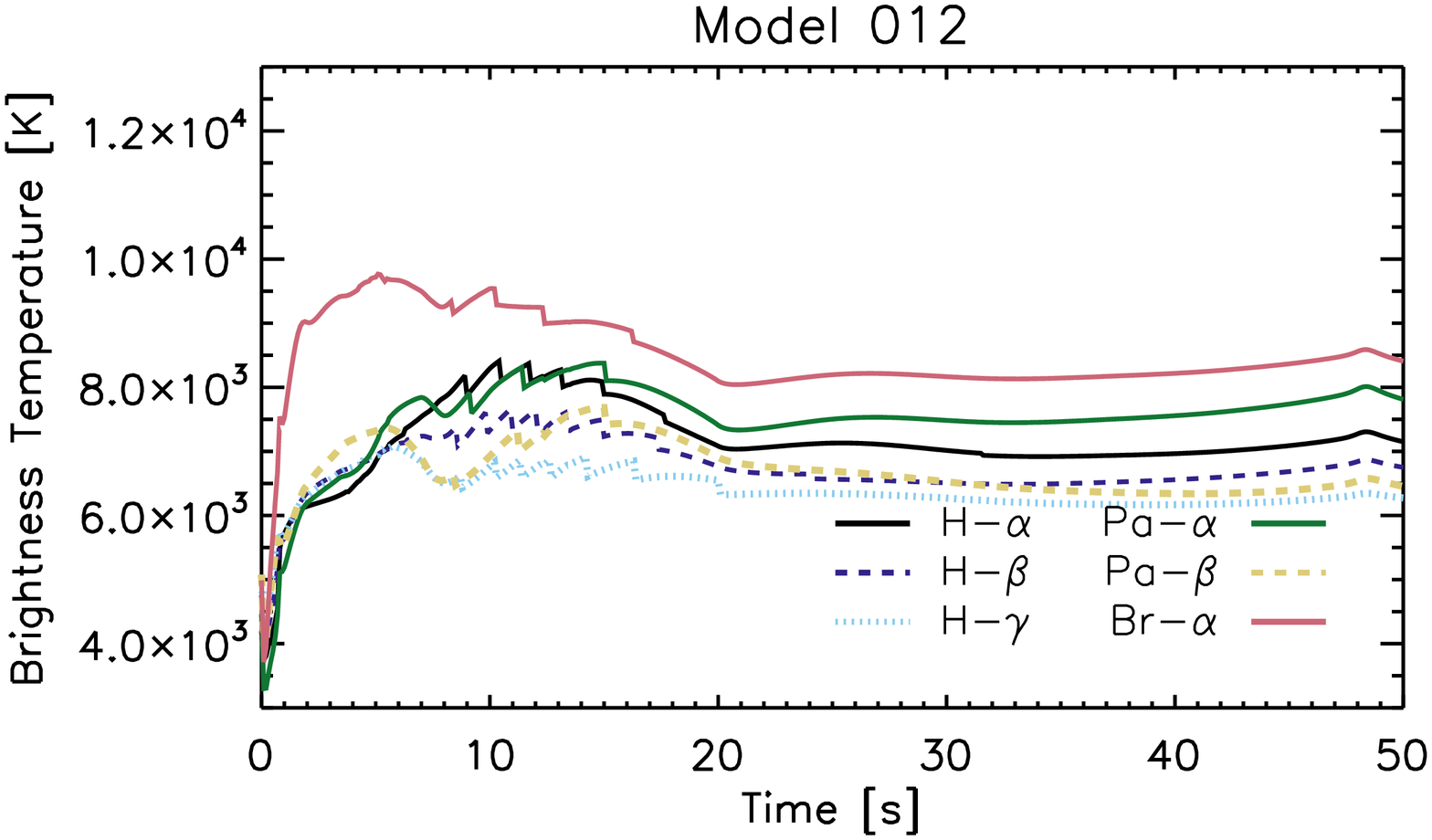}
  \end{minipage}
  \begin{minipage}[b]{0.5\linewidth}
    \centering
    \includegraphics[width=\textwidth]{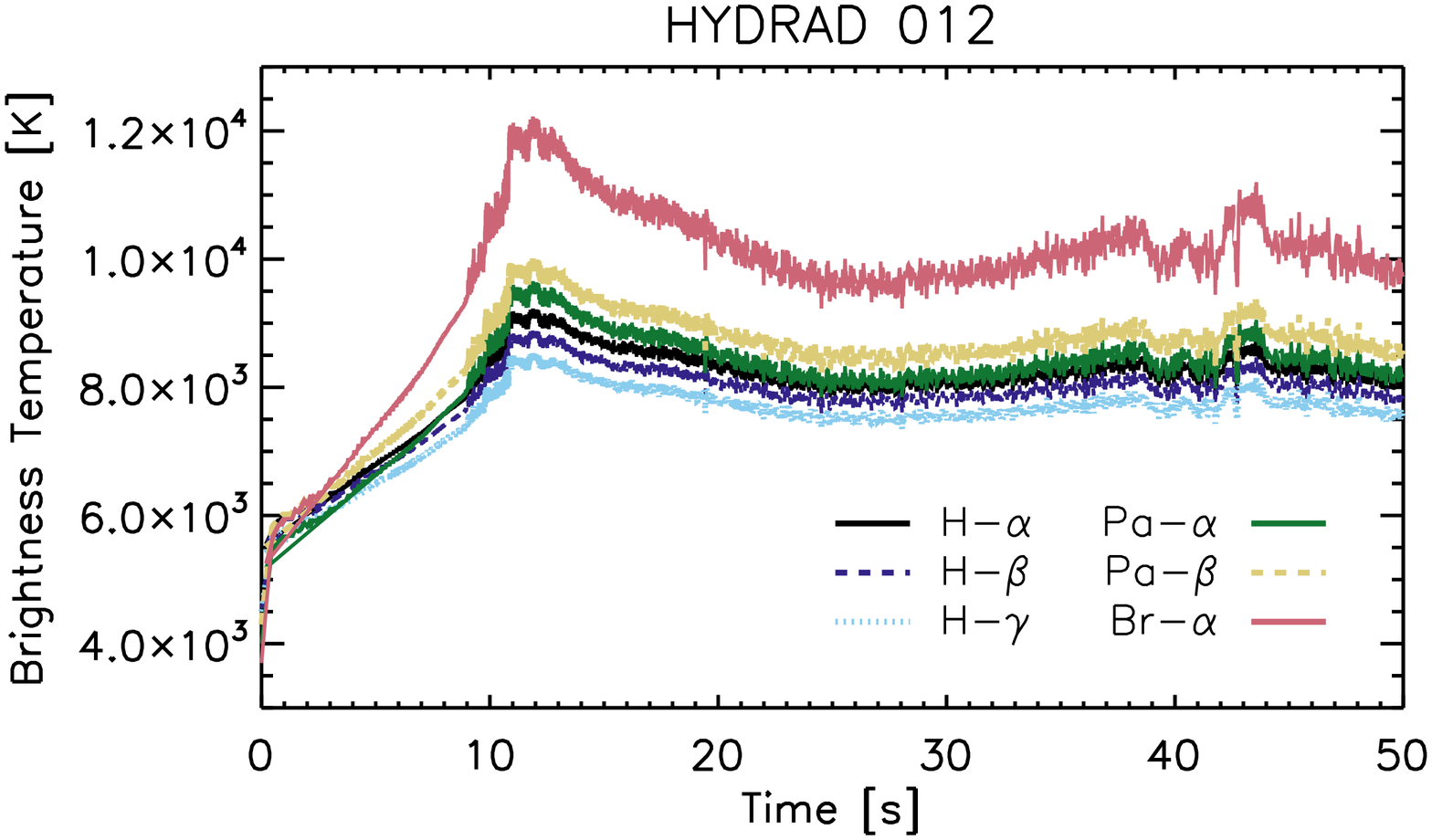}
  \end{minipage}
\caption{The brightness temperature with time for two simulations showing the values in RADYN (left) and HYDRAD (right).  The approximation tends to over-estimate the values, particularly in the case of stronger heating (Model 012).  A full comparison of the hydrodynamics of these simulations is available in Appendix \ref{app:radyn}.  \label{fig:Tb_comparison}}
\end{figure*}

In Figure \ref{fig:Tb_comparison}, we compare the values of the brightness temperatures as a function of time in Models 078 and 012 in the database, as well as those calculated with the scaling relation in HYDRAD using the same simulation parameters.  At most times, the footpoint densities in the simulations are comparable, and so we find that the brightness temperatures are generally over-estimated compared to the values in RADYN, particularly in the case with stronger heating (Model 012) by around 25\% at the peak.  As we show in Appendix \ref{app:radyn}, though, the simulations are in good agreement with regards to densities, temperatures, and velocities, so the difference in brightness temperatures are acceptable.  The slopes could perhaps be adjusted to improve the approximation, but that may not improve all cases uniformly, so we choose to use the above values.

\subsection{Non-thermal Collisional Excitation and Ionization}
\label{subsec:nonthermal}

Collisions from non-thermal electrons with the ambient plasma can, in addition to heating the plasma, drive collisional excitation and ionization of the atoms.  Previous studies have shown that this can have a non-negligible effect on the level populations and line profiles of neutral hydrogen \citep{fang1993,fang2003,kasparova2009}.  \citet{fang1993} found that this effect is most important for excitation and ionization from the ground state of hydrogen, and the effect of non-thermal collisions on excited states is negligible.  

While the electron beam is active, therefore, we include non-thermal collisions in the rate equations.  We follow Equation 22 in \citet{allred2015}:  
\begin{equation}
C_{ij}^{\text{non-thermal}} = \zeta_{ij} \frac{\Lambda_{n}}{n_{e} \Lambda_{i} + n_{H} \Lambda_{n}} \frac{dE}{dt}
\end{equation}
where $\zeta_{12} = 2.94 \times 10^{10}$, $\zeta_{13} = 5.35 \times 10^{9}$, $\zeta_{14} = 1.91 \times 10^{9}$, and $\zeta_{1c} = 1.73 \times 10^{10}$ \citep{fang1993}, $\Lambda_{n}$ and $\Lambda_{i}$ are defined in \citet{emslie1978}, and $\frac{dE}{dt}$ is the energy deposition rate by the beam. 

There are two noticeable effects when comparing simulations with and without this effect: the ionization fraction rises more quickly after the onset of heating, and the plasma ionizes at deeper depths.  Since this affects the electron density at various heights in the chromosphere, chromospheric line profiles are also affected.

\subsection{Level Populations}
\label{subsec:levels}

Once the rate coefficients have been calculated, we are ready to solve for the level populations.  Equation \ref{eqn:levels} represents a set of six equations, to which we add one constraint:
\begin{equation}
\sum_{i} n_{i} = 1
\label{eqn:sum}
\end{equation}
\noindent This simply states that the sum of the fractional populations is 1.

We then have a $7 \times 6$ matrix equation, which can be solved through singular-value decomposition (SVD).  We use the SVD function taken from the \textit{Numerical Recipes} text \citep{press2002} for this purpose.  The matrix is not always well-conditioned, so that SVD is the ideal choice for solving the matrix equation.\footnote{The pseudoinverse $(A^{T} A)^{-1} A^{T}$ in the normal equation could also be used to solve the matrix equation in principle.  Through simple tests, however, we have found SVD to be more robust for this application.}  

There is one caveat: in general, we do not know the electron density prior to solving these equations, and the rate coefficients depend on the electron density.  We therefore must make an initial guess to the electron density and, through a gradient descent, iterate in small steps until the solution has converged.

The total electron density is given by the sum of the free electrons from hydrogen, helium, and metals.  We approximate this using only the first 30 elements, as others are too scarce to contribute significantly.
\begin{equation}
n_{e} = n_{H} \times \Bigg[\frac{n_{HII}}{n_{H}} + \sum_{Z = 2}^{Z = 30} A_{Z} \bigg(\sum_{k = 1}^{k = Z} k\ Y_{k} \bigg) \Bigg]
\end{equation}
\noindent where $Z$ is the atomic number, $A_{Z}$ is the fractional abundance of element number $Z$, $k$ is the ionization stage ranging from singly to fully ionized (\textit{e.g.} $k = 2$ corresponds to doubly ionized), and $Y_{k}$ is the fractional population of ionization stage $k$ of element $Z$.

For trace elements, we solve for the ionization fractions $Y_{k}$ using either an equilibrium calculation that is a function of temperature, or by solving a continuity equation for non-equilibrium ionization states, as detailed in \textit{e.g.} \citet{bradshaw2003,bradshaw2003b}.  The non-equilibrium solver can be used with all, some, or none of the elements, as determined by the user and as appropriate for the study at hand.

The code makes an initial guess for the \ion{H}{2} fraction, calculates the rate coefficients, solves Equations \ref{eqn:levels} and \ref{eqn:sum} for the level populations, and then recalculates the total electron density.  The process iterates in small steps until the electron density has converged within a defined relative tolerance (default of $10^{-6}$).




\subsection{Radiative Losses}
\label{subsec:losses}

In previous versions of HYDRAD, the hydrogen and electron densities were essentially equal in the corona (with only a small correction for trace elements), so that the optically thin radiative losses $E_{R}$ (erg\,s$^{-1}$\,cm$^{-3}$) were approximated by setting $n_{e} = n_{H}$.  We drop that approximation now, and use the more accurate:
\begin{equation}
E_{R} = - n_{e} n_{H} \Lambda(T_{e})
\end{equation}
\noindent where $\Lambda(T_{e})$ is the sum of the emissivity over all ions and all transitions for each ion (more precisely, all in the current version of CHIANTI;  \citealt{delzanna2015}), and the minus sign indicates that the energy is lost from the system.  The loss function $\Lambda$ also depends on the ionization fractions, and therefore also utilizes a non-equilibrium ionization calculation when desired.  

In the chromosphere, we calculate radiative losses following the prescription of \citet{carlsson2012}, which calculates the loss of energy from hydrogen, calcium, and magnesium.  We now use the ionization fraction of hydrogen determined by the above prescription in this calculation.

\subsection{Performance}
\label{subsec:performance}

\begin{figure}
\begin{minipage}[b]{\linewidth}
\centering
\includegraphics[width=\textwidth]{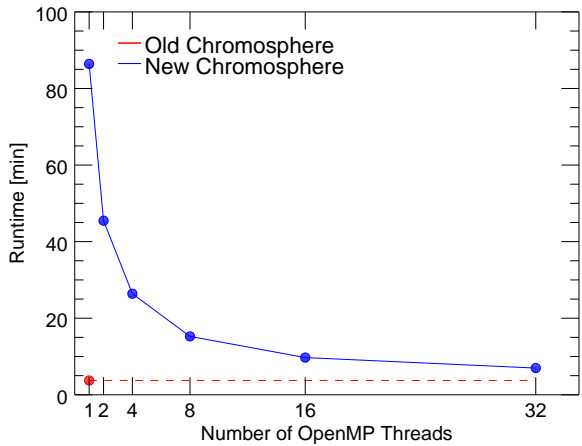}
\end{minipage}
\caption{A comparison of run-times for a simple simulation, using the old version of the chromosphere (red) against the new chromosphere (blue) with various number of cores.  The increase in speed scales as $t \propto N^{-0.7}$, up to at least 32 cores, and gets within a factor of 2 of the run-time using the old chromosphere.}
\label{fig:performance}
\end{figure}

The calculation of the NLTE level populations is a significant computational task.  A few simple tests with various parameters have found that the code slows by a factor of 10--20 compared to previous versions.  To offset these losses, HYDRAD has been recently parallelized with OpenMP.  In Figure \ref{fig:performance}, we briefly show a timing test of the code.  We have run an example simulation of electron beam heating (see the next Section), with energy flux $F_{0} = 3 \times 10^{9}$\,erg\,s$^{-1}$\,cm$^{-2}$, for 10 seconds of heating.  The simulations were only run for 100 seconds of simulation time.  The timing of the simulation run with the older version of the chromosphere is shown in red, compared to the newer version of the chromosphere in blue.  The improvements scale in time $t \propto N^{-0.7}$, where $N$ is the number of cores requested, with improvements up to at least 32 cores.  The resultant timing is within a factor of 2 of that using the original chromosphere.

\section{Modeling}
\label{sec:modeling}

\subsection{Initial Atmosphere}
\label{subsec:initial}

We now examine how these changes affect the dynamics of a coronal loop.  We first examine the hydrostatic profile, and then run a simulation with electron beam heating in order to look closely at the hydrodynamics.  

\begin{figure}
\begin{minipage}[b]{\linewidth}
\centering
\includegraphics[width=\textwidth]{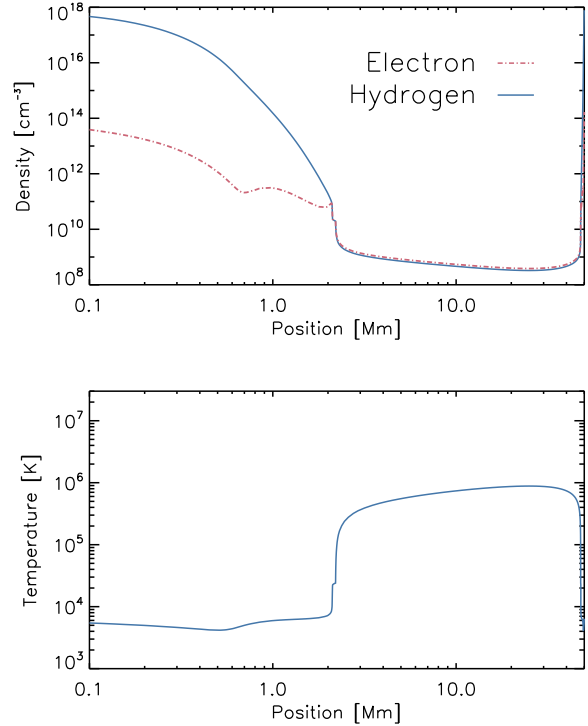}
\end{minipage}
\caption{The hydrostatic density and temperature profiles for a loop of length $2L = 50$\,Mm.  The temperatures are assumed to be initially equilibrated.  Note that this figure (and others in this work) show the full loop, but the x-axis is shown on a logarithmic scale to emphasize the chromosphere.  The electron density in the corona is slightly higher than the hydrogen density due to trace elements, while in the chromosphere it is small due to the large fraction of neutral atoms.}
\label{fig:initial}
\end{figure}

To begin, we calculate the initial conditions for a given coronal loop.  Using the VAL C chromospheric temperature profile \citep{vernazza1981}, we generate a chromospheric density profile by solving the hydrostatic equations, with the above prescription to solve for the electron density along the full loop from footpoint to footpoint, which is assumed to be symmetric in this work in both the geometry and heating.  In Figure \ref{fig:initial}, we show the temperature and density profiles of a loop with length $2L = 50$\,Mm.  The electron and hydrogen temperatures are assumed to be initially equilibrated.  In the corona, the electron density is higher than the hydrogen density due to the electron contribution from trace elements.  In the chromosphere, where the ionization fractions are much lower, the electron density is orders of magnitude lower than the hydrogen density.
  
\begin{figure*}
\begin{minipage}[b]{\linewidth}
\centering
\includegraphics[width=\textwidth]{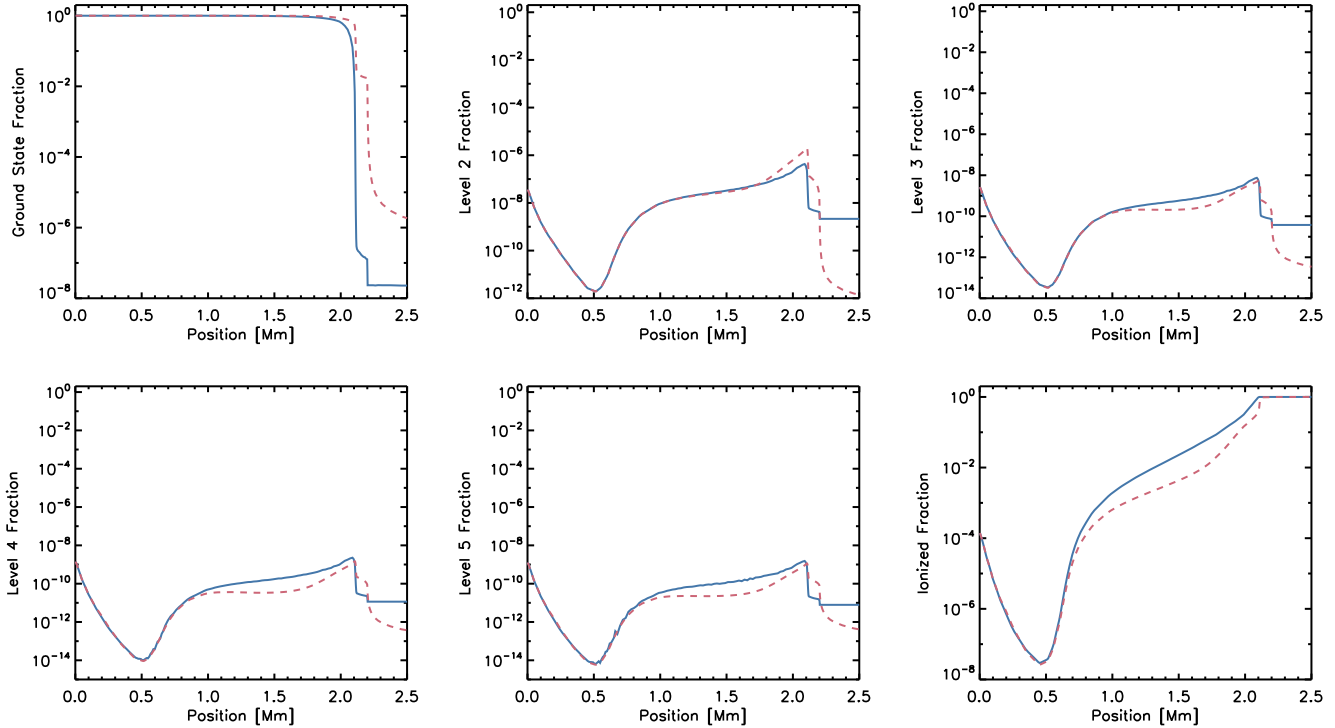}
\end{minipage}
\caption{The hydrostatic hydrogen level populations as calculated by the HYDRAD solver (blue solid), following the description in Section \ref{sec:implementation}, and by RH1.5D (red dashed).  The two methods show good agreement deep in the chromosphere, while the ionized fraction is over-estimated in the upper chromosphere.}
\label{fig:initial_levels}
\end{figure*}

The level populations for the 6-level hydrogen atom in the chromosphere of one of the footpoints are shown in Figure \ref{fig:initial_levels}.  We show the levels as calculated both by the HYDRAD solver (using the \citealt{sollum1999} method), shown as solid lines, as well as the level populations calculated by the RH1.5D code \citep{pereira2015}, shown as dashed lines.  Deep in the chromosphere and nearing the photosphere, the two methods agree almost exactly because the plasma is in collisionally-dominated LTE.  In the chromosphere, however, the radiation field is more significant.  HYDRAD  predicts a higher ionized fraction near the top of the chromosphere, and thus a lower neutral fraction, which in turn means that the electron density is likely over-estimated in the upper chromosphere.  In the corona, the plasma is essentially fully ionized in both cases, though the methods disagree. 

\subsection{Hydrodynamics}
\label{subsec:hydro}

\begin{figure*}
\begin{minipage}[b]{0.5\linewidth}
\centering
\includegraphics[width=\textwidth]{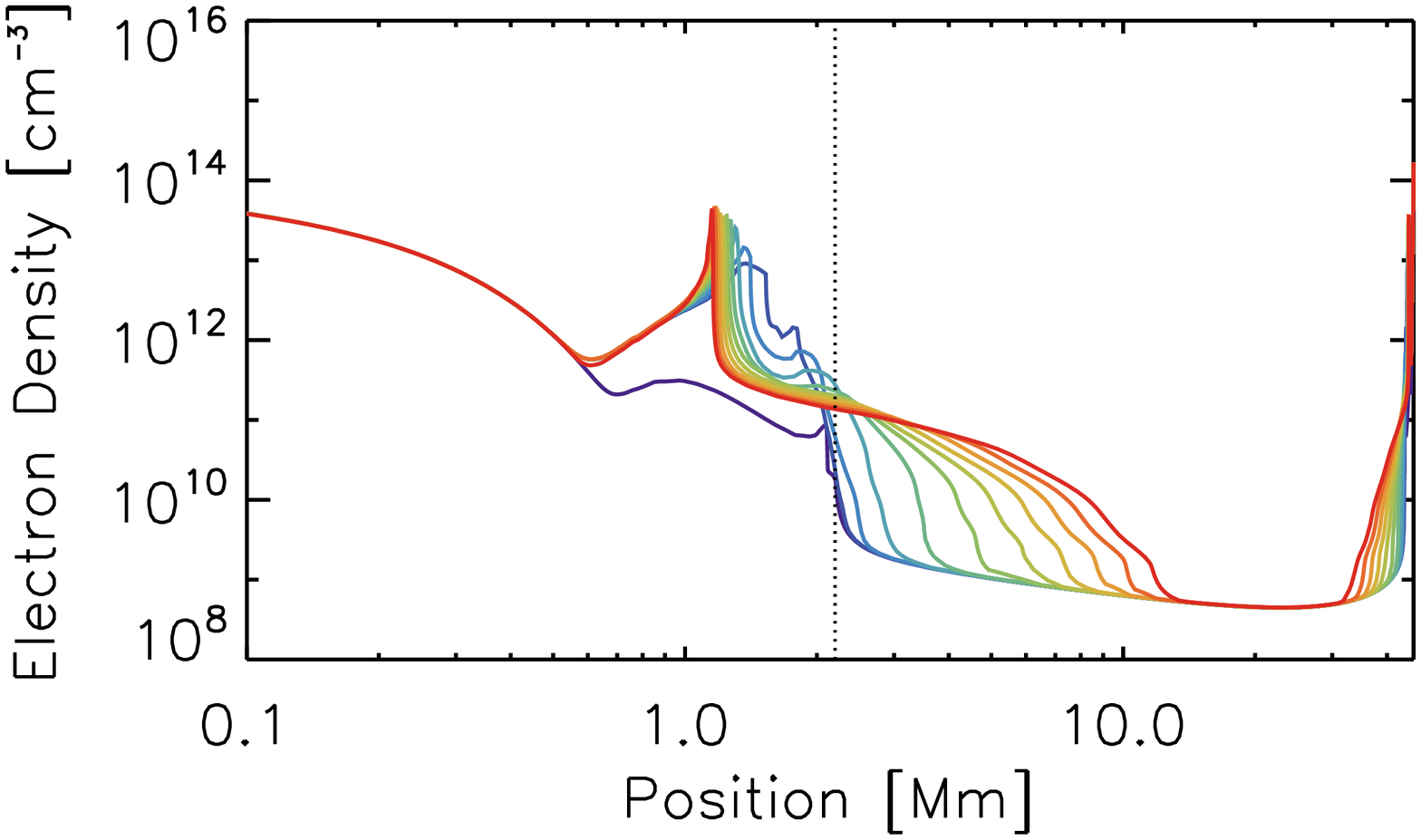}
\end{minipage}
\begin{minipage}[b]{0.5\linewidth}
\centering
\includegraphics[width=\textwidth]{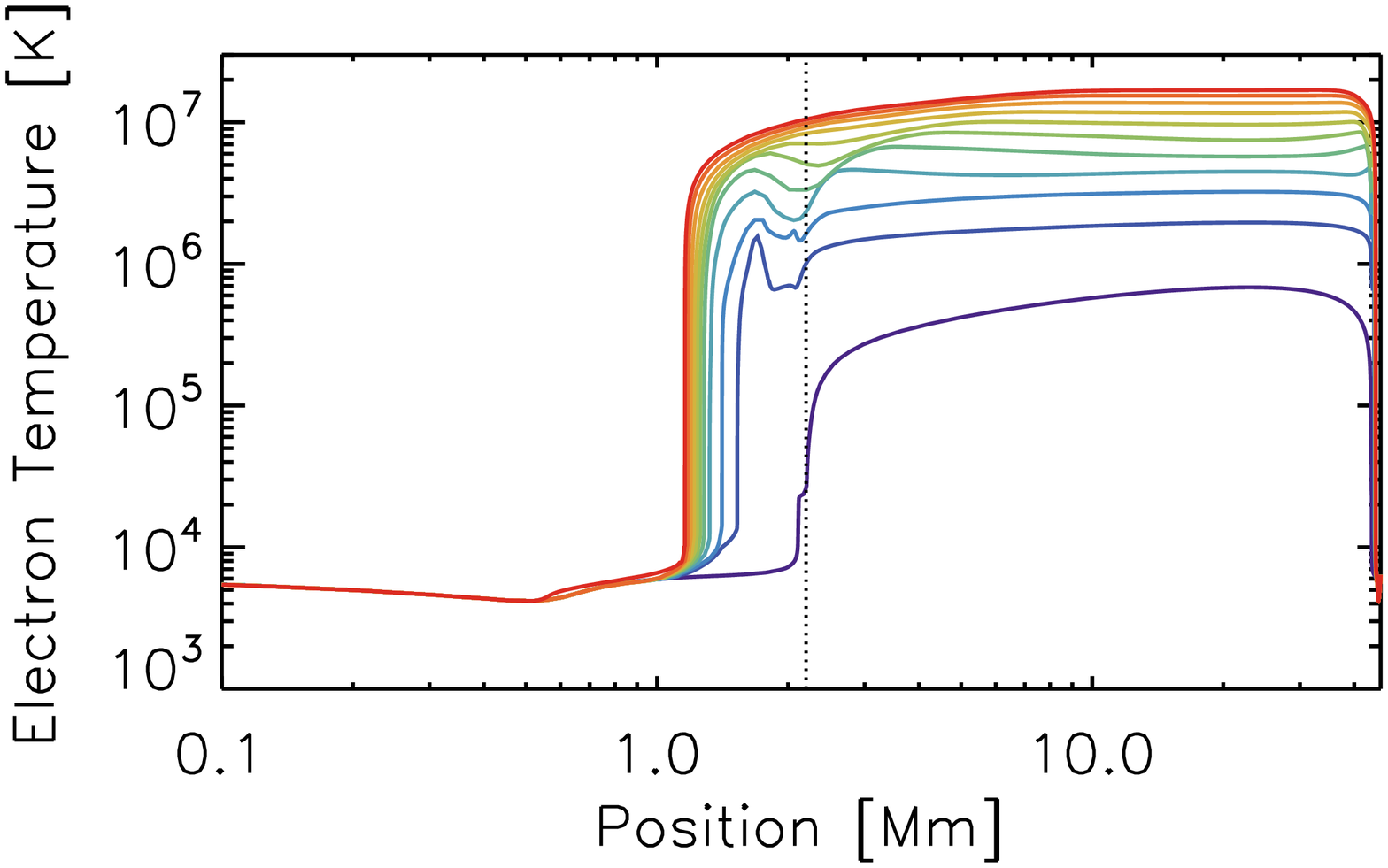}
\end{minipage}
\begin{minipage}[b]{0.5\linewidth}
\centering
\includegraphics[width=\textwidth]{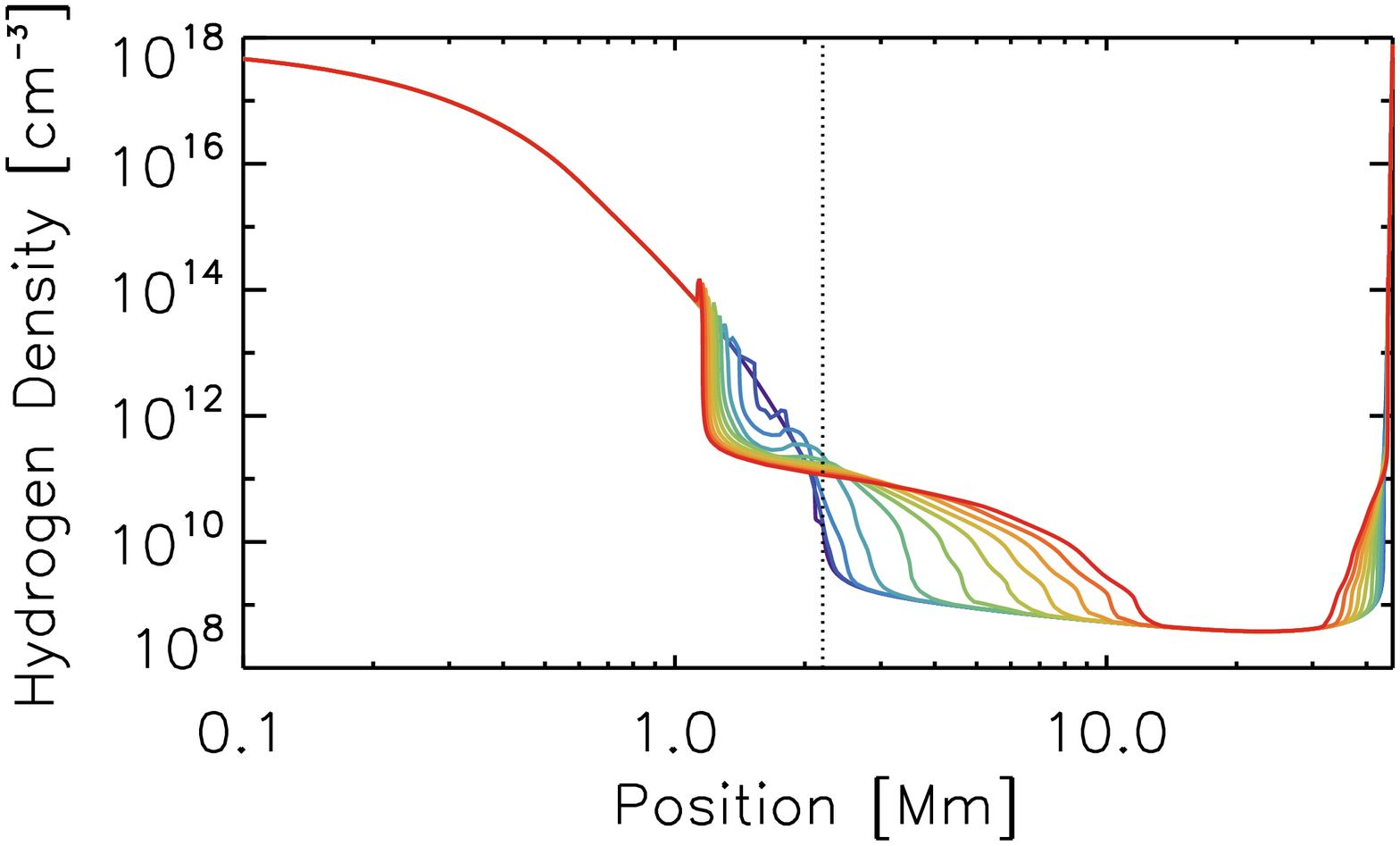}
\end{minipage}
\begin{minipage}[b]{0.5\linewidth}
\centering
\includegraphics[width=\textwidth]{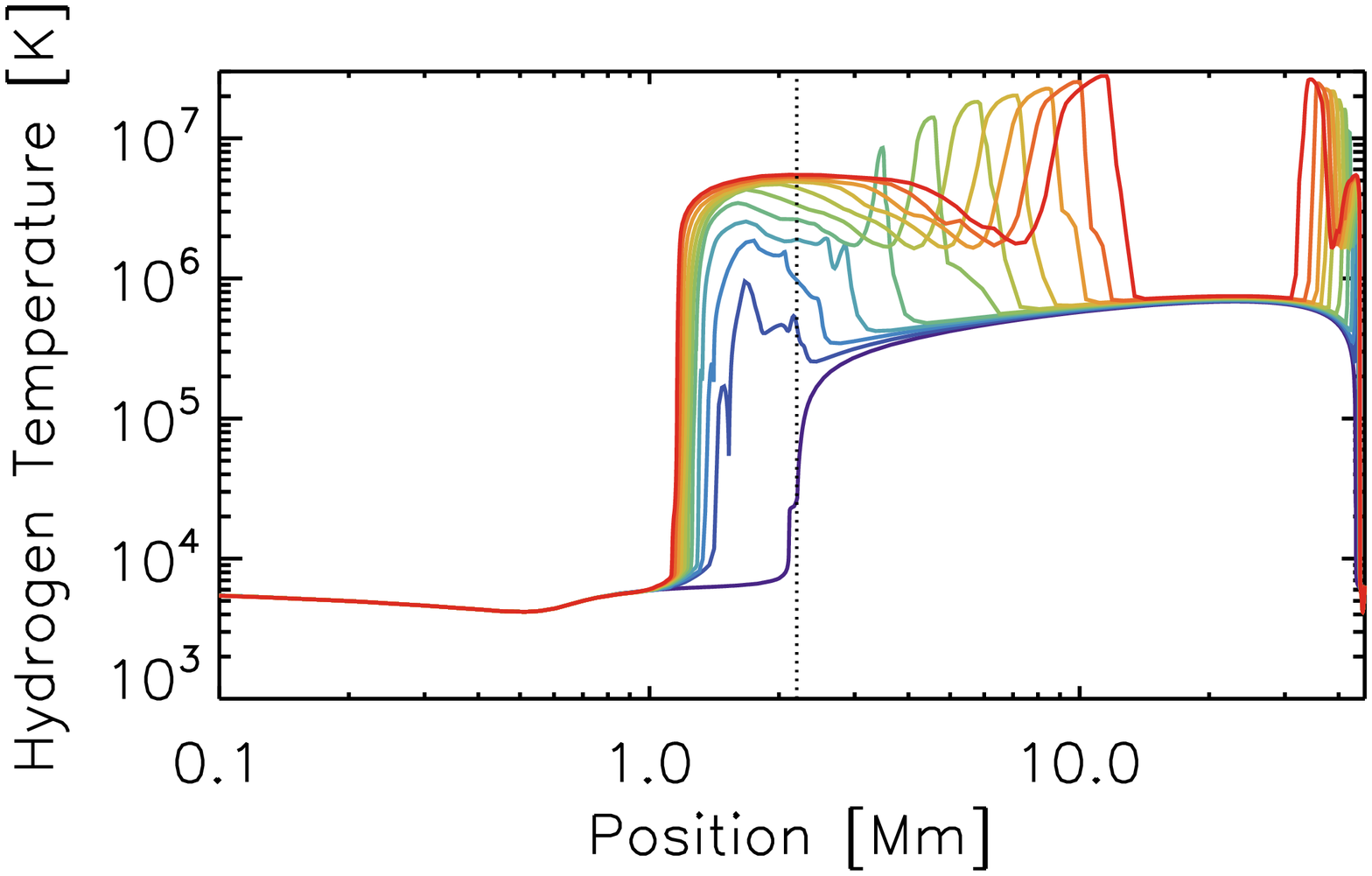}
\end{minipage}
\begin{minipage}[b]{0.5\linewidth}
\centering
\includegraphics[width=\textwidth]{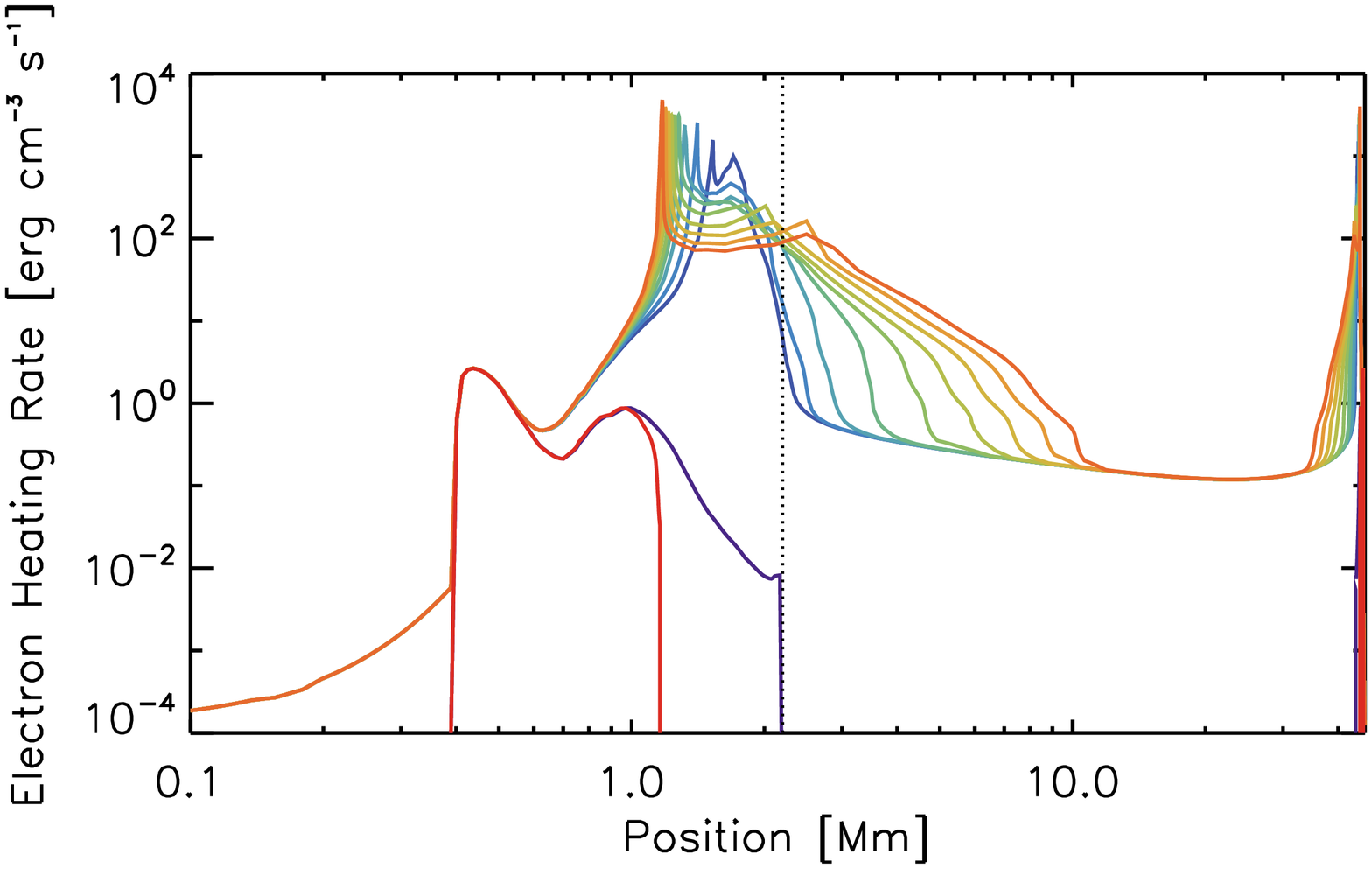}
\end{minipage}
\begin{minipage}[b]{0.5\linewidth}
\centering
\includegraphics[width=\textwidth]{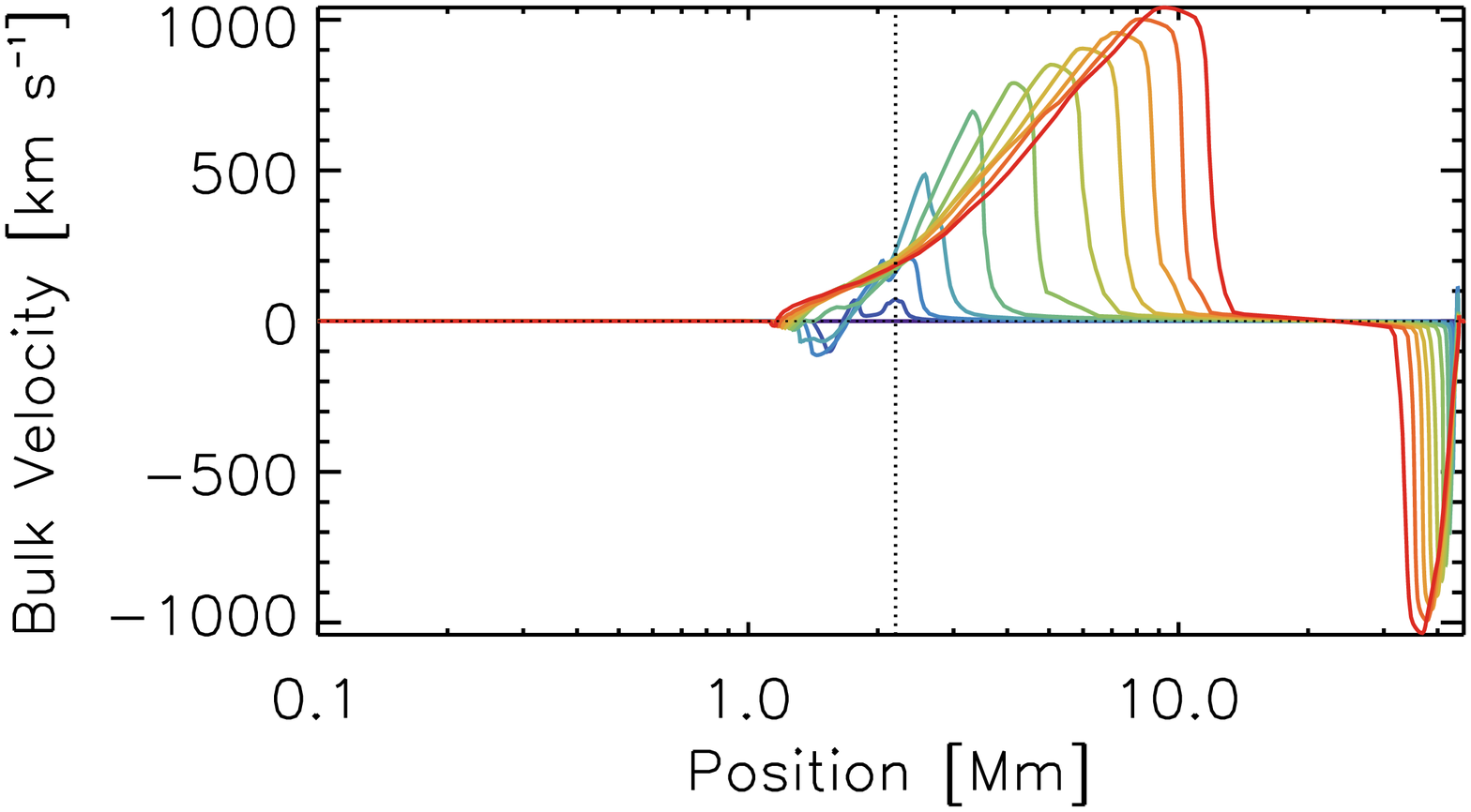}
\end{minipage}
\caption{The hydrodynamics of a loop heated for 10 seconds by an electron beam with $F_{0} = 3 \times 10^{10}$\,erg\,s$^{-1}$\,cm$^{-2}$, sharp low energy cut-off $E_{c} = 15$\,keV, and spectral index $\delta = 5$.  The electron and hydrogen temperatures and densities are shown, along with the rate of energy deposition by the beam (plus the background heating term), and the bulk flow velocity.  All plots show a 1 second cadence, from purple to red.  The dotted black line marks the initial transition region, and velocities traveling to the right are defined as positive.  The heating quickly causes a sharp ionization in the chromosphere, and the increased pressure drives a strong, explosive evaporation event.  Movies of each individual plot are available in the electronic version of the manuscript, showing the first 25 seconds of the simulation.}
\label{fig:hydro}
\end{figure*}

To understand how the parameters vary over time, we now run a dynamic simulation.  We impose heating due to an electron beam \citep{reep2013,reep2016a}, assuming a constant energy flux $F_{0} = 3 \times 10^{10}$\,erg\,s$^{-1}$\,cm$^{-2}$, sharp low energy cut-off $E_{c} = 15$\,keV, and spectral index $\delta = 5$, for a total of 10 seconds.  Figure \ref{fig:hydro} shows the hydrodynamics of that simulation, from top: the electron density, electron temperature, hydrogen density, hydrogen temperature, electron heating rate (including the background heating term), and the bulk flow velocity of the plasma.  All are shown at a 1 second cadence, from purple through red, and the black dotted line marks the initial transition region.  The strong heating event quickly raises the electron temperature at the top of the chromosphere and in the corona (to about 20 MK), which causes the plasma to strongly ionize, liberating electrons from the mostly neutral plasma and increasing the electron density.  The increased pressure in the chromosphere quickly drives a strong and explosive evaporation event, carrying significant amounts of plasma into the corona, as well as driving a slower condensation deeper into the chromosphere.  The hydrogen temperature slowly equilibrates through collisions between the hydrogen and electrons.  

\begin{figure*}
\begin{minipage}[b]{\linewidth}
\centering
\includegraphics[width=\textwidth]{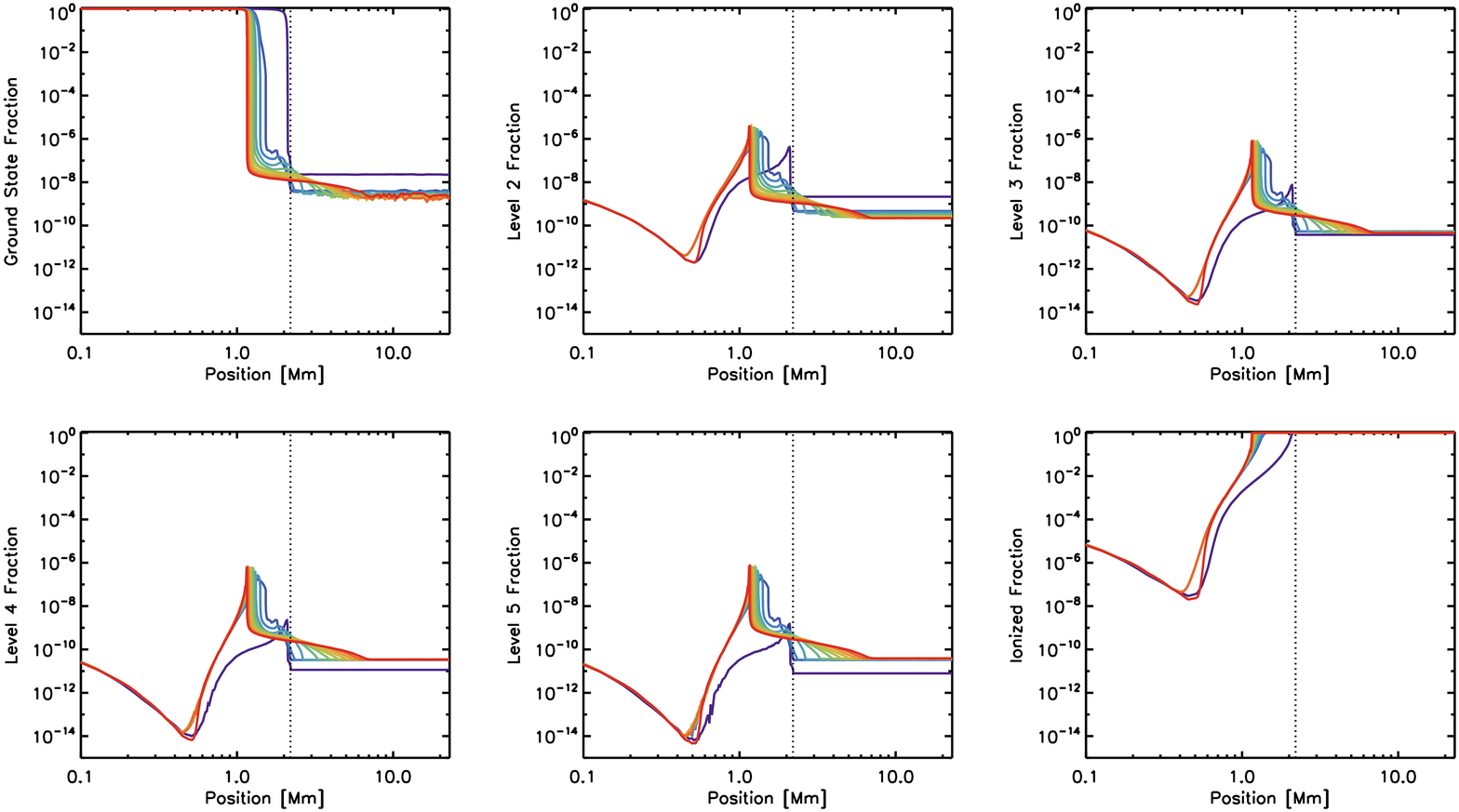}
\end{minipage}
\caption{The evolution of the fractional level populations of hydrogen for the duration of heating during an electron beam heating simulation, shown at 1 second cadence from purple to red.  The dotted black line marks the initial transition region.  The chromosphere quickly becomes fully ionized, reducing the fraction of neutral hydrogen in the ground state significantly, while also exciting the upper levels.  Advection carries neutral hydrogen from the first 5 levels into the corona, though it remains strongly ionized.  A movie of this figure is available in the electronic version of the manuscript, showing the first 25 seconds of the simulation.}
\label{fig:dynamic_levels}
\end{figure*}

For that same simulation, in Figure \ref{fig:dynamic_levels}, we show the evolution of each of the six level populations with time during the heating period, at a 1 second cadence from purple to red.  The heating to the beam quickly raises the temperature of the chromosphere, which causes a sharp rise in the ionized fraction there.  Increased collisional excitation due to the increased electron density also causes sharp spikes in the higher level states to form at the same location.  As chromospheric evaporation begins to carry material into the corona, neutral hydrogen from all levels is advected into the corona, though the hydrogen remains nearly fully ionized.  

\begin{figure*}
\begin{minipage}[b]{\linewidth}
\centering
\includegraphics[width=\textwidth]{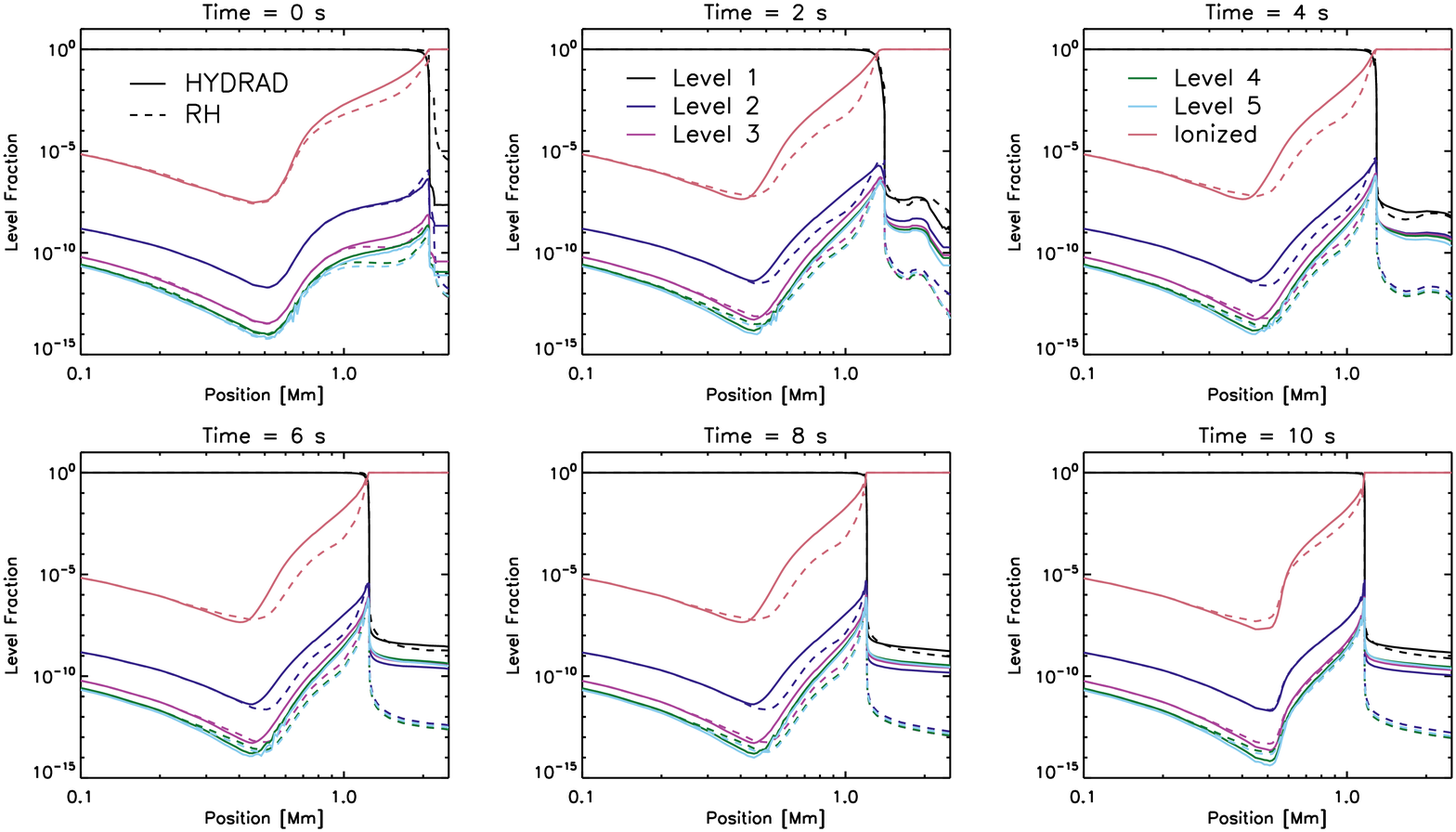}
\end{minipage}
\caption{A comparison of the hydrogen level populations as computed by the method outlined in Section \ref{sec:implementation} (solid) and by RH1.5D (dashed).  The two methods disagree in the upper chromosphere during the heating period, primarily due to the effect of non-thermal ionization and excitation, which RH1.5D does not account for.  The two methods also disagree in the corona, though the change in the resultant electron density is negligible.  Please note that RH1.5D uses the output from HYDRAD as input and assumes a static atmosphere, which is not a self-consistent comparison.  In Appendix \ref{app:radyn}, we therefore include a detailed comparison to another numerical model.}
\label{fig:rh_comp}
\end{figure*}

A natural question is to ask how the approximation compares to a more detailed calculation of the levels.  We have therefore calculated the level populations using a full radiative transfer solution via RH1.5D.  In Figure \ref{fig:rh_comp}, we show a comparison of the level populations at times 0, 2, 4, 6, 8, 10\,s into the simulation as calculated both using the method in Section \ref{sec:implementation} and by RH1.5D itself (which will be used to forward model emissions).  We show the first 2.5\,Mm of the loop (the chromosphere to bottom of the corona), with all six levels.  At time 0, the two methods agree in the deep chromosphere ($\lesssim 1.0$\,Mm), while they disagree in the upper chromosphere by a factor of up to 5 or so (most notably in the ionized fraction).  At later times, the methods diverge in the upper chromosphere, primarily because of the effect of non-thermal collisional ionization and excitation, which drastically alters the populations from their static solutions (which RH1.5D assumes).  When the heating ceases, at time 10\,s, the non-thermal effect also ceases and the two are found to be in reasonable agreement at all heights.  One other important difference to note is that RH1.5D assumes a static atmosphere, meaning that time and advective gradients are not included in the calculations, though the electron densities in the simulation reflect those effects.  There are important differences here that should be kept in mind while examining the results of forward modeling, therefore.  

\section{Forward Modeling}
\label{sec:forward}

The true test of any model is its ability to reproduce observations, so we therefore turn to forward modeling spectral data seen with IRIS.  We consider the 2014 November 19 UT14:14 flare, which was the focus of \citet{warren2016}, and attempt to model emission from \ion{O}{1}, \ion{C}{2}, and \ion{Mg}{2}.  In \citet{reep2016b}, it was found that while \ion{Si}{4} emission could be accurately reproduced with optically thin calculations, the observed strong stationary component of \ion{C}{2} was not found in the model.  \ion{Mg}{2} was found to have bursts in intensity and was weakly red-shifted during the heating period, though the modeling did not attempt to reproduce this line.  

\begin{figure*}[t!]
  \centerline{\includegraphics[angle=90,width=\linewidth]{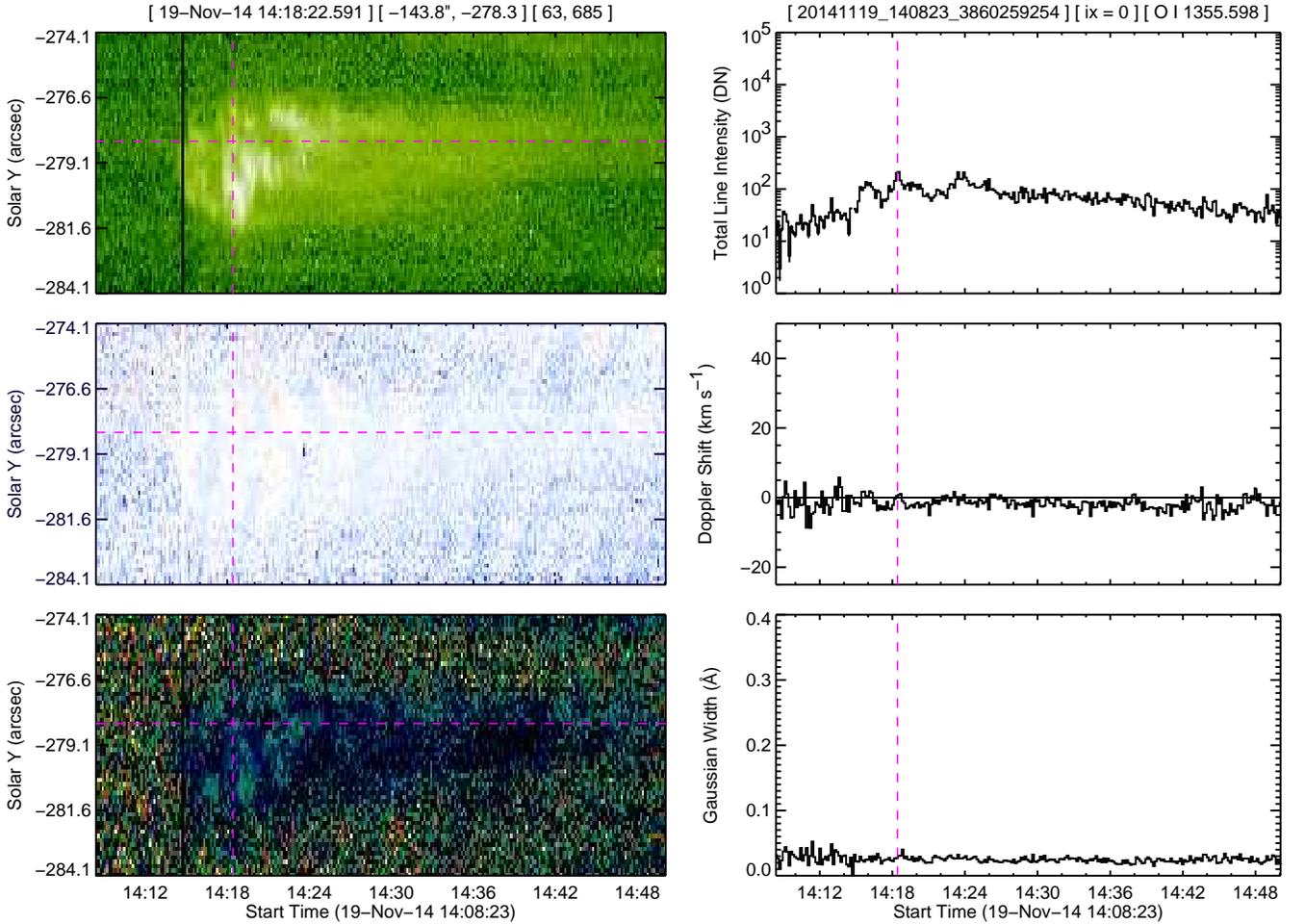}}
\caption{Intensities, Doppler shifts, and Gaussian line widths determined from moments of the \ion{O}{1} 1355.598\,\AA\ line.  The line remains close to stationary at all times relative to the background, with small variations up to about 5\,km\,s$^{-1}$.  The intensity rises by about a factor of 10 up 100--200\,DN, while the line width remains essentially constant.  The left panels show these parameters along the slit as a function of time, while the right panels show the values at one given pixel (marked by the dashed pink line).  Note that the range of line widths displayed here is much narrower than those displayed in Warren et al. (2016).}
\label{fig:oxygen}
\end{figure*}

\citet{warren2016} observed the \ion{O}{1} 1355.598\,\AA\ line, but it was not shown explicitly in the paper so we present it here before synthesizing the line with a forward model.  Figure \ref{fig:oxygen} shows the intensity, Doppler shift, and Gaussian line width, calculated with the moments of the line (as explained in \citealt{warren2016}).  On the left, these quantities are shown as a function of time along the slit, and on the right, at a single pixel marked by the dashed pink lines.  During the event, the line brightens, but remains essentially stationary relative to the background.  There is also no noticeable broadening associated with the brightening.  

Attempts to reproduce this line using optically thin assumptions (\textit{i.e.} the method in \citealt{reep2016b}) result in a line that is consistently red-shifted during the heating period, like \ion{Si}{4} and \ion{C}{2}.  One possible explanation of this is that we have disregarded NLTE and/or opacity effects by using optically thin assumptions.  Therefore, in order to attempt to explain this line, we follow the basic methodology of \citet{reep2016b}: we run many simulations to create a multi-threaded model, from which we calculate light curves and line profiles.  Instead of synthesizing the line with optically thin assumptions, however, we use RH1.5D \citep{pereira2015}, which solves a full radiative transfer calculation at a given time snapshot, including important NLTE and opacity effects.  Though \ion{O}{1} is our primary focus, we also synthesize \ion{C}{2} and \ion{Mg}{2} emission to contrast with observations.

We have therefore run hydrodynamic simulations with HYDRAD of loops subjected to heating by electron beams, following the parameter space of \citet{reep2016b}.  We use energy fluxes ranging from $10^{8}$ to $10^{11}$\,erg\,s$^{-1}$\,cm$^{-2}$, and the RHESSI-derived low energy cut-off $E_{c} = 11$\,keV and spectral index $\delta = 6$ (see \citealt{reep2016b} for the RHESSI data).  We arbitrarily assume a heating duration lasting 10\,s following that paper in order to facilitate the comparisons, but in the more recent paper \citet{reep2018b}, it was found that 10\,s is insufficient to reproduce \ion{Fe}{21} Doppler shift observations seen in much larger flares, which seem to require durations averaging between 50 to 100\,s.  

On all of these simulations, we have run RH1.5D to calculate each line profile for \ion{O}{1} 1355.598\,\AA, \ion{C}{2} 1334.535\,\AA, and \ion{Mg}{2} 2796.354\,\AA.  \ion{Mg}{2} is calculated with partial redistribution (PRD), which is particularly important for this transition \citep{leenaarts2013}, using the `PRD\_ANGLE\_APPROX' scheme in RH1.5D \citep{leenaarts2012}.  We treat the other lines with complete redistribution (CRD).  We ran RH1.5D with five atoms: H, O, C, and Mg as active, and He as passive (\textit{i.e.} only used as a source of background opacity).  We truncated the loop at the size of an IRIS pixel, approximately 240\,km on the sun ($\approx 2.6$\,Mm in field-aligned coordinate with a 46\,Mm loop).

Following \citet{warren2016}, we subtract the continuum near each line, and then calculate the moments of each line.  Because RH1.5D produces intensities in absolute units, we convert to DN by convolving the output with the IRIS response, obtained from the SolarSoftWare IDL routine ``iris\_get\_response.''  In this way, the intensities are directly comparable to the observed values.  All wavelengths are listed in vacuum wavelengths.\footnote{\ion{O}{1} has a rest wavelength of 1355.598\,\AA\ according to both CHIANTI and NIST databases, but the model atom in RH1.5D produces a rest wavelength of 1355.63\,\AA, which amounts to $\approx 7$\,km\,s$^{-1}$ difference.  This has been corrected for in this work.}  For these lines, we assume a 3\,km\,s$^{-1}$ microturbulence value.

\begin{figure*}
  \begin{minipage}[b]{0.31\linewidth}
    \centering
    \includegraphics[width=\linewidth]{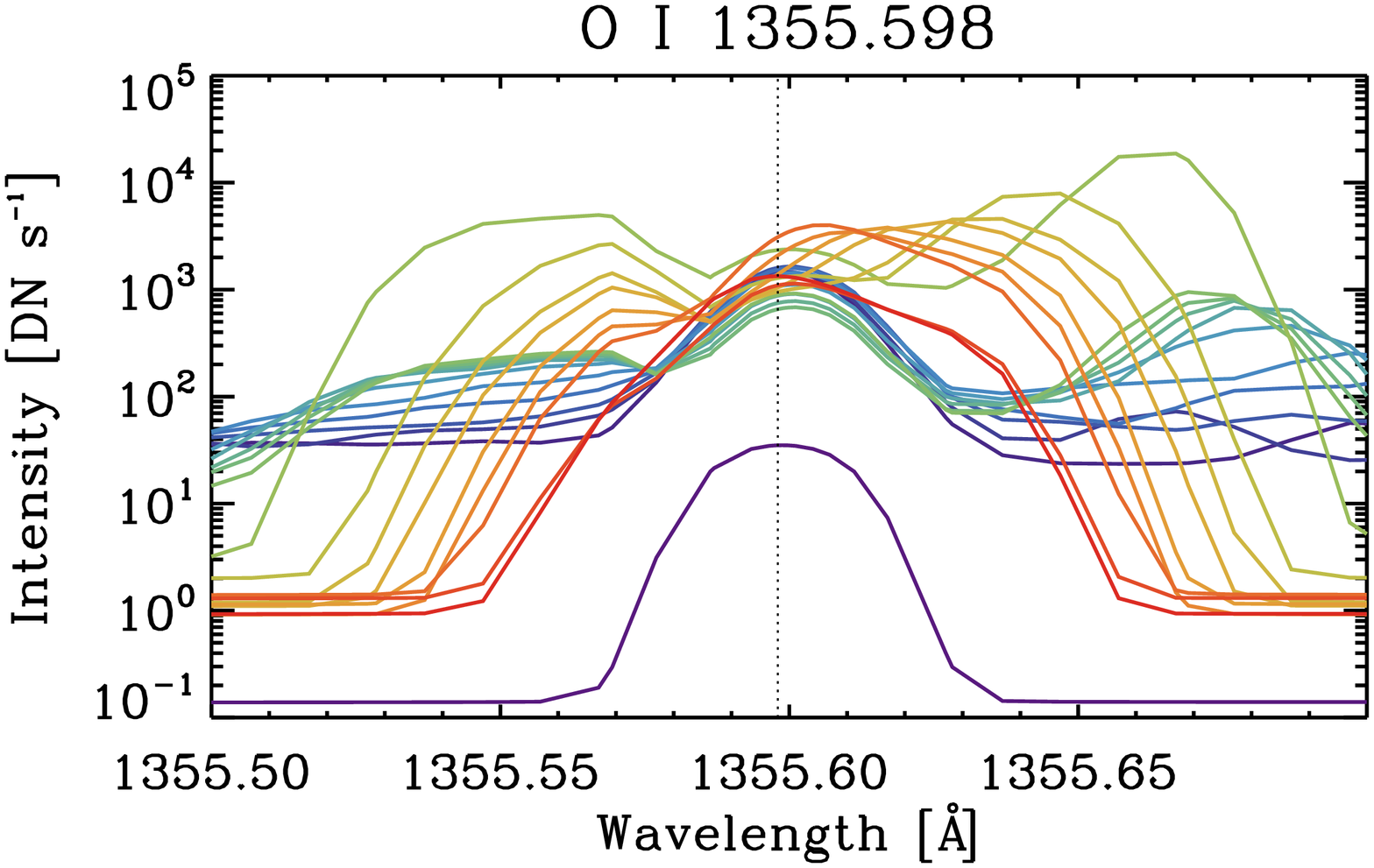}
    \includegraphics[width=\linewidth]{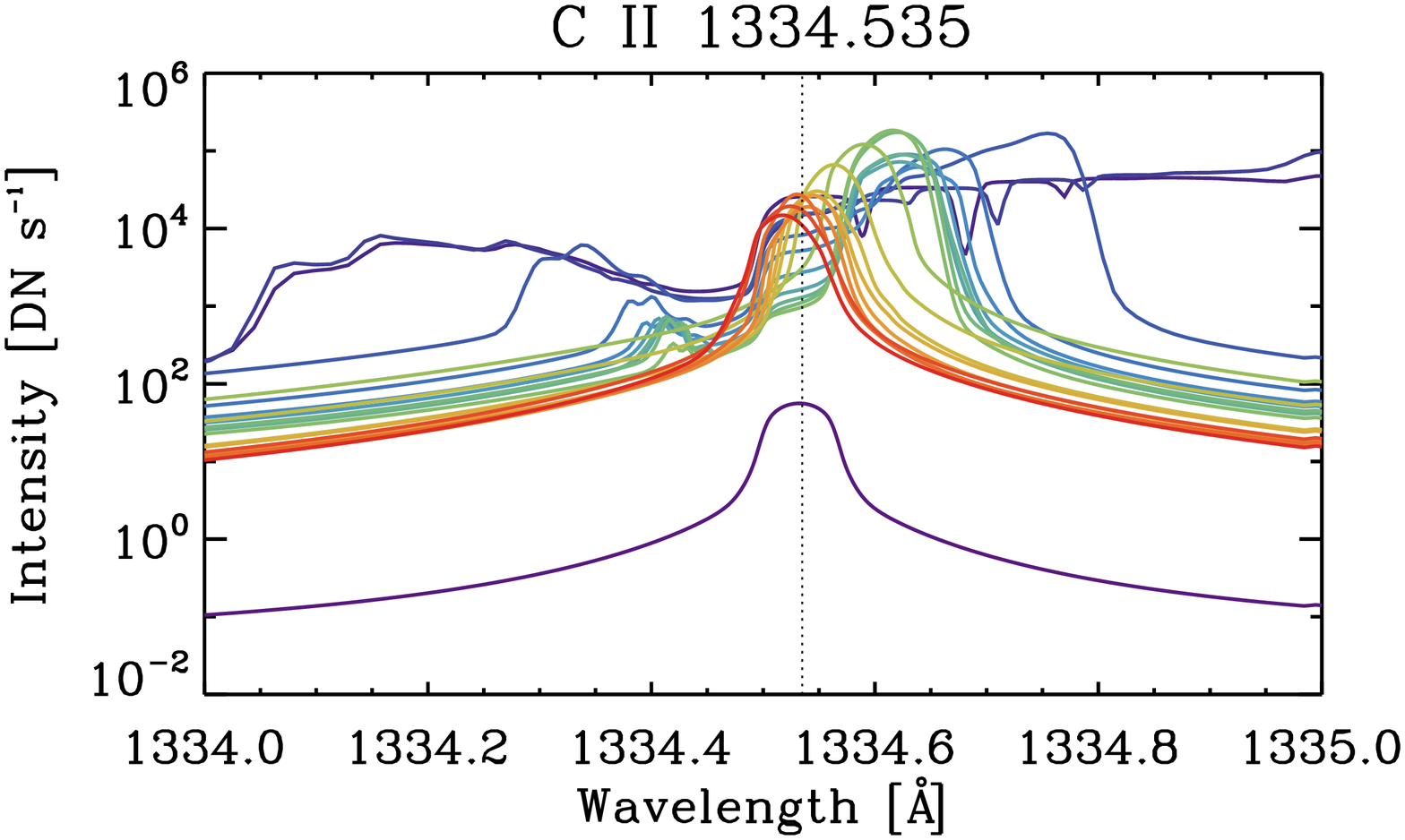}
  \end{minipage}
  \begin{minipage}[b]{0.31\linewidth}
    \centering
    \includegraphics[width=\linewidth]{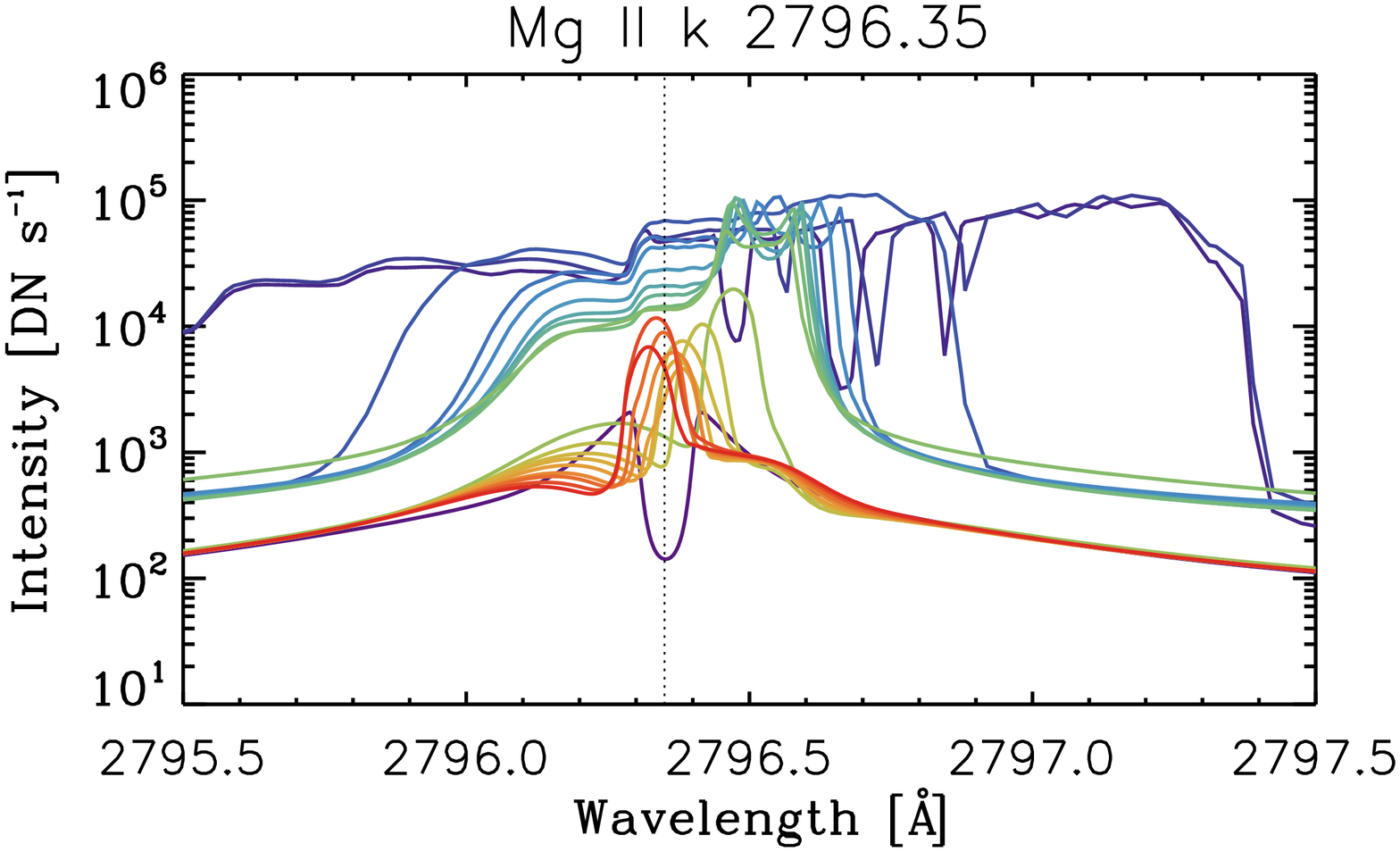}
    \includegraphics[width=\linewidth]{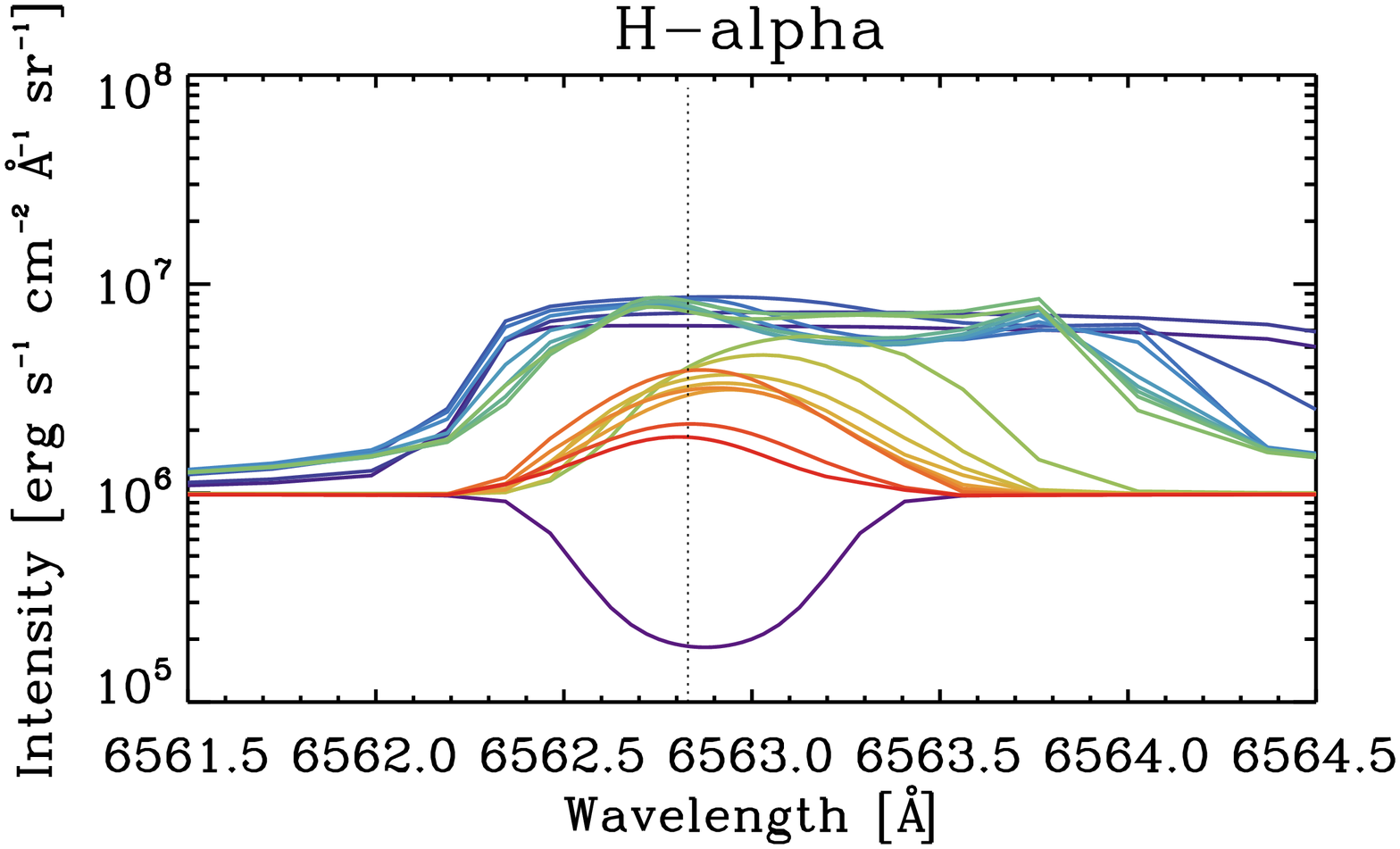}
  \end{minipage}
    \begin{minipage}[b]{0.38\linewidth}
    \centering
    \includegraphics[width=\linewidth]{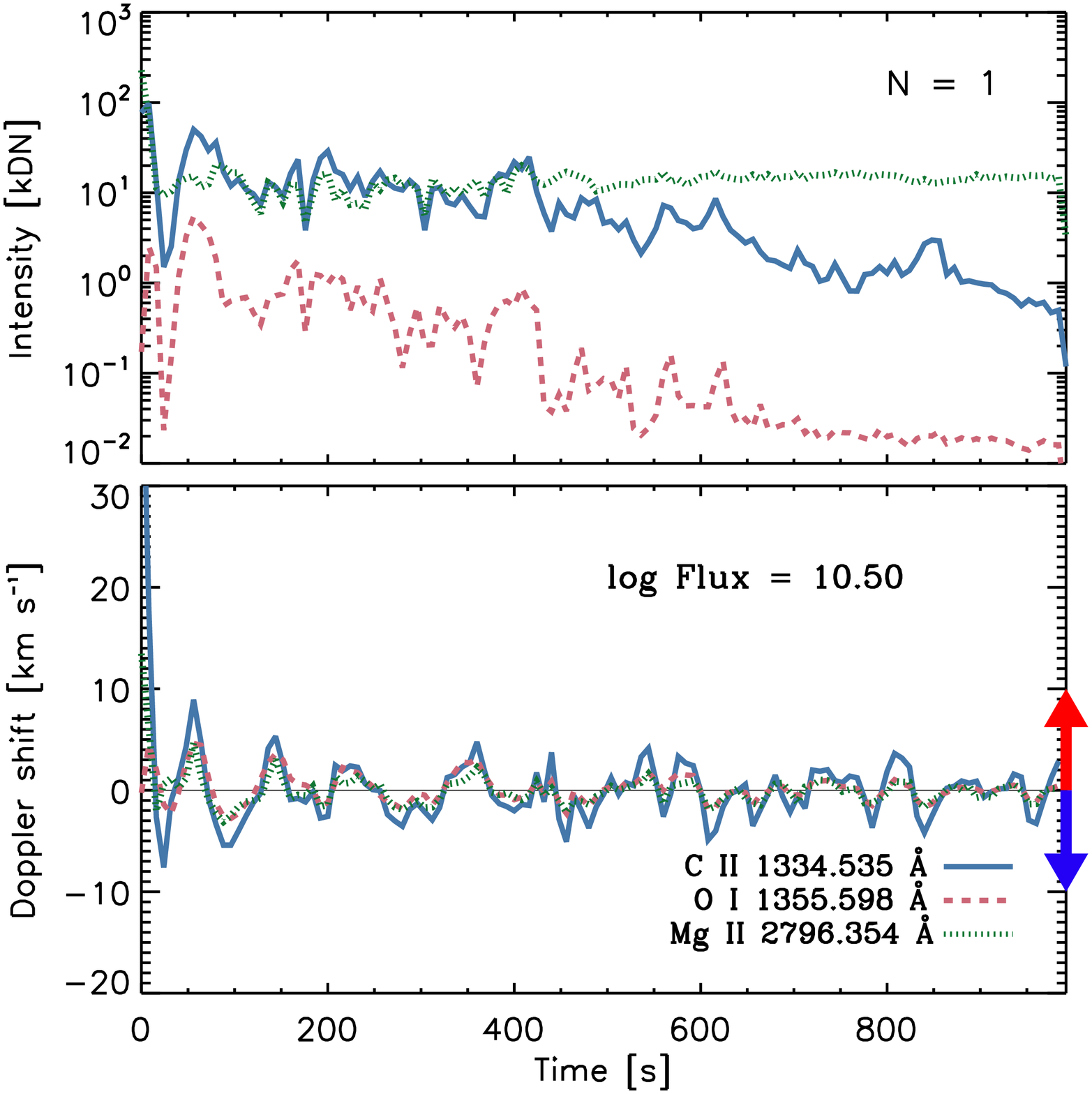}
  \end{minipage}
\caption{Light curves and line profiles for the single thread simulation shown in Figure \ref{fig:hydro}.  At left, the line profiles (\ion{O}{1}, \ion{C}{2}, \ion{Mg}{2}, and $H\alpha$) for the first 20 seconds of the simulation, shown as different colors at a 1 second cadence, ranging from violet through red.  At right, the light curves and Doppler shifts calculated from the moments of each line.  These assume a filling factor of 1.  This simulation is inconsistent with the observations because there is no persistent red-shift in \ion{C}{2}, there is no gradual decay of red-shift in \ion{Mg}{2}, and the intensity of \ion{O}{1} is too large.}
\label{fig:lines}
\end{figure*}

As an example, consider the simulation in the previous section, with energy flux $F_{0} = 3 \times 10^{10}$\,erg\,s$^{-1}$\,cm$^{-2}$, whose hydrodynamics were shown in Figure \ref{fig:hydro}.  We calculate the line profiles and light curves for each of the three lines, shown in Figure \ref{fig:lines}.  The line profiles (\ion{O}{1}, \ion{C}{2}, \ion{Mg}{2}, and $H\alpha$) are shown at a 1 second cadence for the first 20 seconds of the simulation.  Each brightens significantly during the heating period, and begins to dim as cooling begins.  Initially, all four show a red-shift due to a strong red-wing component (up to $\approx 100$\,km\,s$^{-1}$), which most strongly affects the \ion{C}{2} line.  There are two obvious discrepancies with the observational data, however: the \ion{C}{2} red-shifts are short-lived ($\approx 30$\,s), and the \ion{O}{1} line is much brighter (and wider) than observed.  RH1.5D assumes statistical equilibrium and disregards the level populations from HYDRAD, which means that the O I populations could be affected since they are sensitive to charge exchange \citep{lin2015}.  That the \ion{C}{2} red-shifts are so short-lived reiterates a major fault of single loop models, which predict only short-lived chromospheric condensations.  $H\alpha$ is shown in absolute units since it is not observed by IRIS, but its development can be compared to \textit{e.g.} \citet{kuridze2015}.  The relative brightenings are large compared to that work, the line width is too narrow (see \citealt{kowalski2017b} for the likely explanation), and there is no apparent central reversal (compare \citealt{rubiodacosta2015}).    

In order to improve the results, we appeal to the multithreaded model.  We have written an IDL routine that creates a multithreaded line profile as a function of time that can be used to then calculate the light curves and Doppler shifts.  As in \citet{reep2016b} and \citet{reep2018b}, we select the total number of threads $N$ and the average waiting time between new threads $r$ (using a Poisson distribution).  We select energy fluxes from a power law, with index of the energy distribution $\alpha$, and the minimum energy flux on that energy distribution $F_{\text{min}}$.  We choose $N \times r = 600$\,s in all cases, which is the approximate duration of the hard X-ray burst.  The low energy cut-off and spectral index of the injected electron distribution were derived from RHESSI data (shown in \citealt{reep2016b}).  Finally, because we assume that all threads are rooted within one pixel, we divide the total intensity by $N$, which equivalently says that each has a cross-sectional area equal to the pixel area divided by $N$.  We wish to stress that because there are multiple random variables, the results can change even using the same parameters, but the trends remain essentially unchanged.

\begin{figure*}
  \begin{minipage}[b]{0.32\linewidth}
    \centering
    \includegraphics[width=2.2in]{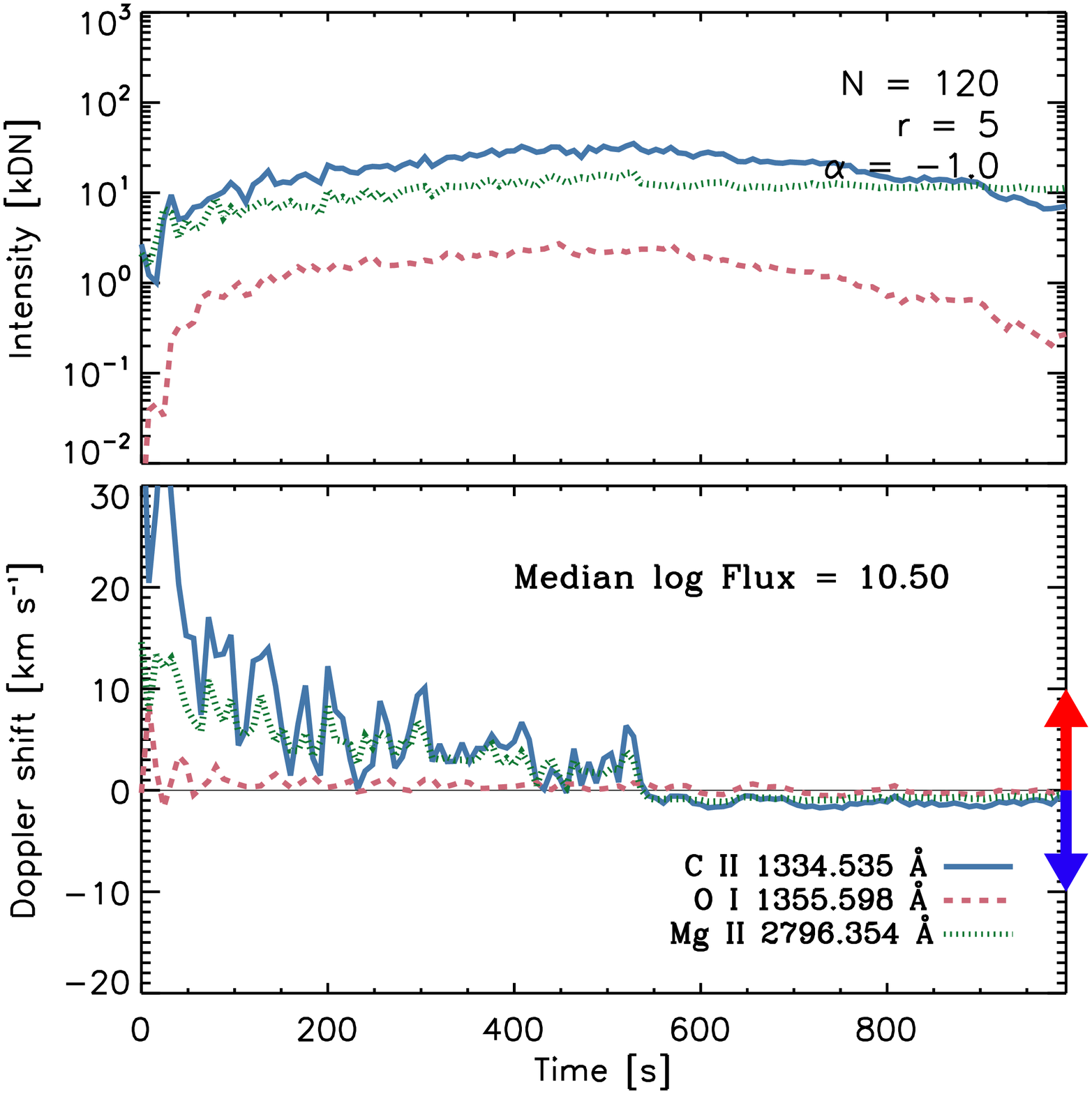}
  \end{minipage}
  \begin{minipage}[b]{0.32\linewidth}
    \centering
    \includegraphics[width=2.2in]{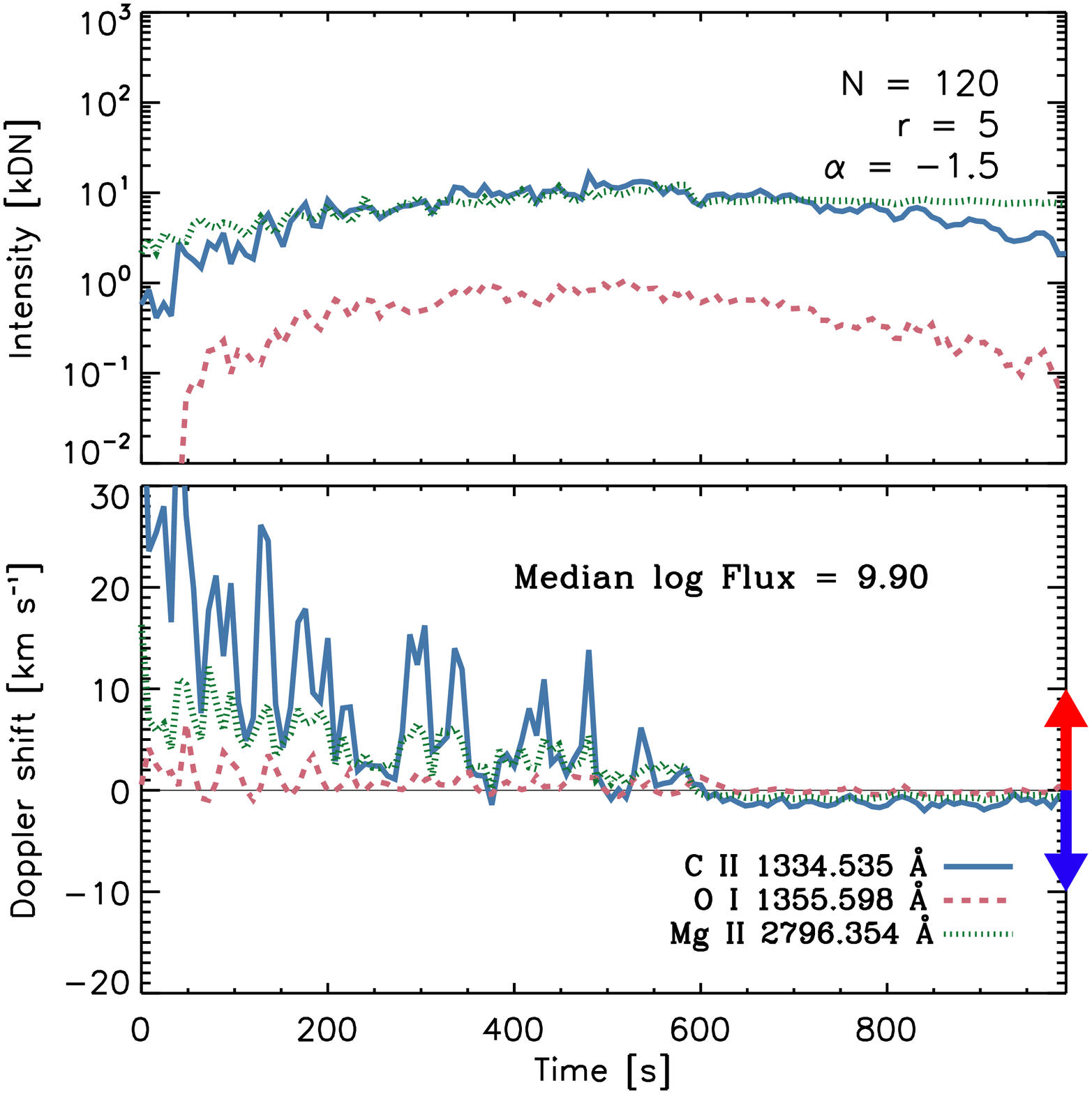}
  \end{minipage}
    \begin{minipage}[b]{0.32\linewidth}
    \centering
    \includegraphics[width=2.2in]{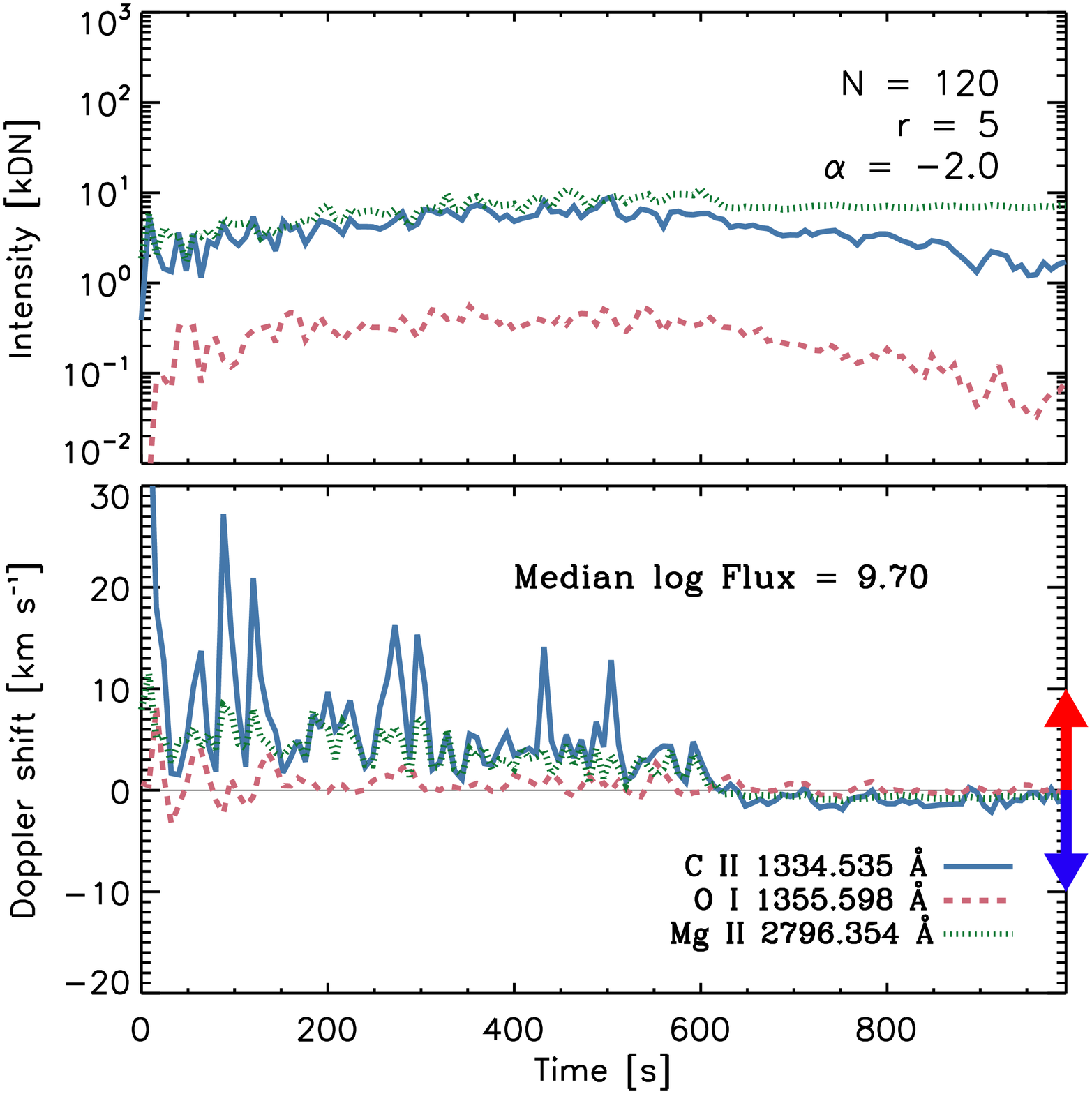}
  \end{minipage}
\caption{Multithreaded light curves and Doppler shifts for the three spectral lines under consideration, synthesized by RH1.5D, using $N = 120$ threads, $r = 5$\,s per thread on average, and $F_{min} = 3 \times 10^{9}$\,erg\,s$^{-1}$\,cm$^{-2}$.  From left to right, each uses a spectral index of the power law distribution of energies $\alpha = -1, -1.5, -2$.  Red-shifts are defined as positive.  The Doppler shifts are in general agreement with the observations: \ion{O}{1} is essentially stationary in all cases, while \ion{C}{2} is strongly red-shifted during the heating period, and \ion{Mg}{2} weakly red-shifted.  The intensities of \ion{C}{2} and \ion{Mg}{2} are roughly consistent with the observed values, but the intensity of \ion{O}{1} is larger than observed.}
\label{fig:bestfit}
\end{figure*}

We begin with parameters that were deemed a good fit in \citet{reep2016b}: $N = 120$ threads, $r = 5$\,s per thread, and $F_{\text{min}} = 3 \times 10^{9}$\,erg\,s$^{-1}$\,cm$^{-2}$.  Figure \ref{fig:bestfit} shows three cases with spectral indices $\alpha = -1, -1.5, -2$.  In all three cases, \ion{O}{1} is stationary, with only small bursts of red-shifts (2 -- 3\,km\,s$^{-1}$), but its intensity is too bright by about a factor of 10.  \ion{C}{2} shows strong red-shifts that begin with the onset of heating, up to 30\,km\,s$^{-1}$, gradually decaying in magnitude, until the red-shifts finally cease after the heating period.  Its intensity grows smoothly, reaching peaks of between 10--100\,kDN, slightly lower than the observed peak of 100\,kDN (though the average value was closer to 10\,kDN).  \ion{Mg}{2} behaves similarly, forming a strong initial red-shift ($\approx 10$\,km\,s$^{-1}$) that decays gradually over the heating period, averaging values less than 5\,km\,s$^{-1}$.  Its intensity rises slightly, to levels of 10-20\,kDN, without much variation.  The peak intensity in all cases is smaller than the observed value, which reached as high as 100\,kDN, with a background level of about 10\,kDN.  In general, the behavior shows good agreement with the observed trends, though the intensities vary and do not agree completely with the observations.  In particular, the ratio of \ion{O}{1} to \ion{C}{2} is approximately constant, and the parameters only seem to reproduce one or the other at a given time.  This may be due to the assumptions of the beam heating (\textit{e.g.} fixed cut-off energy on all threads, short duration heating, \textit{etc.}), due to the assumed values of $N$, $\alpha$, or $F_{\text{min}}$, or perhaps due to assumptions about the initial atmosphere.  The beam heating parameters (low energy cut-off and spectral index), however, were taken from fits to the RHESSI data for this event, so it is unlikely that they are the issue.

\begin{figure*}
  \begin{minipage}[b]{0.32\linewidth}
    \centering
    \includegraphics[width=2.2in]{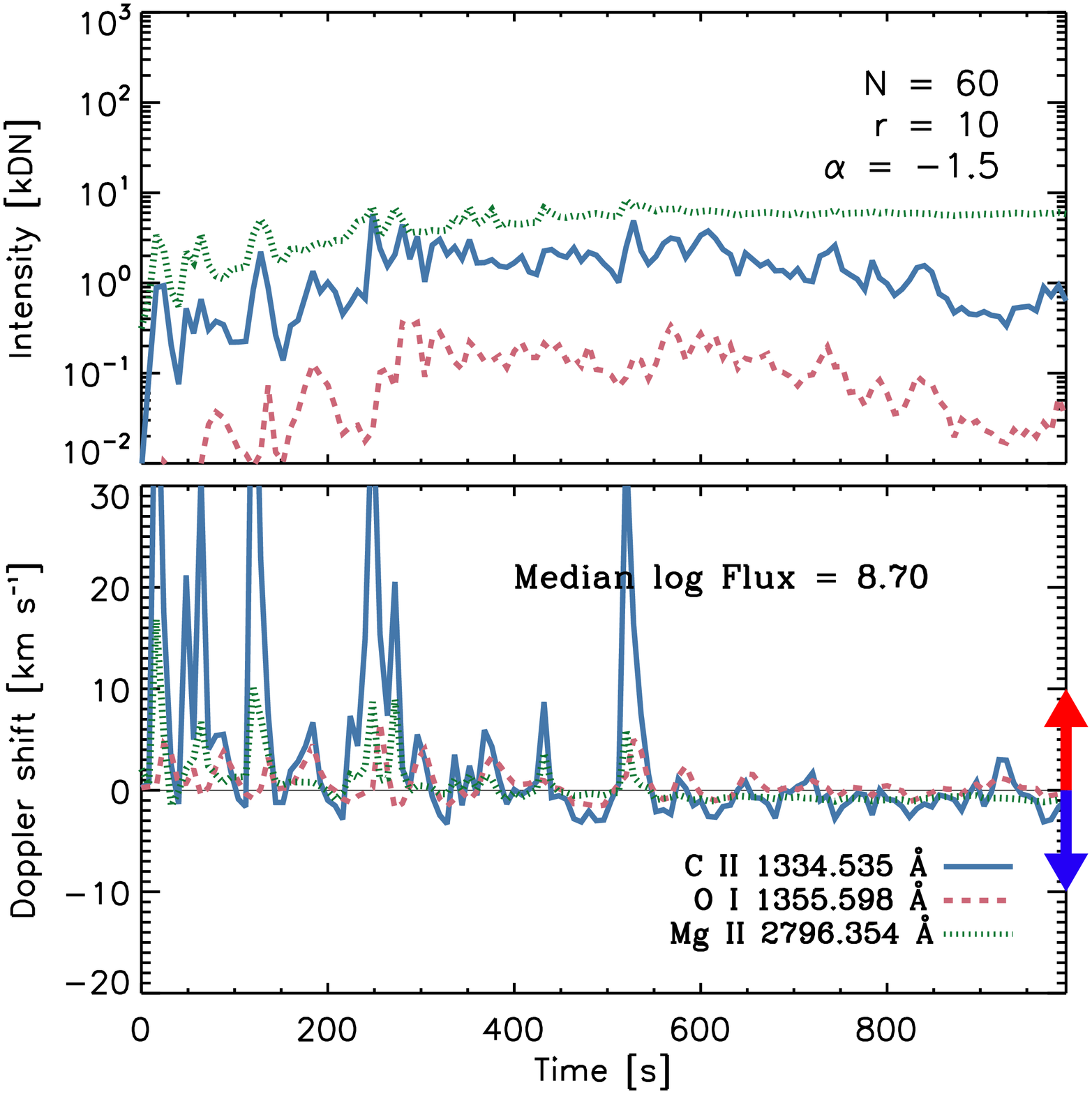}
  \end{minipage}
  \begin{minipage}[b]{0.32\linewidth}
    \centering
    \includegraphics[width=2.2in]{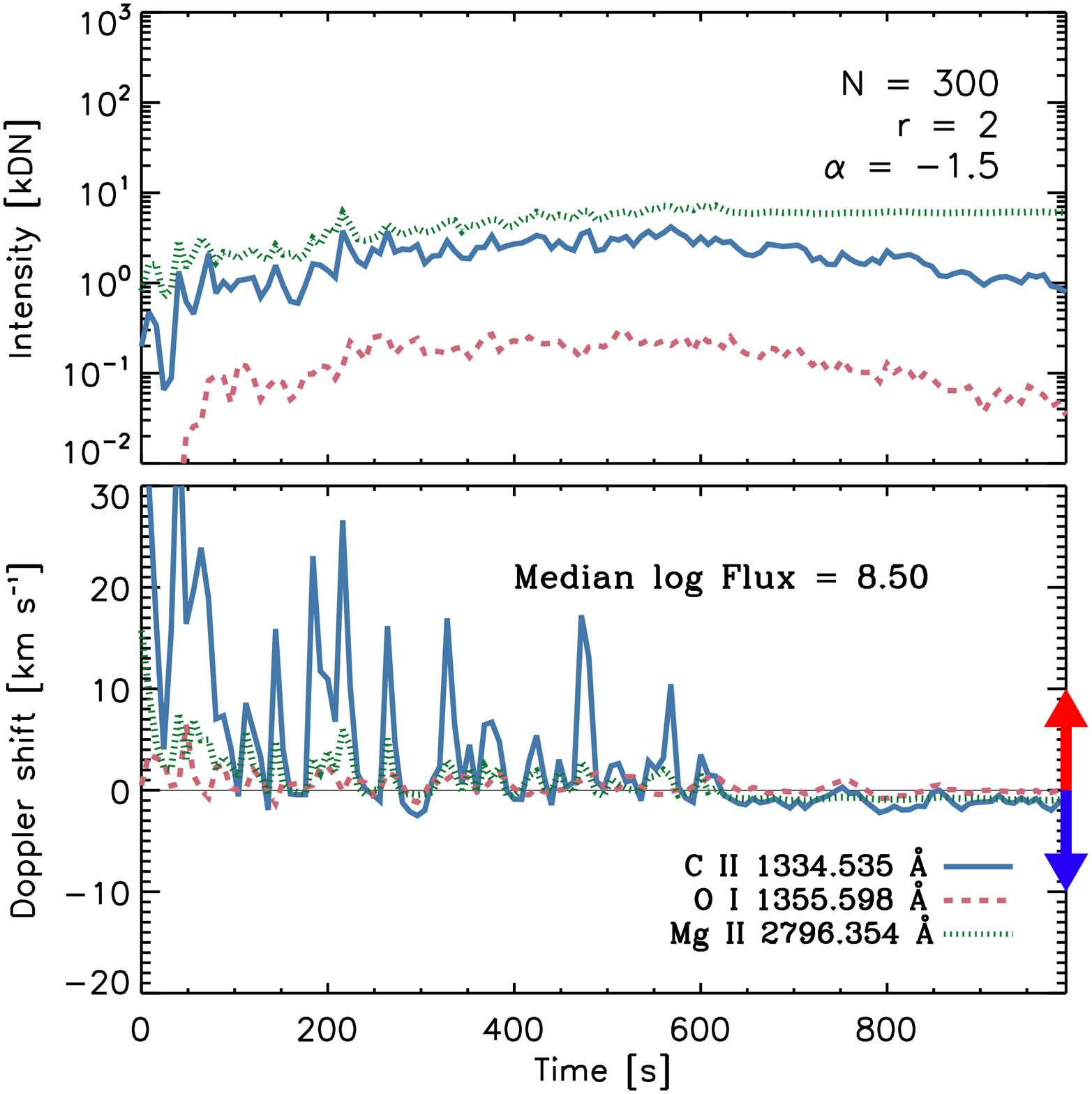}
  \end{minipage}
    \begin{minipage}[b]{0.32\linewidth}
    \centering
    \includegraphics[width=2.2in]{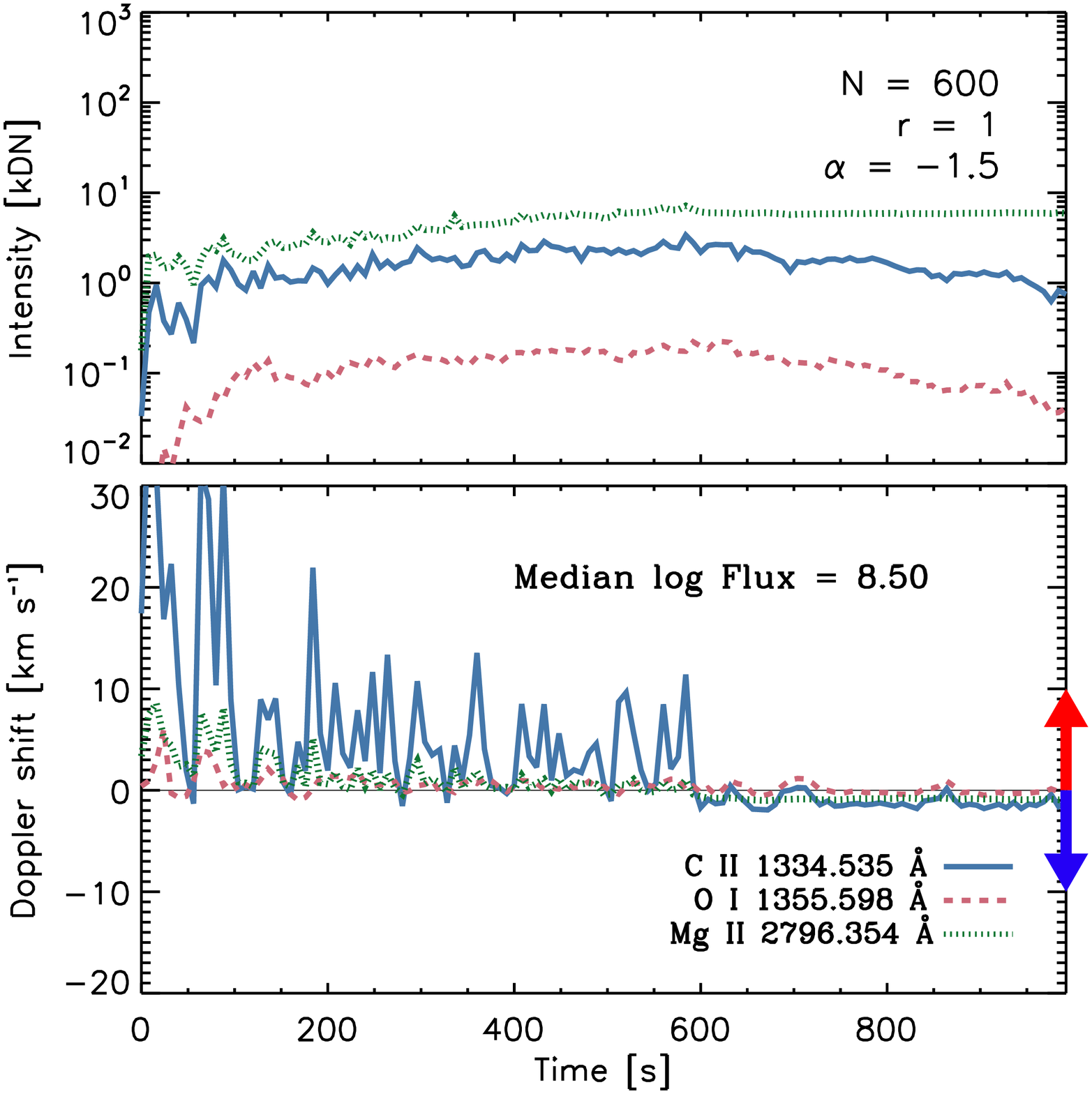}
  \end{minipage}
    \begin{minipage}[b]{0.32\linewidth}
    \centering
    \includegraphics[width=2.2in]{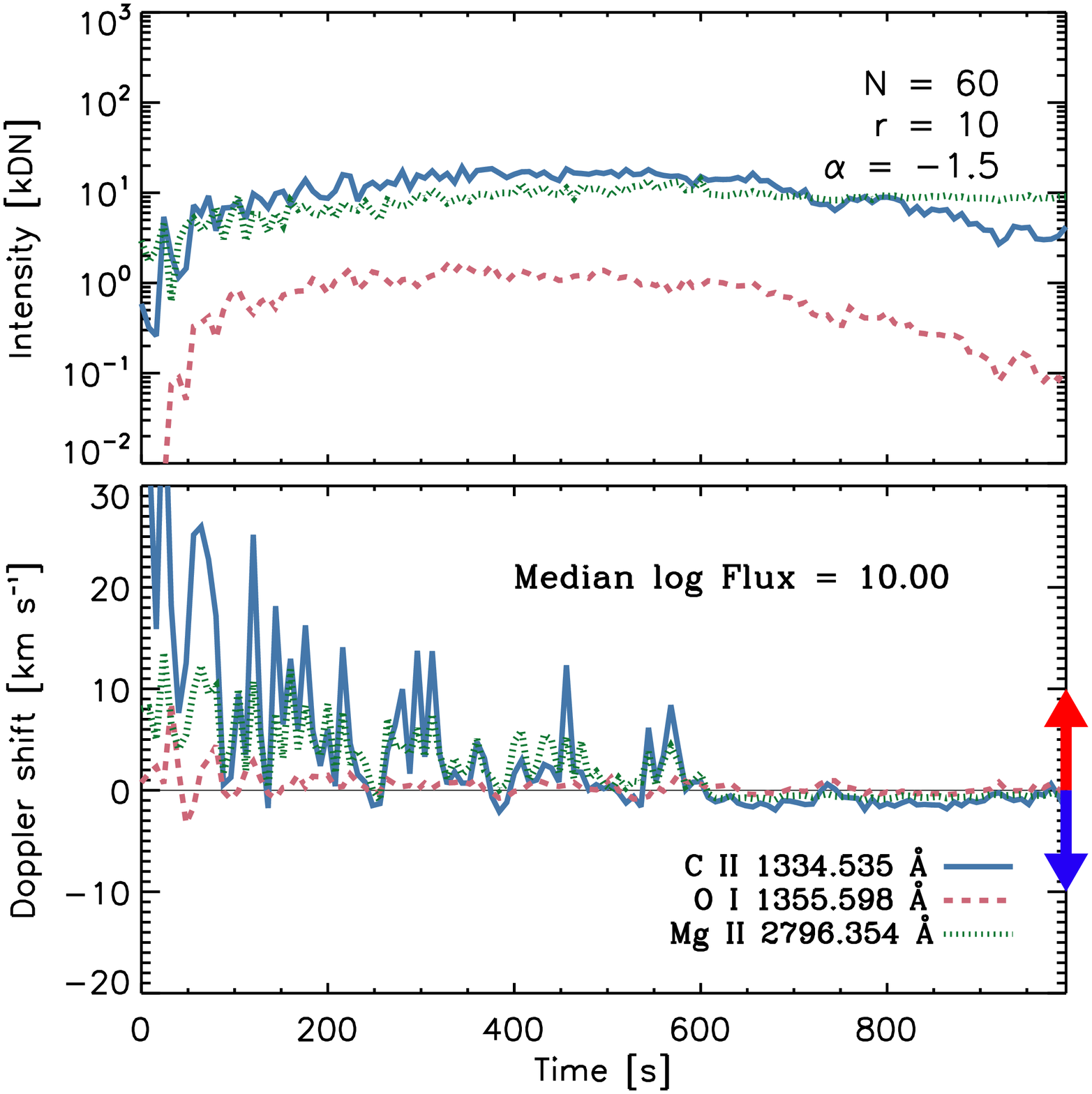}
  \end{minipage}
  \begin{minipage}[b]{0.32\linewidth}
    \centering
    \includegraphics[width=2.2in]{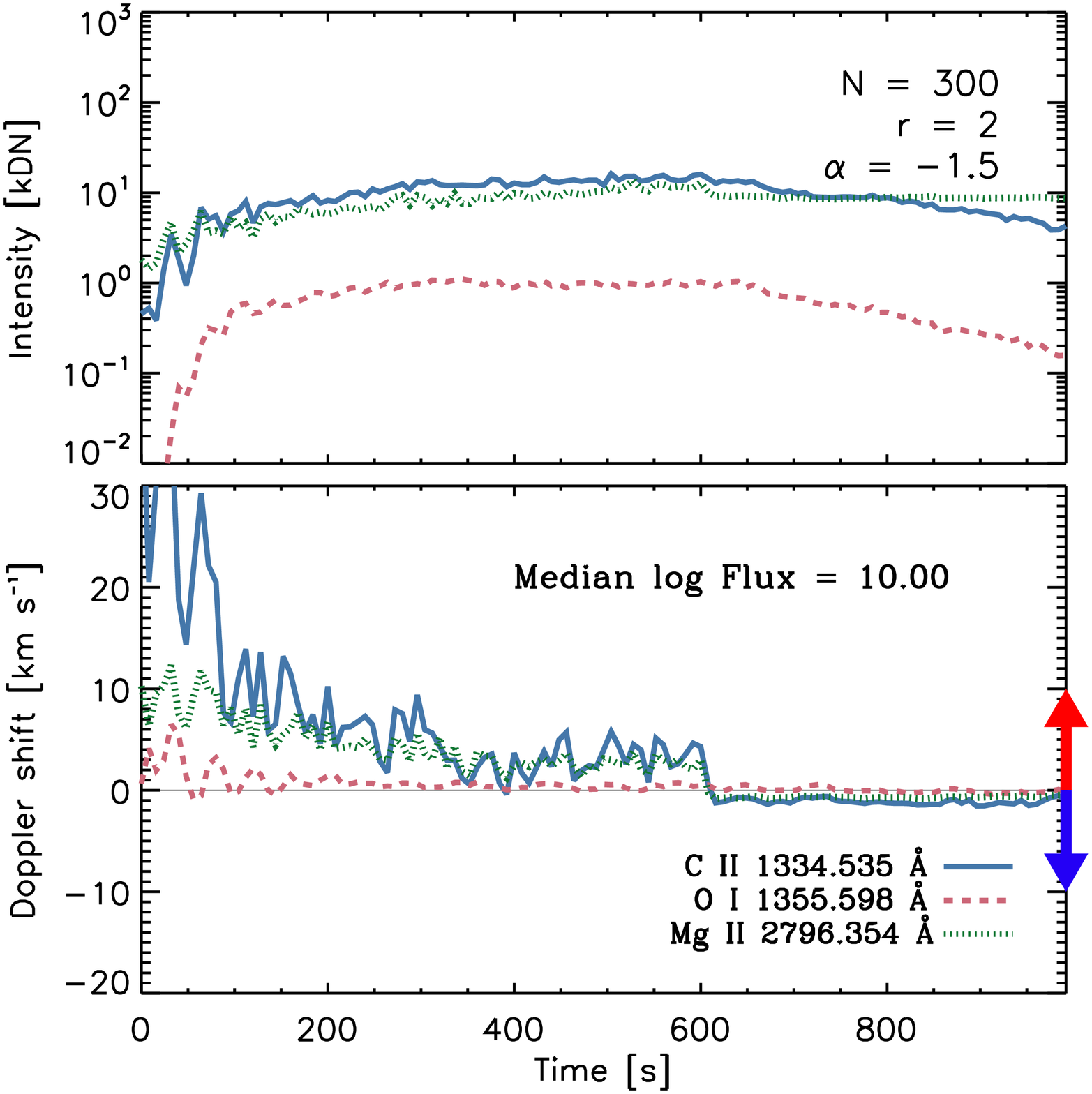}
  \end{minipage}
    \begin{minipage}[b]{0.32\linewidth}
    \centering
    \includegraphics[width=2.2in]{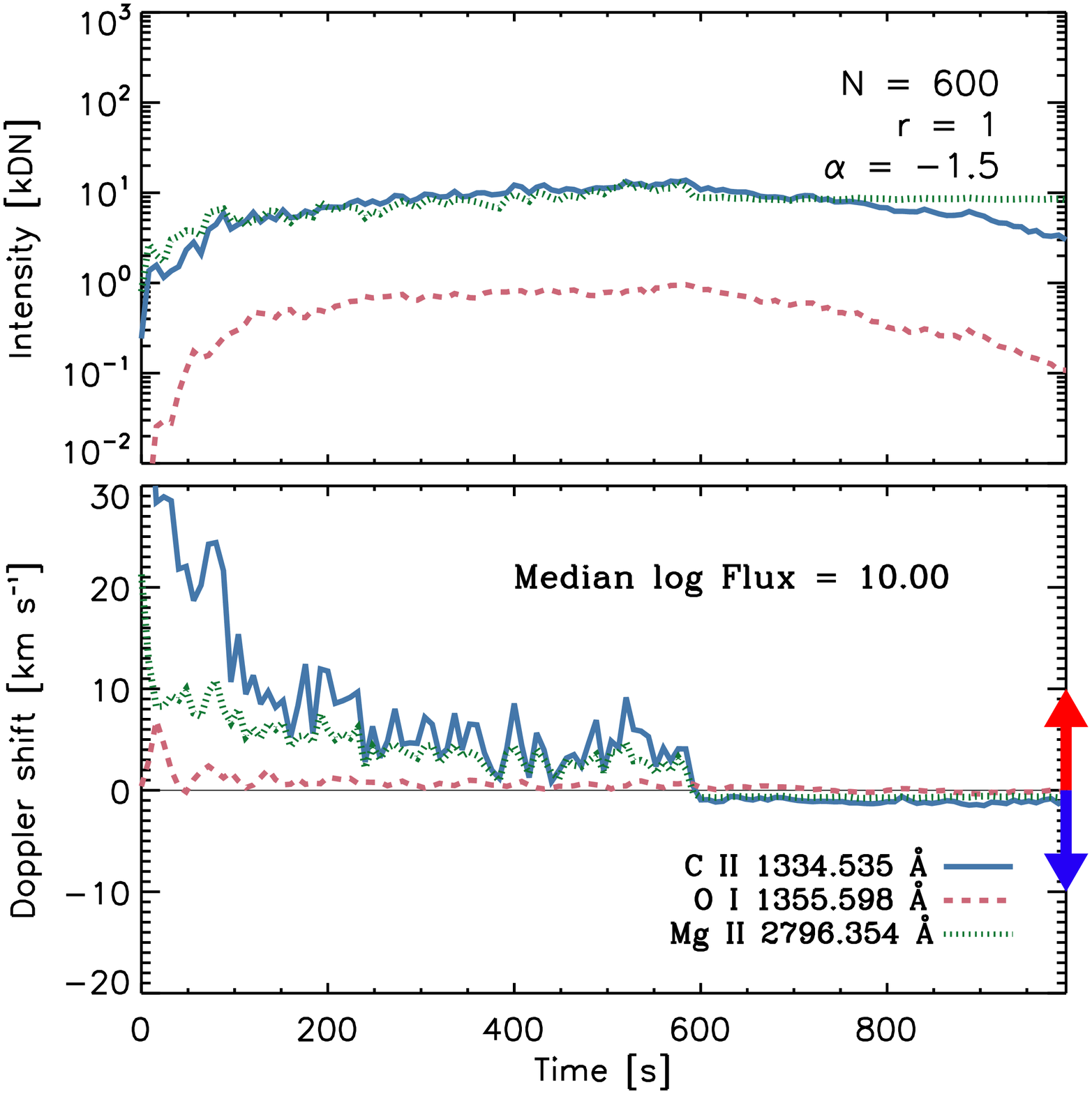}
  \end{minipage}
\caption{Similar to Figure \ref{fig:bestfit}.  The top row shows $F_{\text{min}} = 10^{8}$ and the bottom $3 \times 10^{9}$\,erg\,s$^{-1}$\,cm$^{-2}$, while the columns show $N = 60, 300, 600$ threads, respectively.  As with the previous figure, the results are mostly consistent with observations, in that the persistent red-shifts in \ion{C}{2} are consistent with a large number of threads with high median energy flux, while \ion{O}{1} is close to stationary, but the problem remains that \ion{O}{1} is too bright.  }
\label{fig:parameter_space}
\end{figure*}

Figure \ref{fig:parameter_space} shows a bit more of the parameter space: with $N = 60$, 300, 600 (left, center, right columns, respectively) and $F_{min} = 10^{8}$ and $3 \times 10^{9}$\,erg\,s$^{-1}$\,cm$^{-2}$ (top and bottom rows).  As in \citet{reep2016b}, we find that the persistent red-shifts seen in \ion{C}{2} are consistent with a large number of threads, with high median energy flux.  \ion{O}{1} generally shows little or no shift ($\lesssim 5$\,km\,s$^{-1}$), while \ion{Mg}{2} has shifts up to 10\,km\,s$^{-1}$, which gradually decays with time, which is consistent with observations.  As in Figure \ref{fig:bestfit}, we find that the ratio of intensities between \ion{C}{2} and \ion{O}{1} is approximately constant, such that either \ion{C}{2} is too dim or \ion{O}{1} is too bright, again suggesting that the initially assumed atmosphere differs from actual solar conditions.  

While it seems that the basic model can reproduce many of the observed features of the event, it seems likely that the initial chromospheric density and temperature profile are not consistent with actual solar conditions, suggesting that perhaps a tuning of the atmosphere would better reproduce observables.  For example, a recent study by \citet{ishikawa2018} found that Hanle effect diagnostics depend strongly on the choice of atmospheric model.  With the high cadence and high spatial resolution observations of lines such as Lyman-$\alpha$ (\textit{e.g.} \citealt{ishikawa2017}), it may be possible to directly tune the initial atmosphere for events under examination in the future.  We therefore briefly examine how altering the chromospheric model affects the synthesized line profiles, as compared to observations.

\begin{figure*}
    \begin{minipage}[b]{\linewidth}
    \centering
    \includegraphics[width=\linewidth]{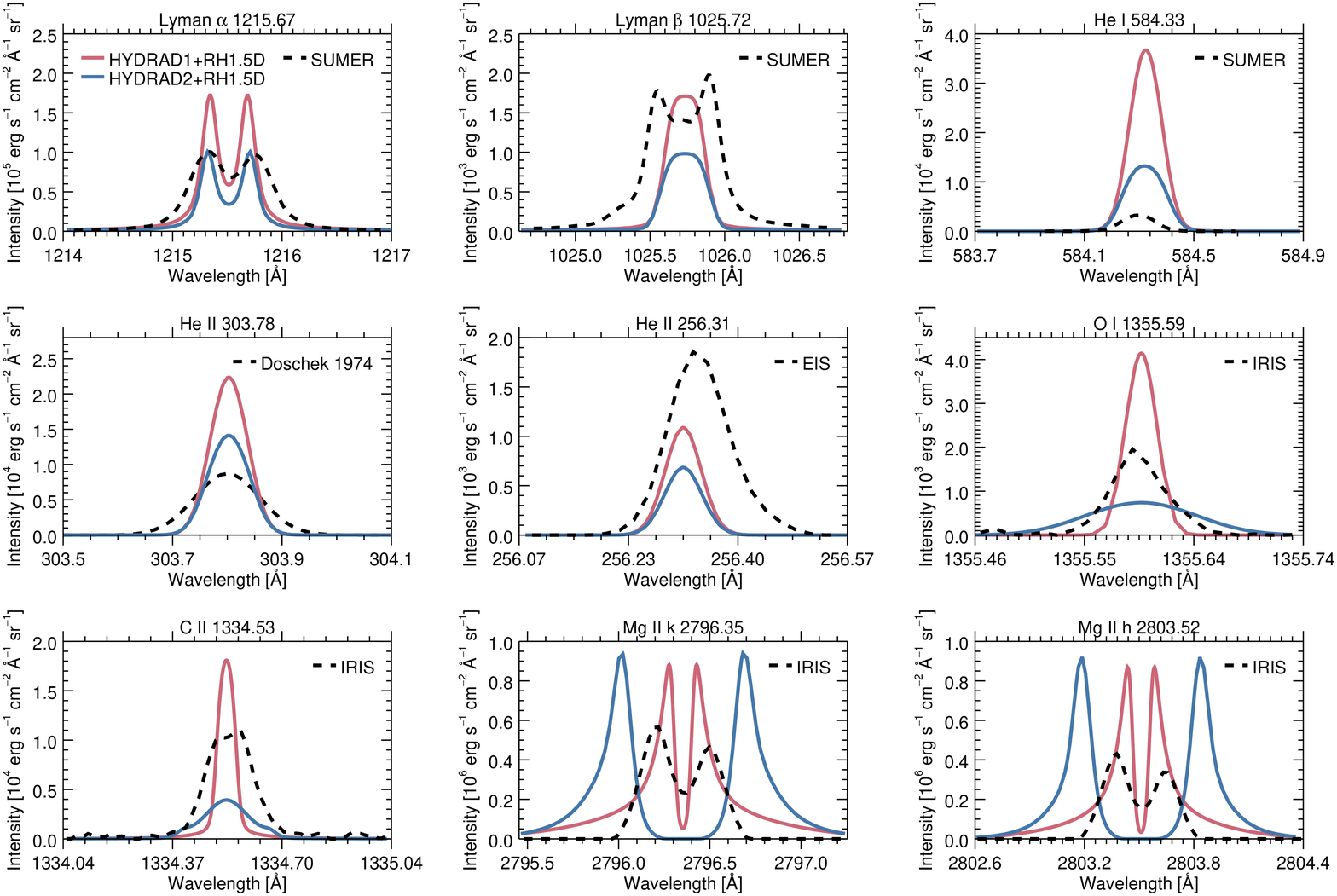}
  \end{minipage}
\caption{A comparison of the line profiles synthesized with RH1.5D using two density profiles against quiet sun observations from SUMER, the rocket flight reported in \citet{doschek1974}, EIS, and IRIS, as labeled.  From left to right, top to bottom: Lyman-$\alpha$, Lyman-$\beta$, \ion{He}{1} 584\,\AA, \ion{He}{1} 304\,\AA, \ion{He}{2} 256\,\AA, \ion{O}{1} 1355.6\,\AA, \ion{C}{2} 1334.4\,\AA, \ion{Mg}{2} k 2796.4\,\AA, and \ion{Mg}{2} h 2803.5\,\AA.  The lines show varying degrees of agreement, but it is clear that not all of the lines can be reproduced simultaneously.}
\label{fig:line_comparisons}
\end{figure*}

In Figure \ref{fig:line_comparisons}, we show 9 transition region and chromospheric lines that can be used as diagnostics of the atmospheric density and temperature profile (left to right, top to bottom): Lyman-$\alpha$, Lyman-$\beta$, \ion{He}{1} 584\,\AA, \ion{He}{1} 304\,\AA, \ion{He}{2} 256\,\AA, \ion{O}{1} 1355.6\,\AA, \ion{C}{2} 1334.4\,\AA, \ion{Mg}{2} k 2796.4\,\AA, and \ion{Mg}{2} h 2803.5\,\AA.  In this figure, the Lyman lines and \ion{Mg}{2} were calculated in PRD while the others were done with CRD.  The solid pink curves show the spectral lines as synthesized by RH1.5D using the VAL C temperature profile, with the default density profile derived from HYDRAD and no microturbulence, while the solid blue curves show a case where we have decreased the footpoint density by a factor of 1.5 and solved the hydrostatic equations again to produce a new density profile that more closely matches the peak intensity of Lyman-$\alpha$, with an assumed microturbulence of 6\,km\,s$^{-1}$.  The dashed black curves show example quiet sun profiles measured by SUMER, a rocket flight reported by \citet{doschek1974}, EIS, or IRIS, as indicated in the plot.  We have convolved each synthesized case with the instrumental line width of each respective instrument.

It is clear that \textit{none} of the model atmospheres reproduces any of the observed profiles in all of the lines simultaneously.  In the solid blue case, where we have reduced the footpoint density and re-solved the hydrostatic equations to improve agreement with the observed intensity in Lyman-$\alpha$, we generally find that the agreement of lines that form at other heights can either improve or worsen.  To emphasize this, we show the wavelength-integrated intensities (over the range in the plots) in Table \ref{table:intensities}, which demonstrates that none are in particularly good agreement.  We have also included the integrated intensities synthesized with RH1.5D from the FAL C \citep{fontenla1993} model atmosphere as a basis for comparison.  This suggests that the temperature profile is inaccurate, or perhaps that a hydrostatic profile may never accurately reproduce the chromosphere lines.  It is likely that we also need a turbulence value as a function of height in the chromosphere to better match the observed broadening.
\begin{table*}
\centering
\footnotesize
\begin{tabular}[t]{ c  l  c  c  }
Line, $\lambda_{0}$ & Source &  $\int I_{\lambda} d\lambda$ & Ratio \\
\hline
 \AA & & \scriptsize erg\,s$^{-1}$\,cm$^{-2}$\,sr$^{-1}$ \footnotesize & \\ 
\hline
Ly-$\alpha$ 1215.67 & SUMER      & $8.51 \times 10^{4}$ & 1.00 \\
 & FAL C         & $3.89 \times 10^{4}$ & 0.45 \\
 & HYDRAD 1 & $8.33 \times 10^{4}$ & 0.98 \\
 & HYDRAD 2 & $4.89 \times 10^{4}$ & 0.57 \\
\hline
Ly-$\beta$ 1025.72 & SUMER        & $1.07 \times 10^{3}$ & 1.00 \\
 & FAL C           & $4.09 \times 10^{2}$ & 0.38 \\
 &  HYDRAD 1  & $5.55 \times 10^{2}$ & 0.52 \\
 &  HYDRAD 2  & $3.52 \times 10^{2}$ & 0.33 \\
\hline
\ion{He}{1} 584.33 & SUMER        & $5.02 \times 10^{2}$ & 1.00 \\\
 & FAL C           & $3.41 \times 10^{2}$ & 0.68 \\
 &  HYDRAD 1  & $5.32 \times 10^{3}$ & 10.60 \\
 &  HYDRAD 2  & $3.52 \times 10^{3}$ &  7.01 \\
\hline
\ion{He}{2} 303.78 & Doschek et al. & $1.31 \times 10^{3}$ & 1.00 \\\
 & FAL C           & $8.92 \times 10^{2}$ & 0.68\\
 &  HYDRAD 1  & $1.89 \times 10^{3}$ & 1.44 \\
 &  HYDRAD 2  & $1.31 \times 10^{3}$ & 1.00 \\
\hline
\ion{He}{2} 256.31 & EIS               & $2.16 \times 10^{2}$ & 1.00 \\\
 & FAL C           & $3.47 \times 10^{1}$ & 0.16 \\
 &  HYDRAD 1  & $7.48 \times 10^{1}$ & 0.35 \\
 &  HYDRAD 2  & $4.86 \times 10^{1}$ & 0.22 \\
\hline

\ion{O}{1} 1355.598 & IRIS               & $1.11 \times 10^{2}$ & 1.00 \\\
 & FAL C           & $3.14 \times 10^{1}$ & 0.28 \\
 &  HYDRAD 1  & $1.36 \times 10^{2}$ & 1.23 \\
 &  HYDRAD 2  & $8.53 \times 10^{1}$ & 0.77 \\
\hline
\ion{C}{2} 1334.535 & IRIS               & $2.10 \times 10^{3}$ & 1.00 \\\
  & FAL C           & $2.14 \times 10^{2}$ & 0.10 \\
  &  HYDRAD 1  & $1.28 \times 10^{3}$ & 0.61 \\
  &  HYDRAD 2  & $7.10 \times 10^{2}$ & 0.34 \\
\hline
\ion{Mg}{2} k 2796.35 & IRIS               & $2.13 \times 10^{5}$ & 1.00 \\\
 & FAL C           & $1.54 \times 10^{5}$ & 0.72\\
 &  HYDRAD 1  & $3.31 \times 10^{5}$ & 1.55\\
 &  HYDRAD 2  & $3.89 \times 10^{5}$ & 1.83 \\
\hline
\ion{Mg}{2} h 2803.52 & IRIS               & $1.41 \times 10^{5}$ & 1.00 \\\
 & FAL C           & $1.13 \times 10^{5}$ & 0.80 \\
 &  HYDRAD 1  & $2.71 \times 10^{5}$ & 1.92 \\
 &  HYDRAD 2  & $3.18 \times 10^{5}$ & 2.26\\
\end{tabular}
\caption{The integrated intensities of each line, calculated for each case in Figure \ref{fig:line_comparisons} as well as for the FAL C atmosphere \citep{fontenla1993}.  The final column shows the ratio of the integrated intensity to that observed.  \label{table:intensities}}
\end{table*}

In order to better constrain future simulations, we suggest that the pre-flare atmosphere (time 0 in the simulations) should be tuned to pre-flare observations.  \citet{reep2016a} reached similar conclusions concerning the modeling of X-ray source heights in a flare.  While the chromosphere is inherently dynamic, having a good initial agreement between the real and model atmosphere improves the confidence that our initial assumptions about the atmospheric profile do not adversely affect the hydrodynamic and forward modeling results.  Spectral polarimetric inversion models have been shown to produce chromospheric profiles \citep{socasnavarro1998,socasnavarro2000}, determining temperature, turbulence, velocity, and magnetic field strength as functions of depth.  In a forthcoming paper, we plan to develop a similar method to therefore tune the pre-flare atmosphere used in simulations to give good agreement with pre-flare observations.  We expect that future instruments such as DKIST or Solar-C may prove fruitful in this regard.

\section{Conclusions}
\label{sec:conclusions}

In this work, we have examined the importance of NLTE effects on the formation of light curves and Doppler shifts.  In general, a model of the dynamic chromosphere requires a detailed treatment to determine ionization fractions and level populations.  Solving the radiative transfer equation in general is one of the most computationally demanding tasks in astrophysics, primarily due to its non-local nature.

In solar flares, there are many indications that there is sub-structuring at spatial resolutions below those of current instrumentation.  This led to the rise of multithreaded models, where many unresolved loops are assumed to be rooted within a single pixel.  Originally, this type of modeling was invoked by \citet{hori1997,hori1998} to explain the large stationary component of \ion{Ca}{19} seen in flare observations with Yohkoh/BCS, where strong blue-shifts were expected from single loop modeling of evaporation flows.  Later papers addressed other problems with long duration cooling of soft X-ray lightcurves \citep{reeves2002,warren2005,warren2006,reep2017}, late phase heating \citep{reeves2007,qiu2016,zhu2018}, or spectral line considerations \citep{rubiodacosta2017,kowalski2017a}.  Interestingly, despite tremendous advances in spatial resolution, there are still indications that the basic flaring loop is unresolved.  For example, the long-lasting red-shifts in \ion{Si}{4} seen by IRIS during many flares can be explained by multithreaded modeling \citep{reep2016b}, but only if there are more than 60 loops rooted within a single pixel, suggesting observations do not come anywhere near resolving the basic filamentation of flares.

In order to tackle both of these computational challenges, we have implemented an approximation to the radiative transfer equations into HYDRAD that gives a fast and reasonable solution to the level populations of hydrogen, using a six-level atom.  In turn, we have improved the calculation of the electron density across the chromosphere and the corona, which more precisely determines radiative losses and dynamic processes such as evaporation speeds.  The code is computationally light enough that many simulations of loops with a wide parameter space can be run in a modest amount of time, which is particularly important for multithreaded modeling.  Furthermore, the code has been parallelized with good scaling up to at least 32 cores, offsetting the loss in computational time due to the NLTE calculations.

In order to test this model, we have then developed a multithreaded model of a flaring event seen by IRIS.  In the observations, it was found found that the \ion{O}{1} remained approximately stationary, while \ion{Si}{4} and \ion{C}{2} showed long-lasting red-shifts ($\approx 60$\,min), and \ion{Mg}{2} formed red-shifted, which gradually decayed \citep{warren2016}.  A multi-threaded model was able to reproduce \ion{Si}{4} closely, both in terms of intensity and Doppler shifts, but faltered with \ion{C}{2} and \ion{O}{1} \citep{reep2016b}.  In this paper, we revisited those simulations with the improved chromospheric model, and with the RH1.5D code, recalculated these lines along with \ion{Mg}{2}, finding fair agreement.  \ion{O}{1} was found to be approximately stationary, \ion{C}{2} strongly red-shifted for long periods of time, and \ion{Mg}{2} forms red-shifted with a gradually decaying speed.  The absolute intensities could be reproduced in either \ion{O}{1} or \ion{C}{2}, but not both simultaneously with this model, however.  These results confirm the importance of NLTE effects on the formation of these lines, even in a multithreaded model, but they do not settle all of the issues.

A close examination of the initial atmosphere reveals that there is, in general, poor agreement with observations of the quiet sun and pre-flare observations.  The assumption of either VAL C or FAL C model does not reproduce all chromospheric lines simultaneously, and likely requires modification.  Unfortunately, simply scaling the density and assumed microturbulence to better match any individual observed line does not necessarily improve the correspondence of other lines with observations.  It is likely that it is also necessary to alter the shape of the temperature profile to improve the fit.  Further, it is implausible that multiple events are described by the same initial atmosphere, and in general we should not assume the same atmospheric profile for all events.  We therefore suggest that pre-flare atmospheres used in simulations be fine tuned to pre-flare observations of events under study (\textit{e.g.} \citealt{reep2016a}), and that future instrumentation (\textit{e.g.} DKIST or Solar-C), combined with spectral inversion methods \citep{socasnavarro2000}, can assist in that endeavor.  We plan a future study to address the importance and plausibility of such a method. 
\appendix

\addcontentsline{toc}{section}{Appendices}
\renewcommand{\thesubsection}{\Alph{subsection}}

\subsection{RADYN Comparison}
\label{app:radyn}

In this appendix, we briefly compare the method of approximating level populations against the commonly used RADYN model.  We do not enumerate all of the differences in physics and numerics here, though it would be a useful exercise to show a comparison of all the strengths and weaknesses of commonly used models such as these two.  Our primary purpose here is to compare the chromospheric electron density that results from the Sollum method compared to a treatment that solves the full radiative transfer equation.

We show two simulations of electron beam heating, taken from the F-CHROMA website\footnote{\url{https://star.pst.qub.ac.uk/wiki/doku.php/public/solarmodels/start}}, model numbers 078 and 012, which had electron beams with low energy cut-off $E_{c} = 10$\,keV, spectral index $\delta = 8$, and peak energy flux $F_{0} = 3 \times 10^{9}$ and $3 \times 10^{10}$\,erg\,s$^{-1}$\,cm$^{-2}$, heated with a triangular profile over 20\,s.  The loop length is 22\,Mm in total (11 Mm in RADYN, which only solves half the loop), and the simulations were run for 50\,s.  We copied RADYN's chromospheric temperature profile into HYDRAD, set the foot-point density to agree with RADYN, and then allowed HYDRAD to solve for the hydrostatic initial conditions.  Because the two codes have different assumed background heating functions and radiative loss functions, the initial conditions do not agree perfectly, but are fairly close at all heights.  Non-equilibrium ionization was used for only helium and calcium in HYDRAD for this comparison, as RADYN solves these in non-equilibrium as well.

\begin{figure*}
\begin{minipage}[b]{0.32\linewidth}
\centering
\includegraphics[width=\textwidth]{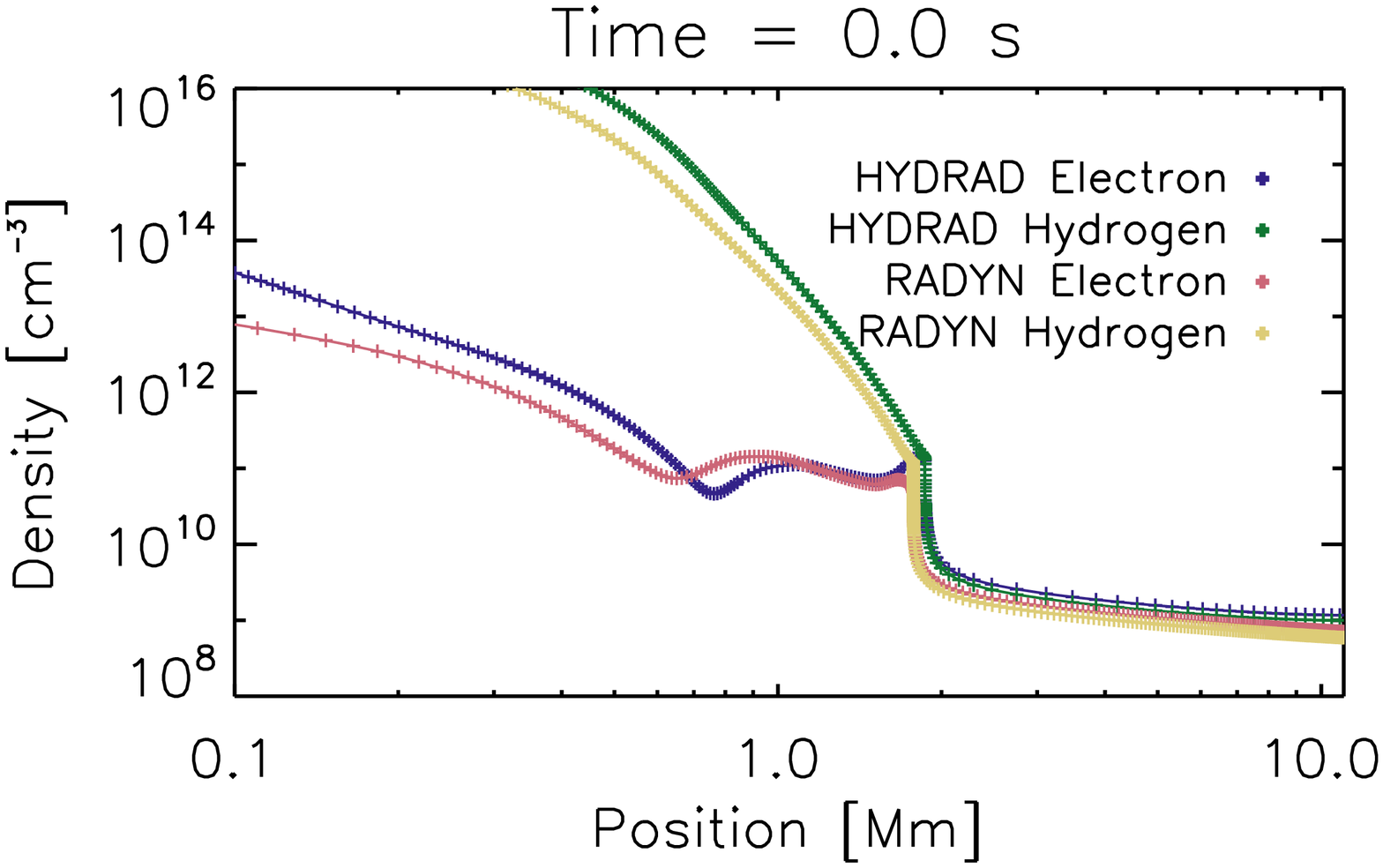}
\end{minipage}
\begin{minipage}[b]{0.32\linewidth}
\centering
\includegraphics[width=\textwidth]{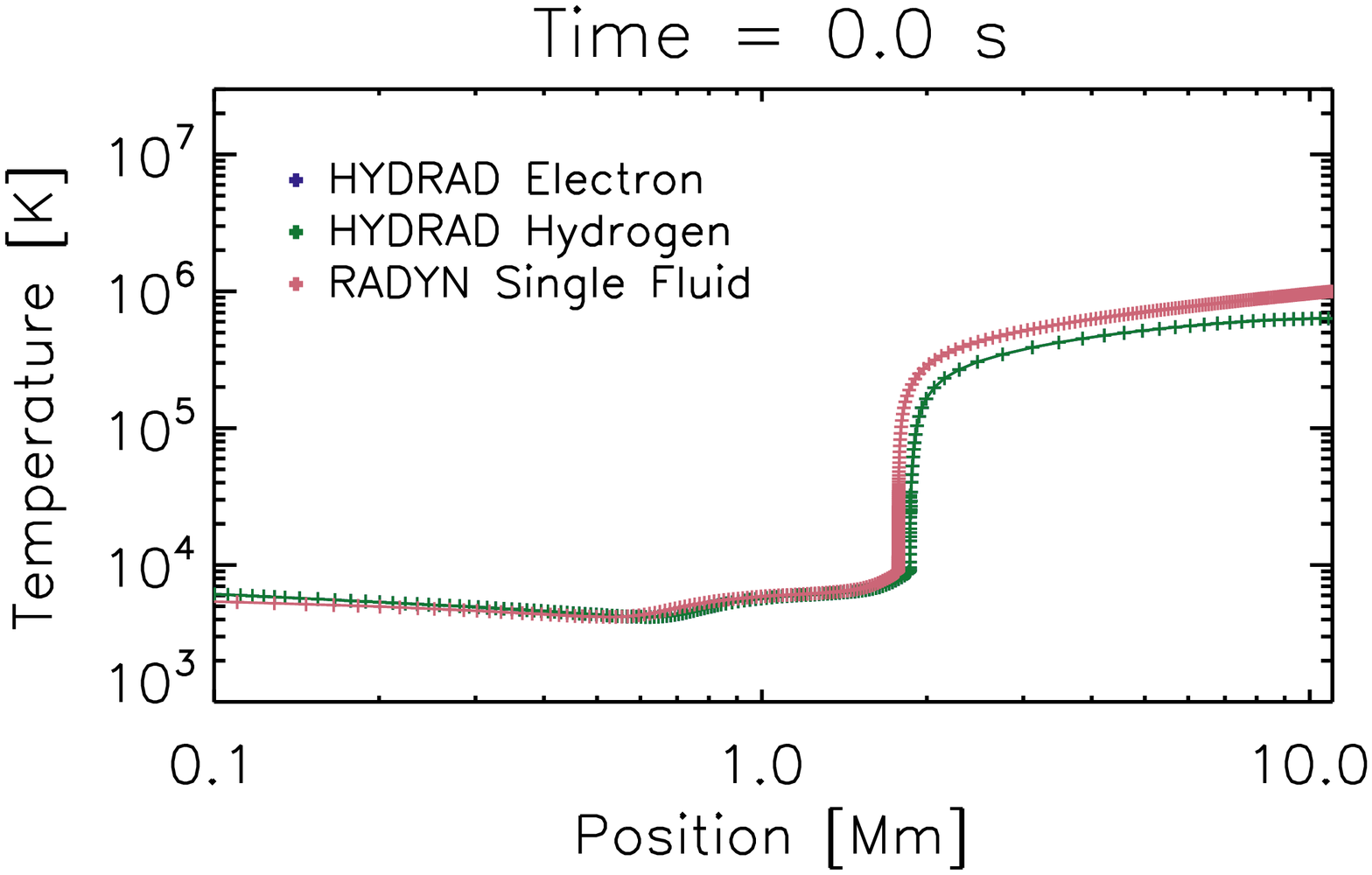}
\end{minipage}
\begin{minipage}[b]{0.32\linewidth}
\centering
\includegraphics[width=\textwidth]{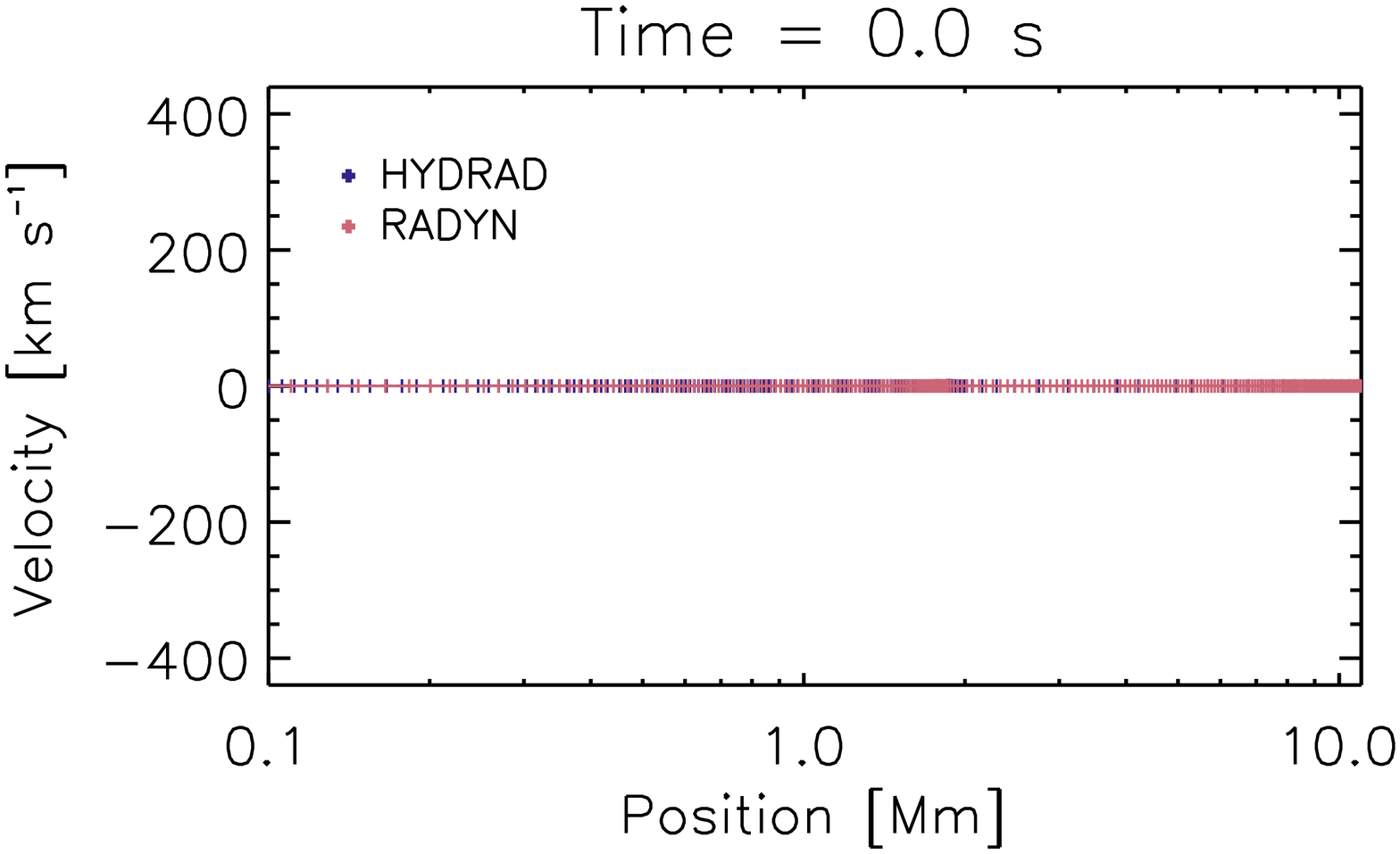}
\end{minipage}
\begin{minipage}[b]{0.32\linewidth}
\centering
\includegraphics[width=\textwidth]{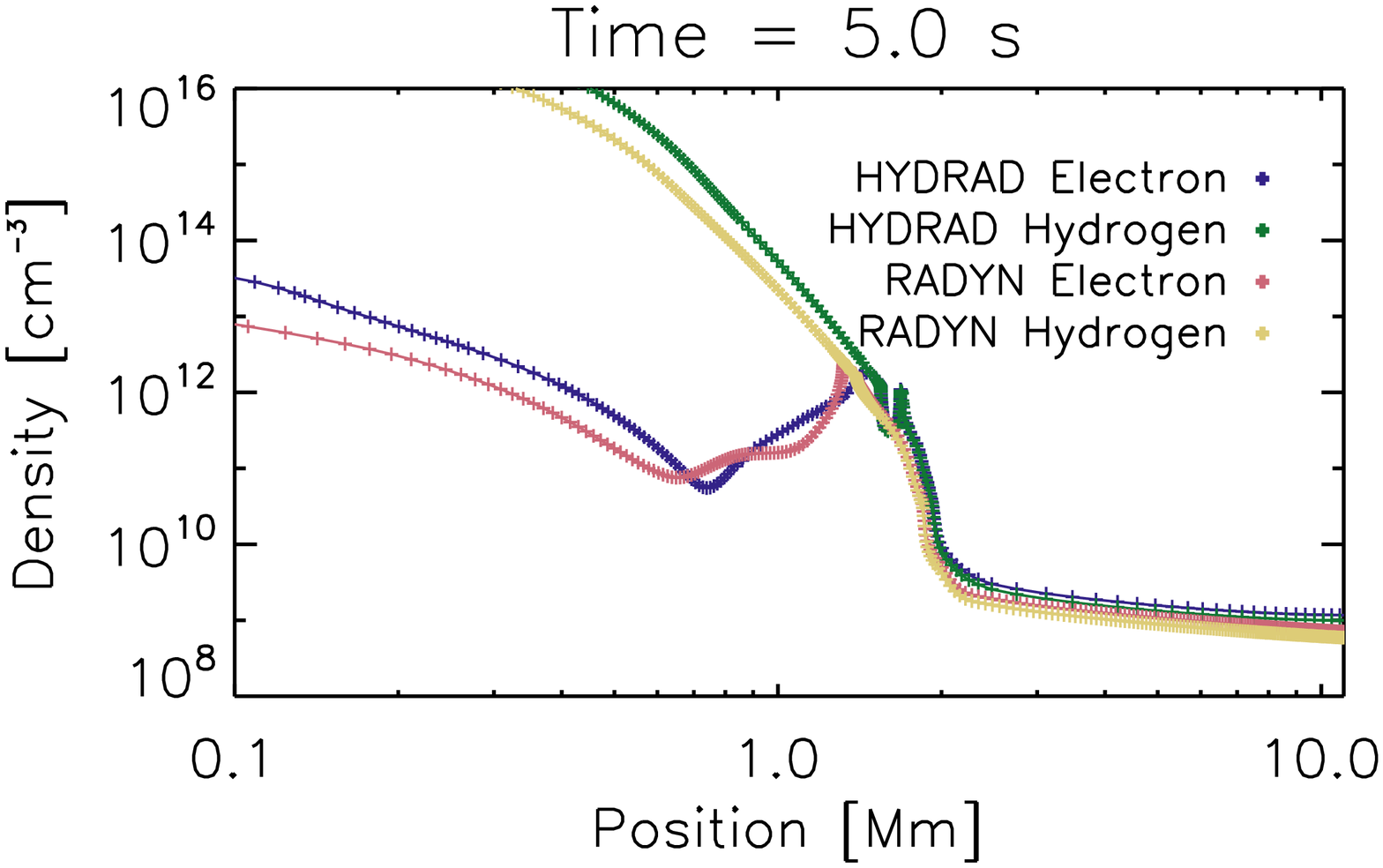}
\end{minipage}
\begin{minipage}[b]{0.32\linewidth}
\centering
\includegraphics[width=\textwidth]{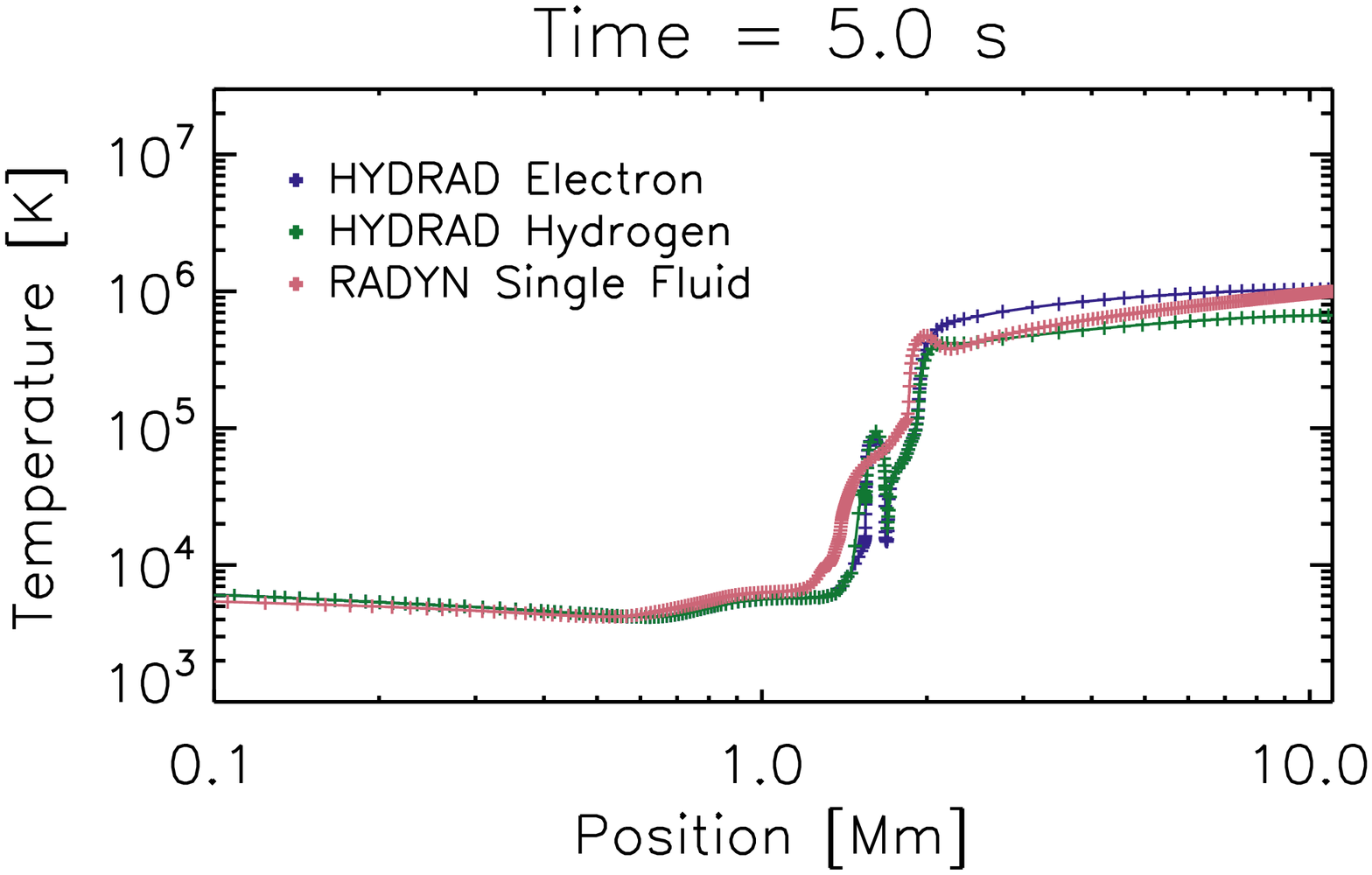}
\end{minipage}
\begin{minipage}[b]{0.32\linewidth}
\centering
\includegraphics[width=\textwidth]{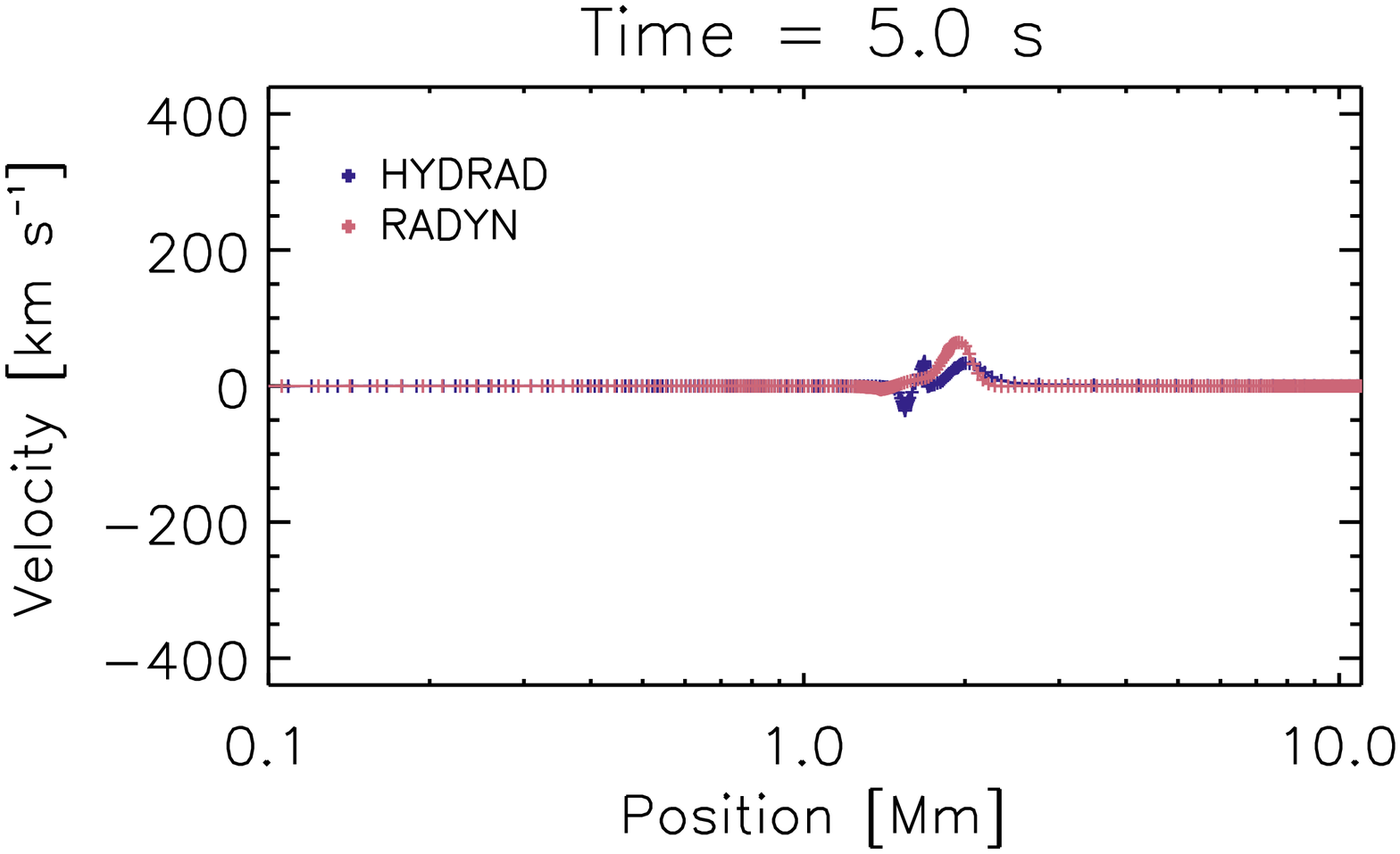}
\end{minipage}
\begin{minipage}[b]{0.32\linewidth}
\centering
\includegraphics[width=\textwidth]{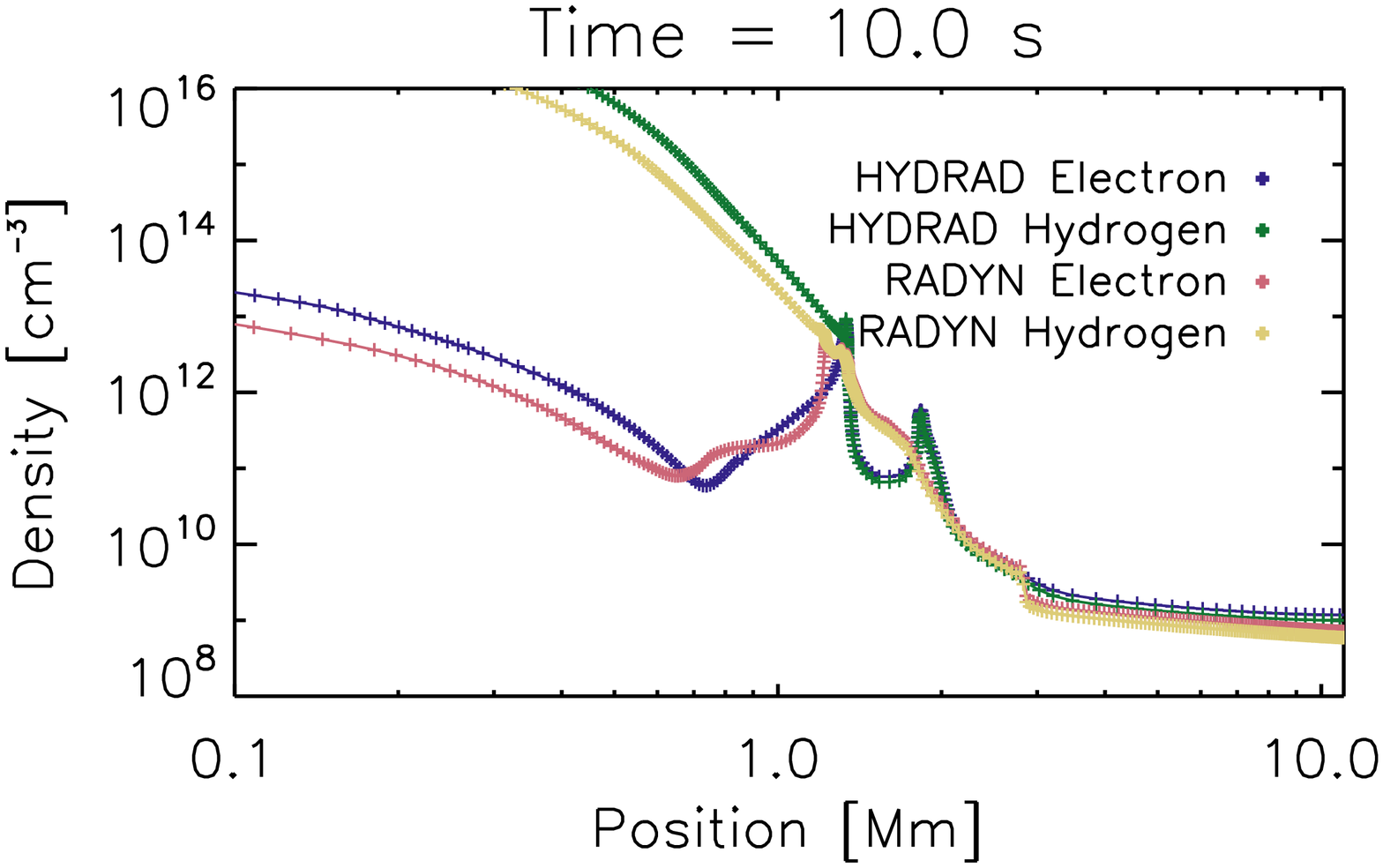}
\end{minipage}
\begin{minipage}[b]{0.32\linewidth}
\centering
\includegraphics[width=\textwidth]{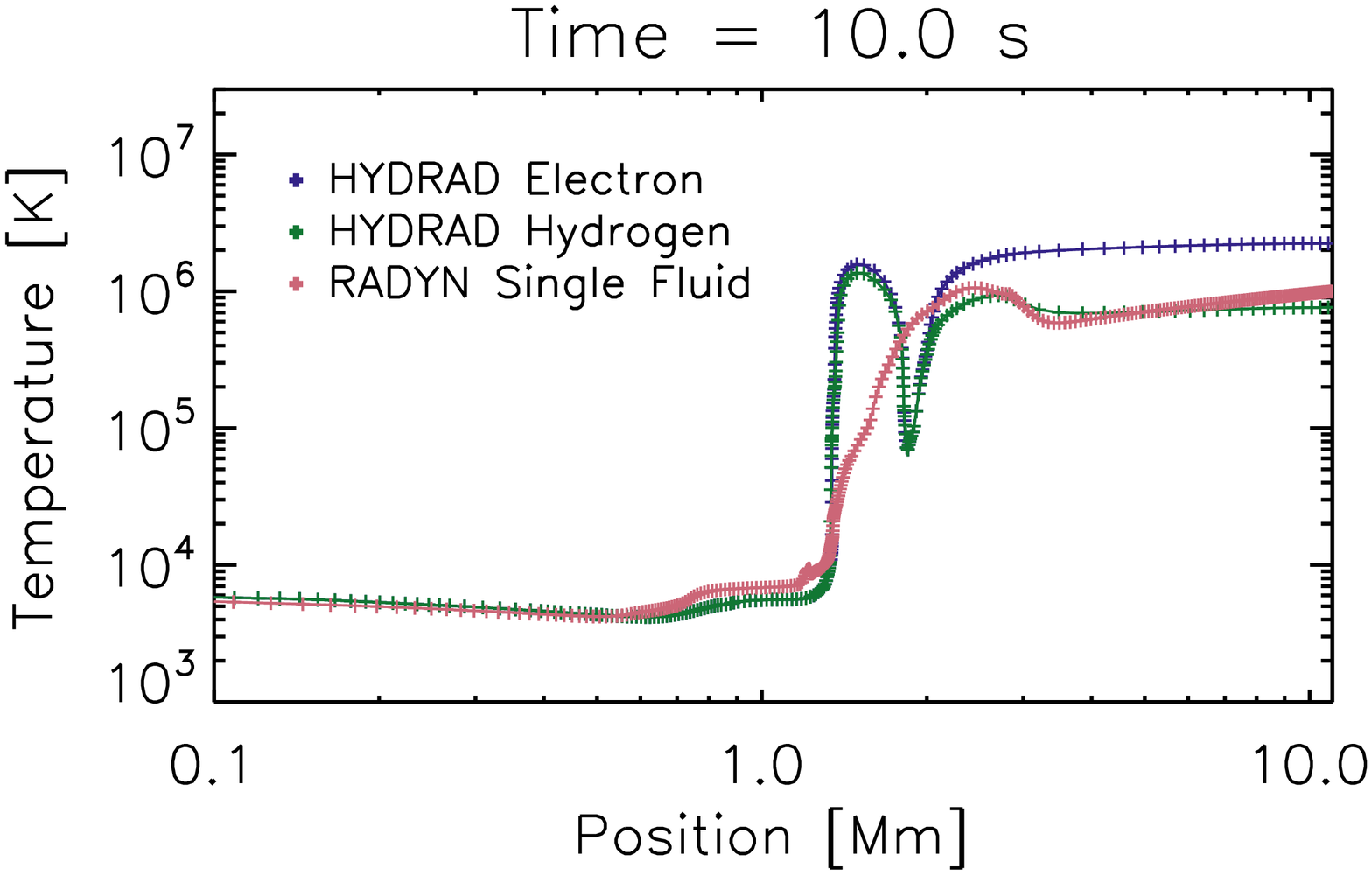}
\end{minipage}
\begin{minipage}[b]{0.32\linewidth}
\centering
\includegraphics[width=\textwidth]{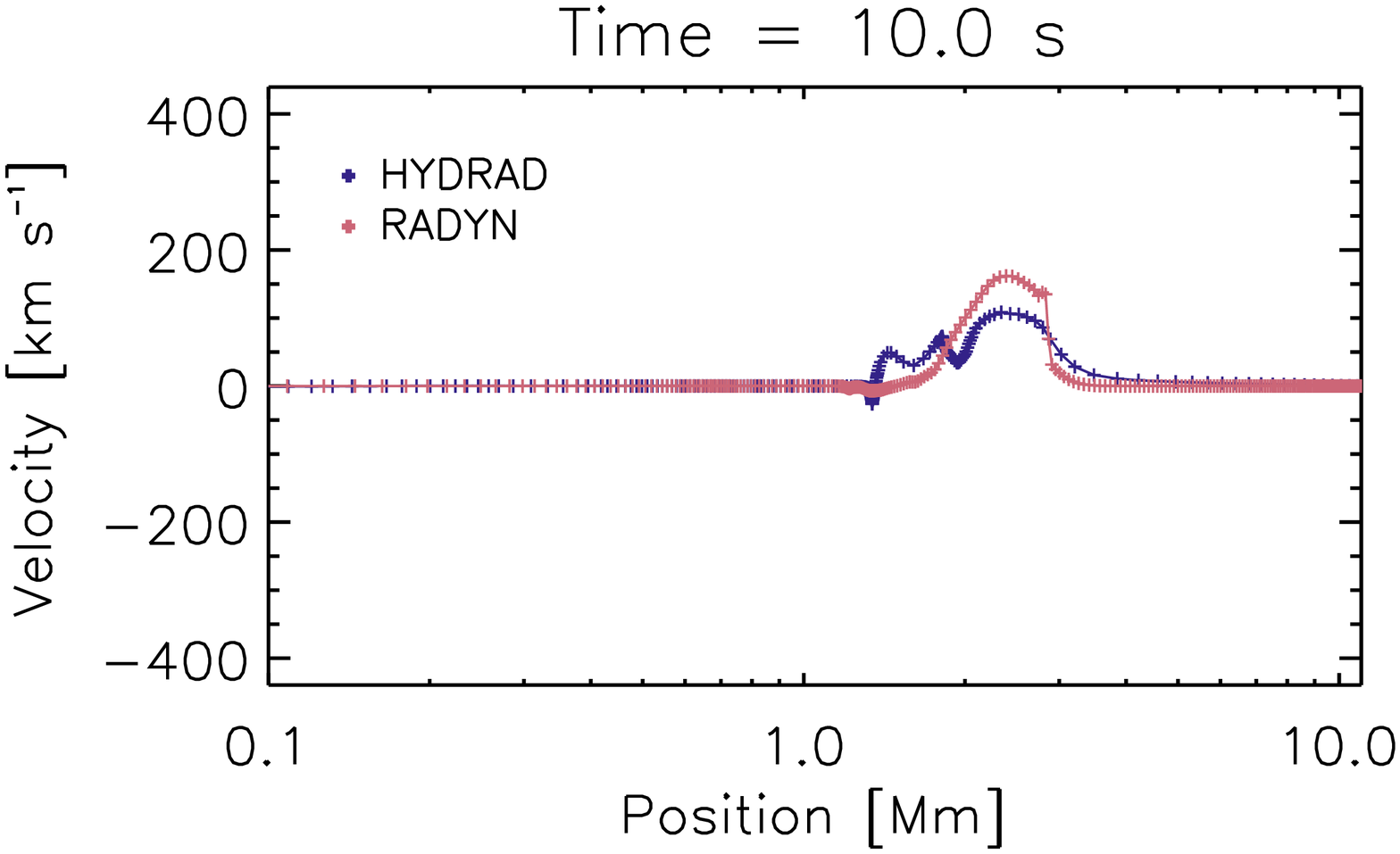}
\end{minipage}
\begin{minipage}[b]{0.32\linewidth}
\centering
\includegraphics[width=\textwidth]{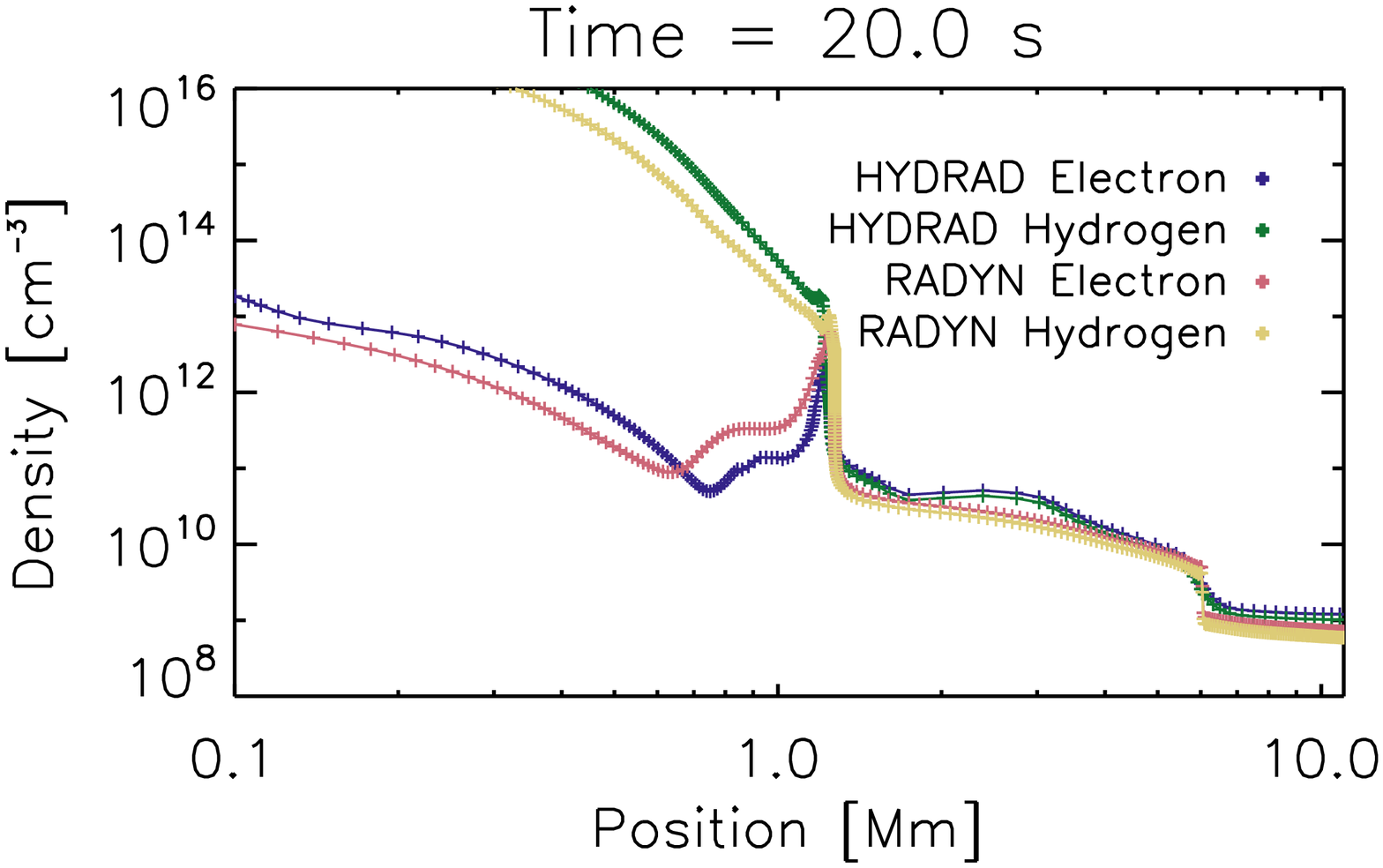}
\end{minipage}
\begin{minipage}[b]{0.32\linewidth}
\centering
\includegraphics[width=\textwidth]{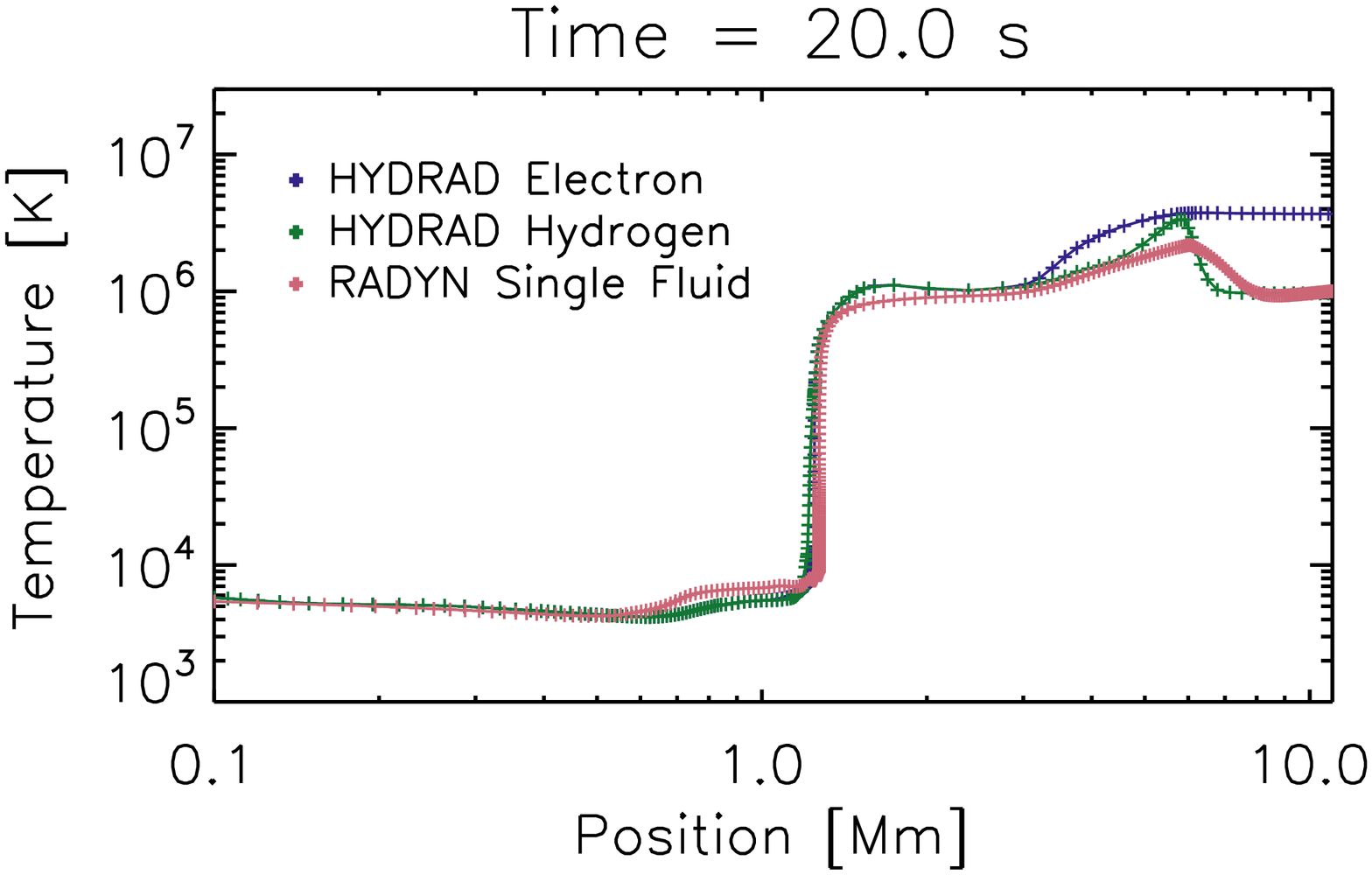}
\end{minipage}
\begin{minipage}[b]{0.32\linewidth}
\centering
\includegraphics[width=\textwidth]{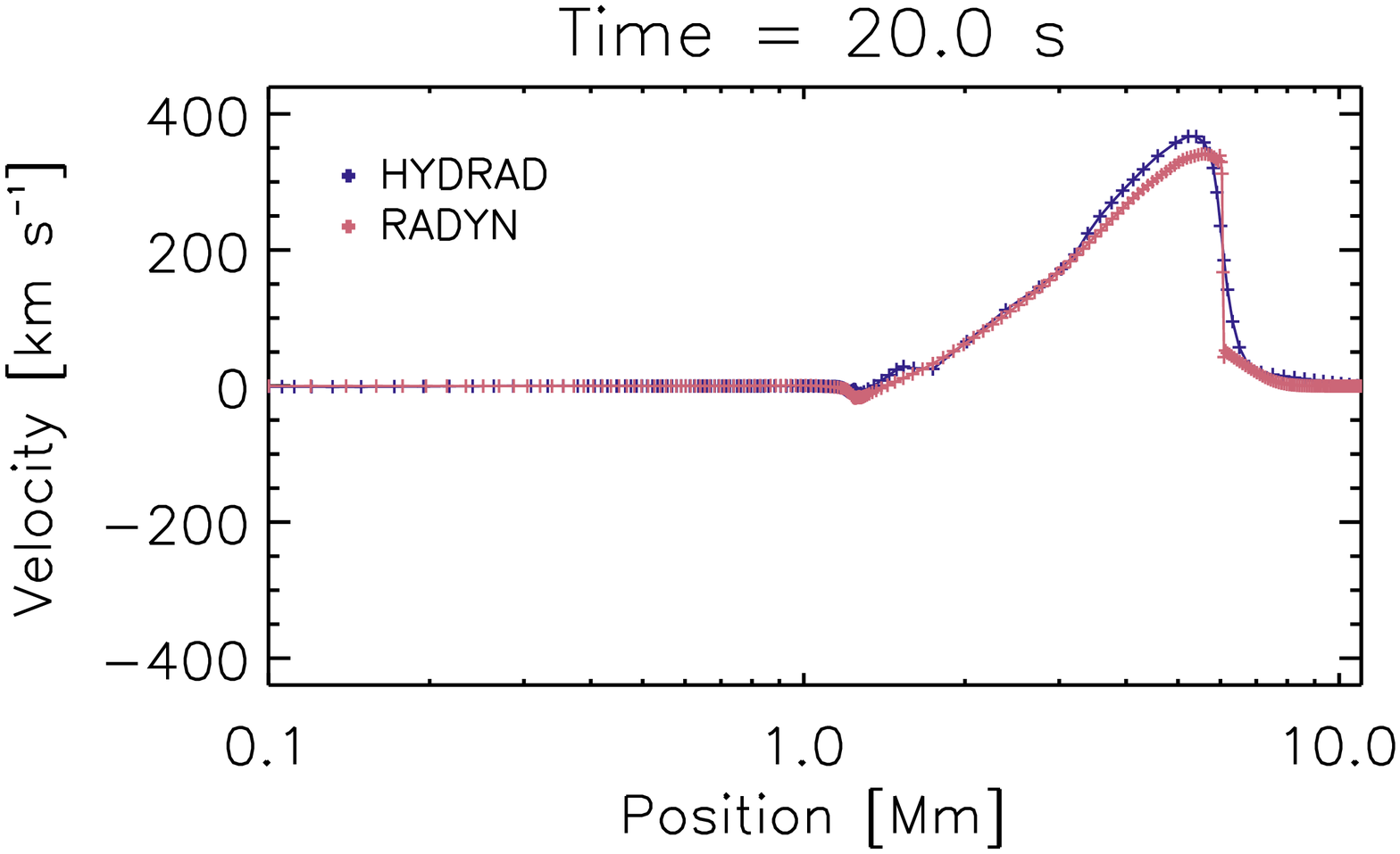}
\end{minipage}
\begin{minipage}[b]{0.32\linewidth}
\centering
\includegraphics[width=\textwidth]{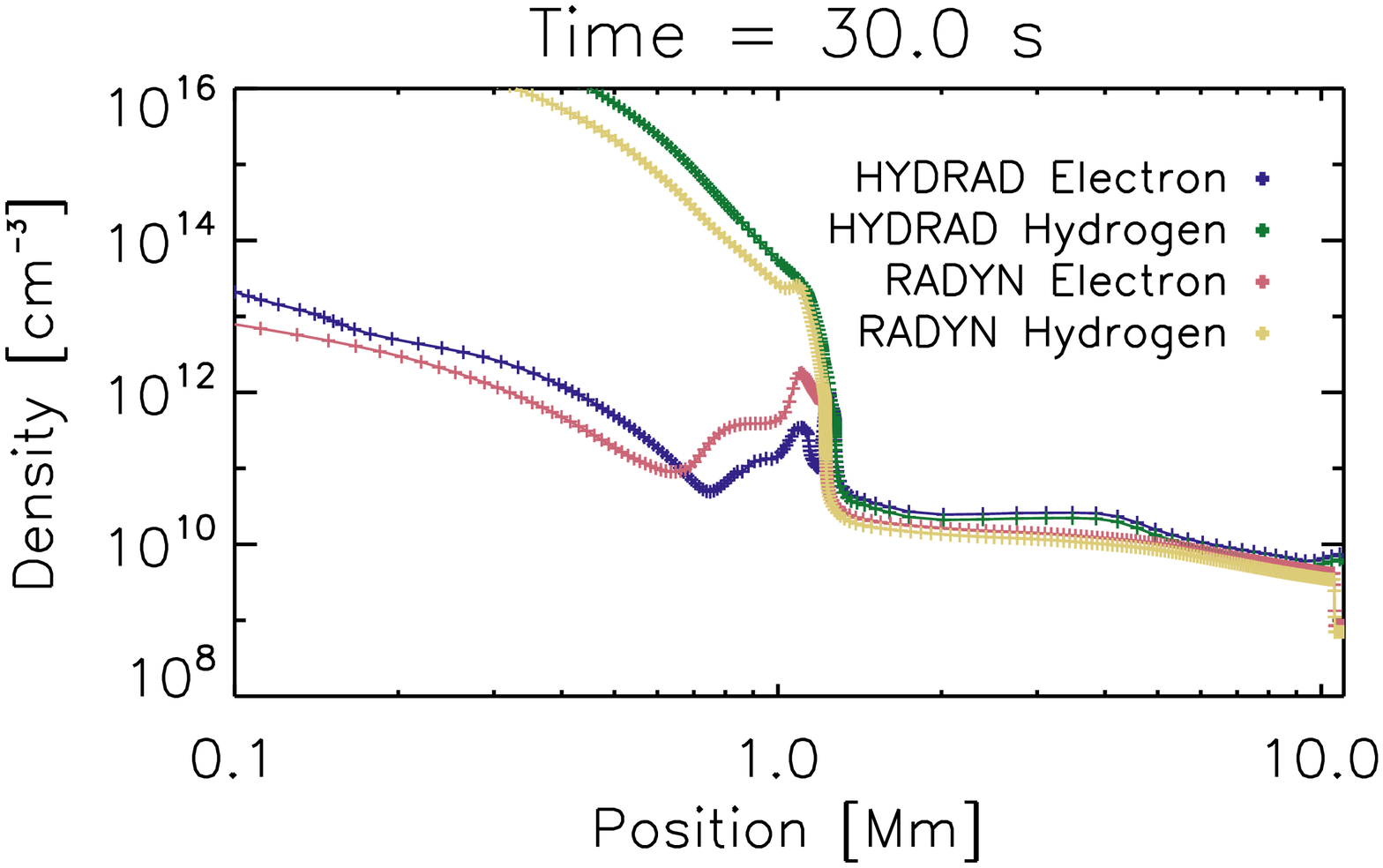}
\end{minipage}
\begin{minipage}[b]{0.32\linewidth}
\centering
\includegraphics[width=\textwidth]{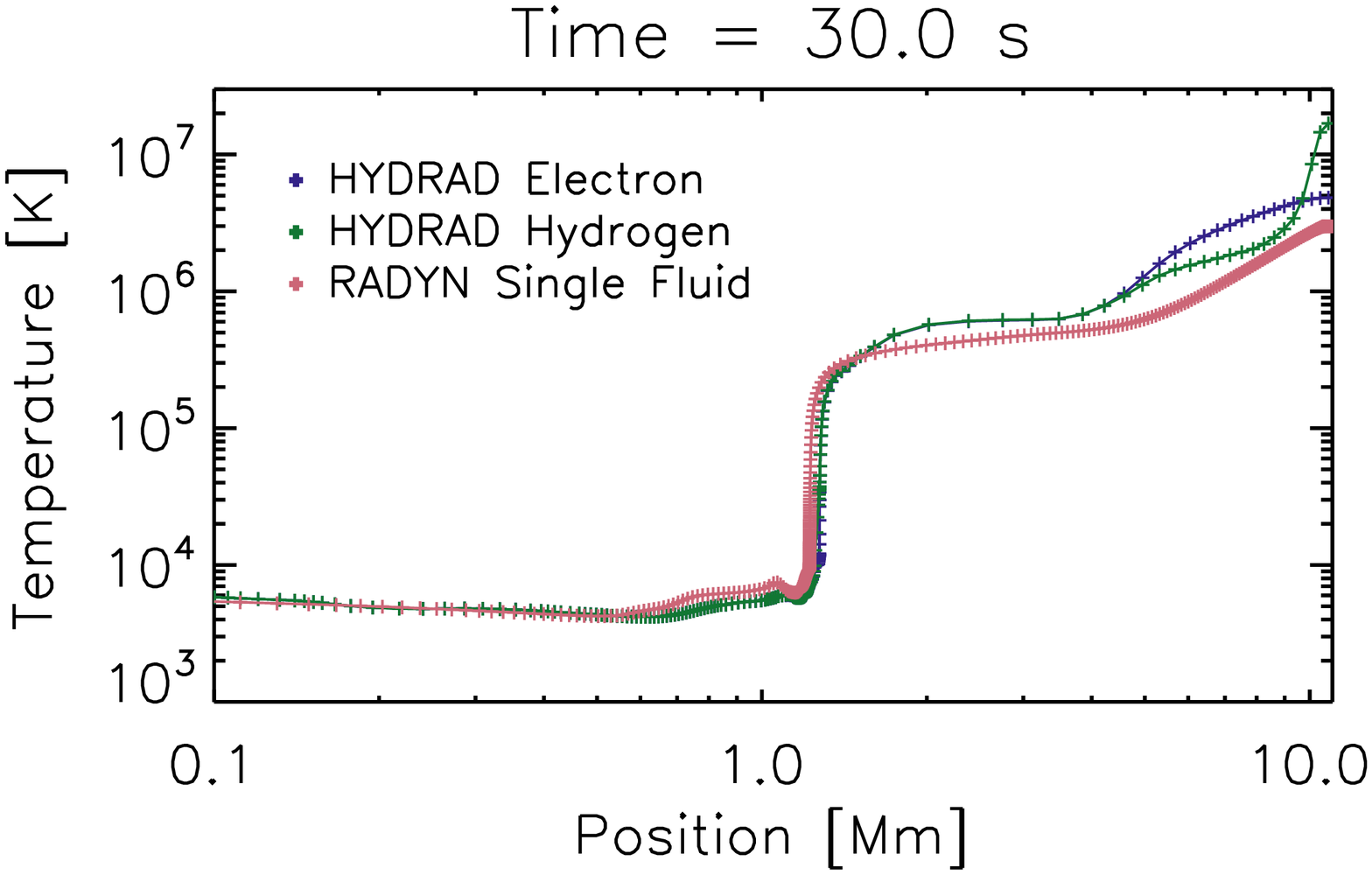}
\end{minipage}
\begin{minipage}[b]{0.32\linewidth}
\centering
\includegraphics[width=\textwidth]{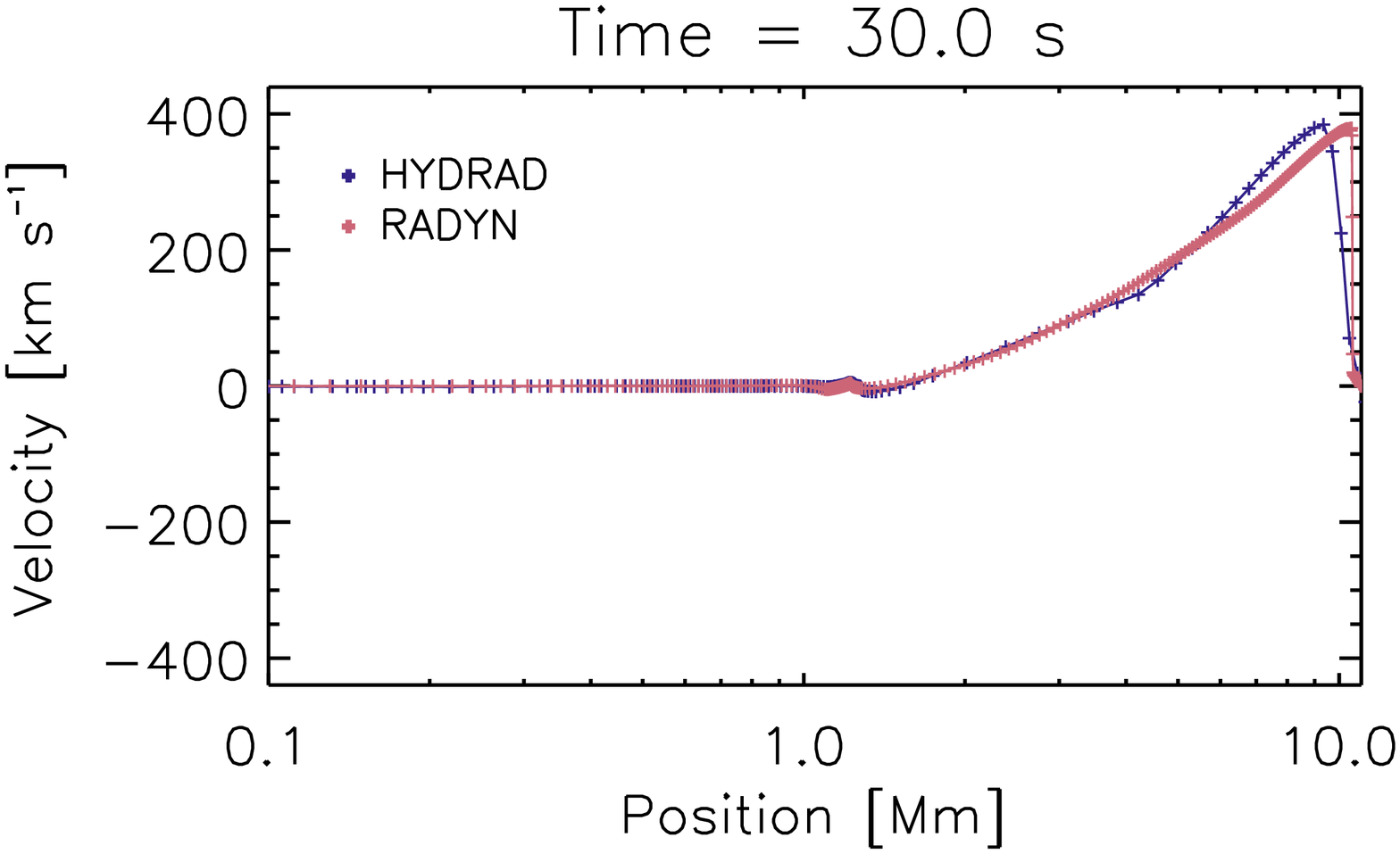}
\end{minipage}
\caption{A comparison between HYDRAD and RADYN, using model 078 from the F-CHROMA website.  From left, the densities, temperatures, and bulk velocities in the simulations at selected times.  We find excellent agreement between the two, in general, though there are differences.  Movies of these figures are available in the electronic version of this paper with the full duration of the simulation at 0.1\,s cadence, along with a movie showing the evolution of the hydrogen level populations.  }
\label{fig:078}
\end{figure*}

In Figure \ref{fig:078}, we show the comparison of the weaker case, model 078 at times 0, 5, 10, 20, 30\,s into the simulation.  In the online version, we have provided movies of the comparisons that show all time steps from 0--50\,s at 0.1\,s cadence, along with a movie comparing the hydrogen level populations.  The left column shows the electron and hydrogen densities along the loop.  Due to the differences in the initial conditions, HYDRAD is slightly denser in the chromosphere, particularly near the photosphere.  The electron density evolution shows good agreement until the cooling period, when HYDRAD's chromosphere cools more quickly than RADYN's, so that its ionized fraction falls more rapidly.  The middle column shows the temperatures along the loop.  RADYN's temperature (single fluid) is found to be closer to HYDRAD's hydrogen temperature in its temporal and spatial evolution than to the electron temperature.  Finally, the right column shows the bulk velocity in the two codes, which agrees closely at all times.  Considering the significant differences in physics and numerics between the codes, we consider the overall agreement to be excellent.

\begin{figure*}
\begin{minipage}[b]{0.32\linewidth}
\centering
\includegraphics[width=\textwidth]{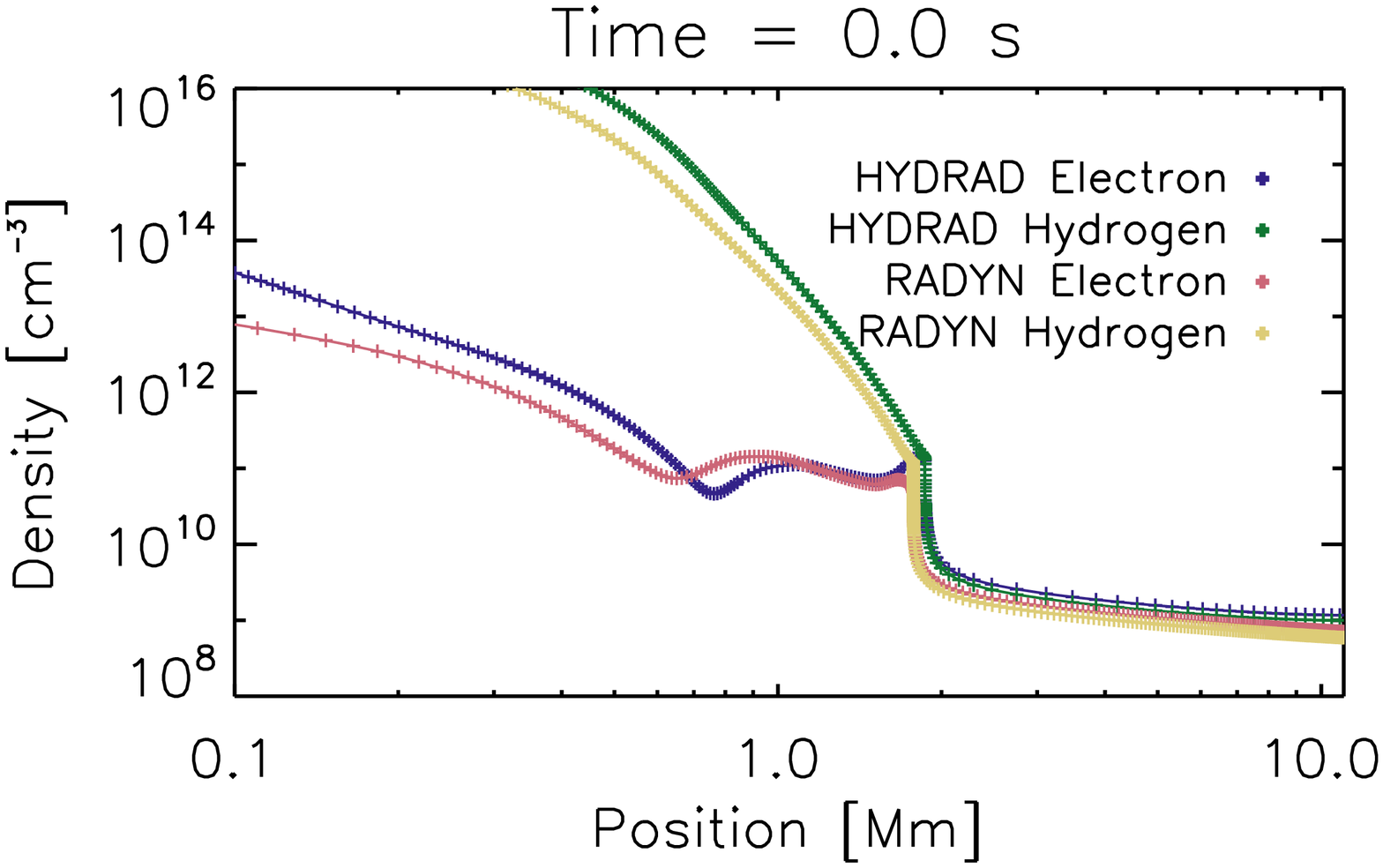}
\end{minipage}
\begin{minipage}[b]{0.32\linewidth}
\centering
\includegraphics[width=\textwidth]{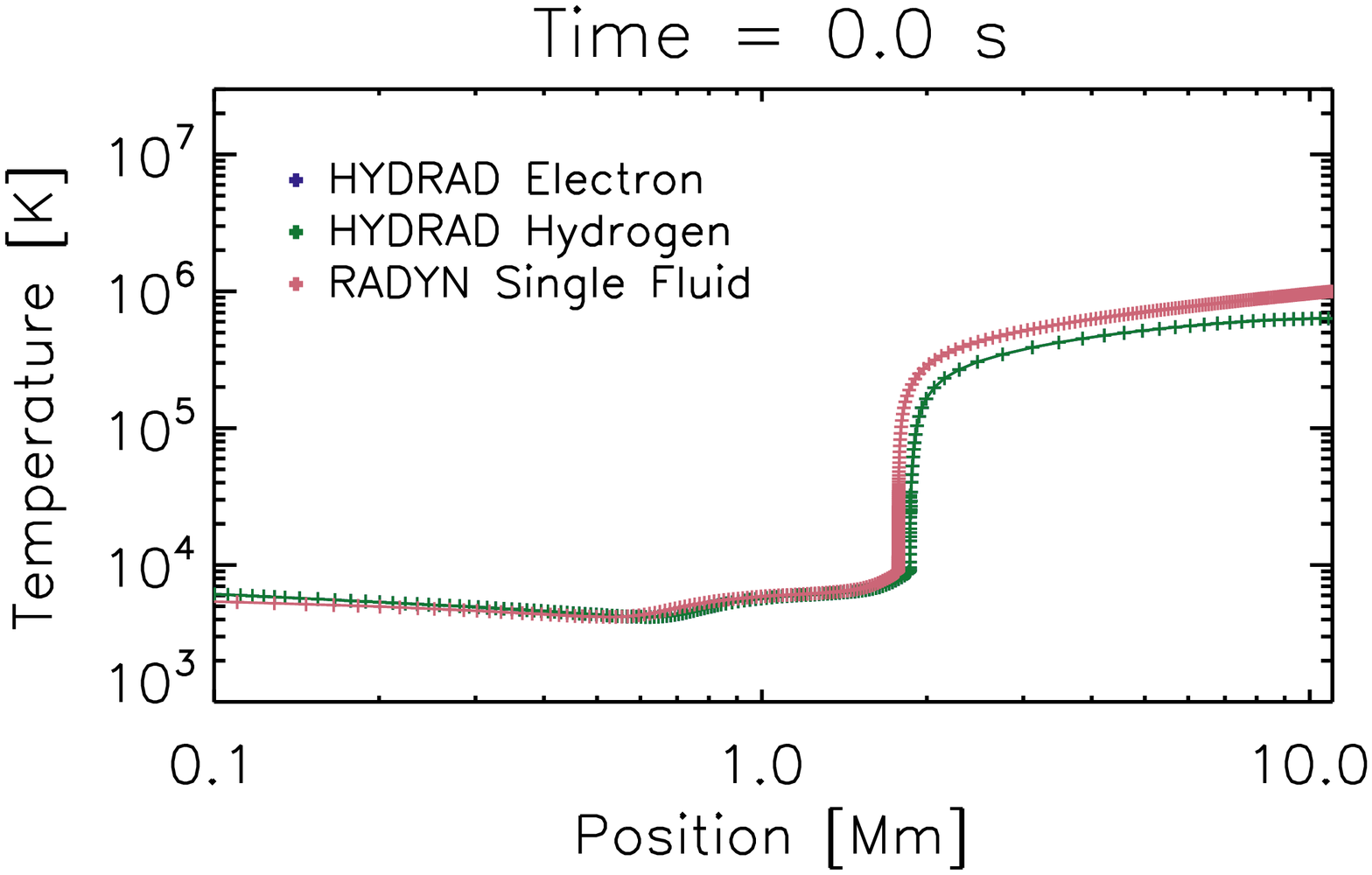}
\end{minipage}
\begin{minipage}[b]{0.32\linewidth}
\centering
\includegraphics[width=\textwidth]{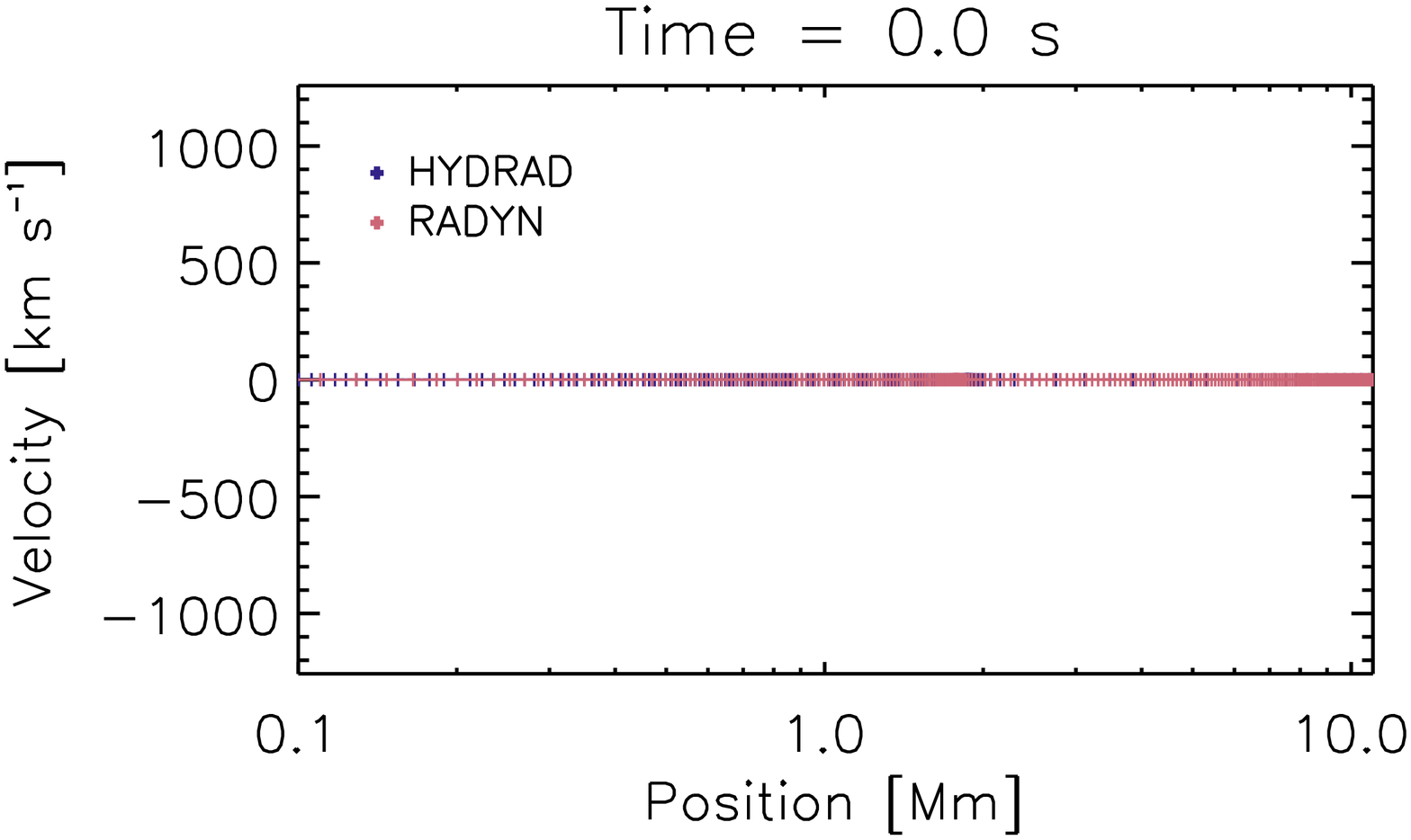}
\end{minipage}
\begin{minipage}[b]{0.32\linewidth}
\centering
\includegraphics[width=\textwidth]{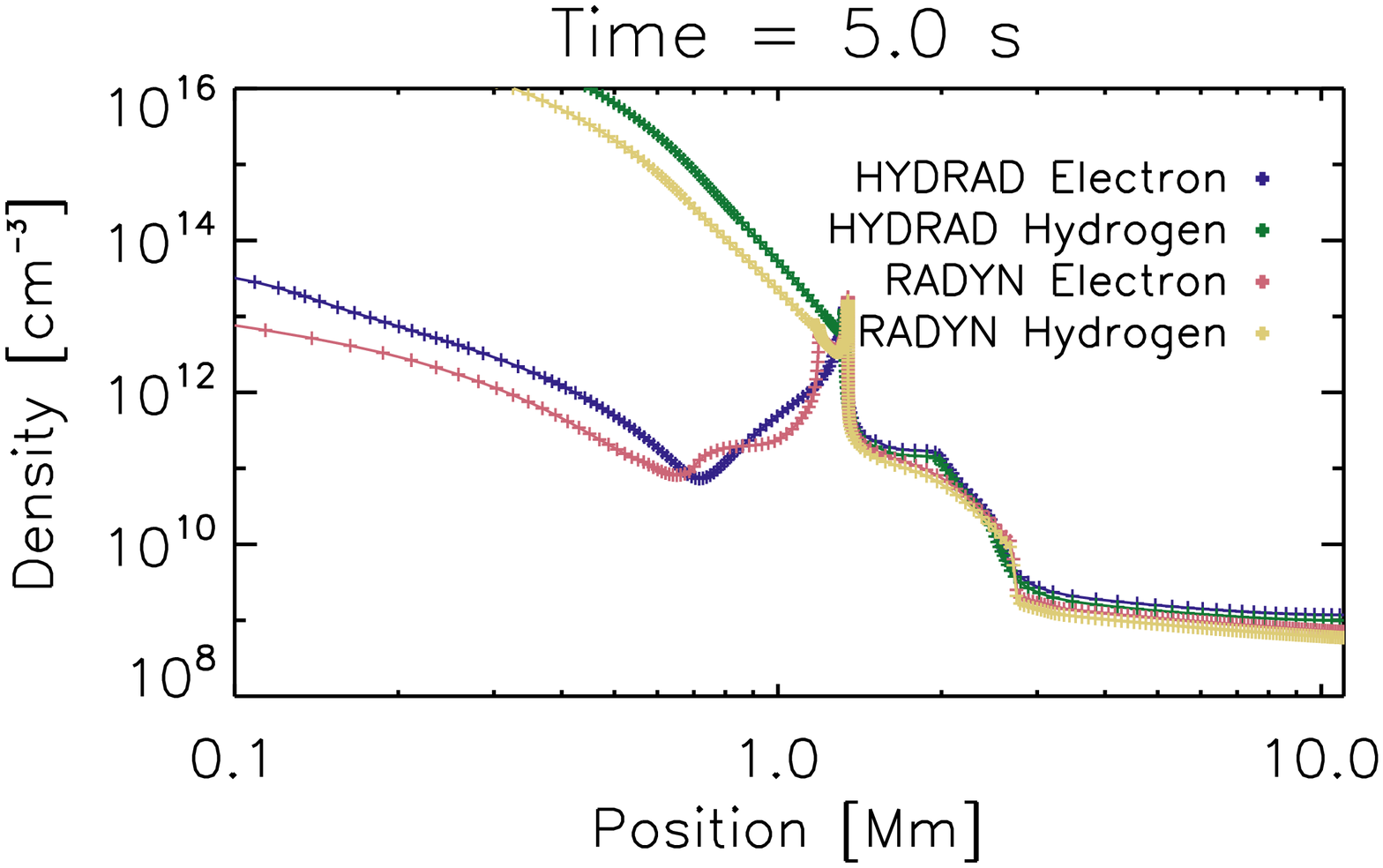}
\end{minipage}
\begin{minipage}[b]{0.32\linewidth}
\centering
\includegraphics[width=\textwidth]{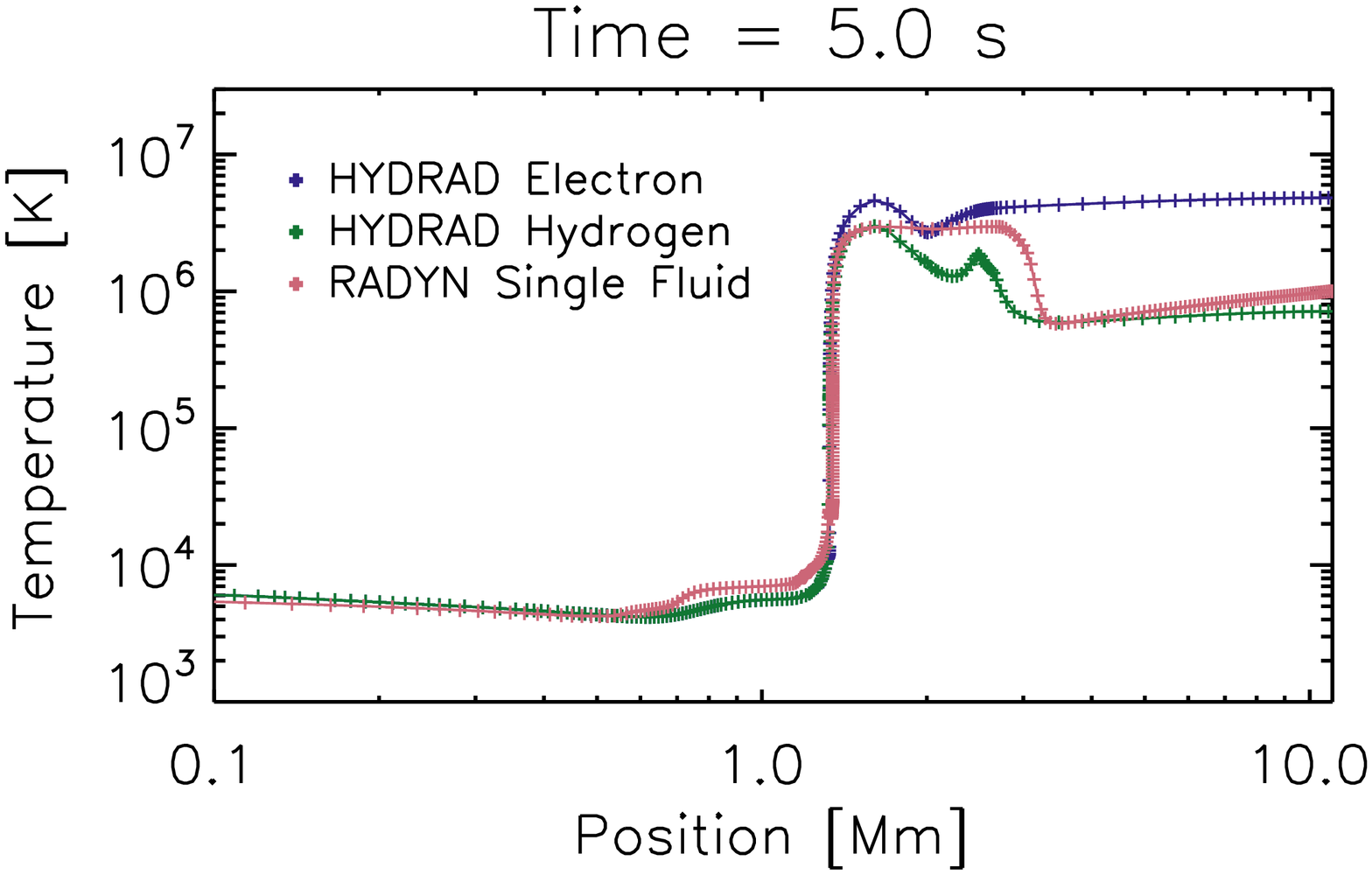}
\end{minipage}
\begin{minipage}[b]{0.32\linewidth}
\centering
\includegraphics[width=\textwidth]{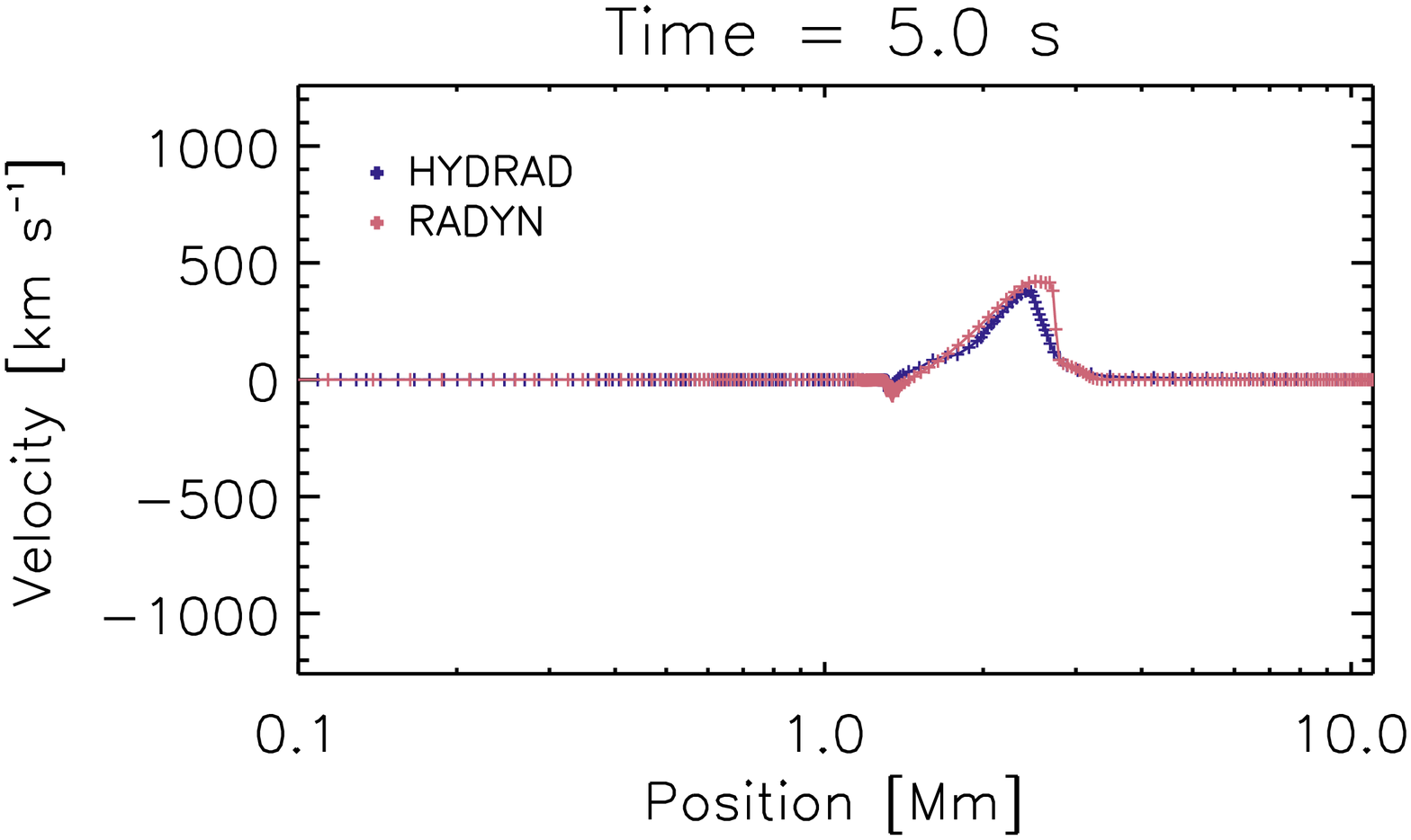}
\end{minipage}
\begin{minipage}[b]{0.32\linewidth}
\centering
\includegraphics[width=\textwidth]{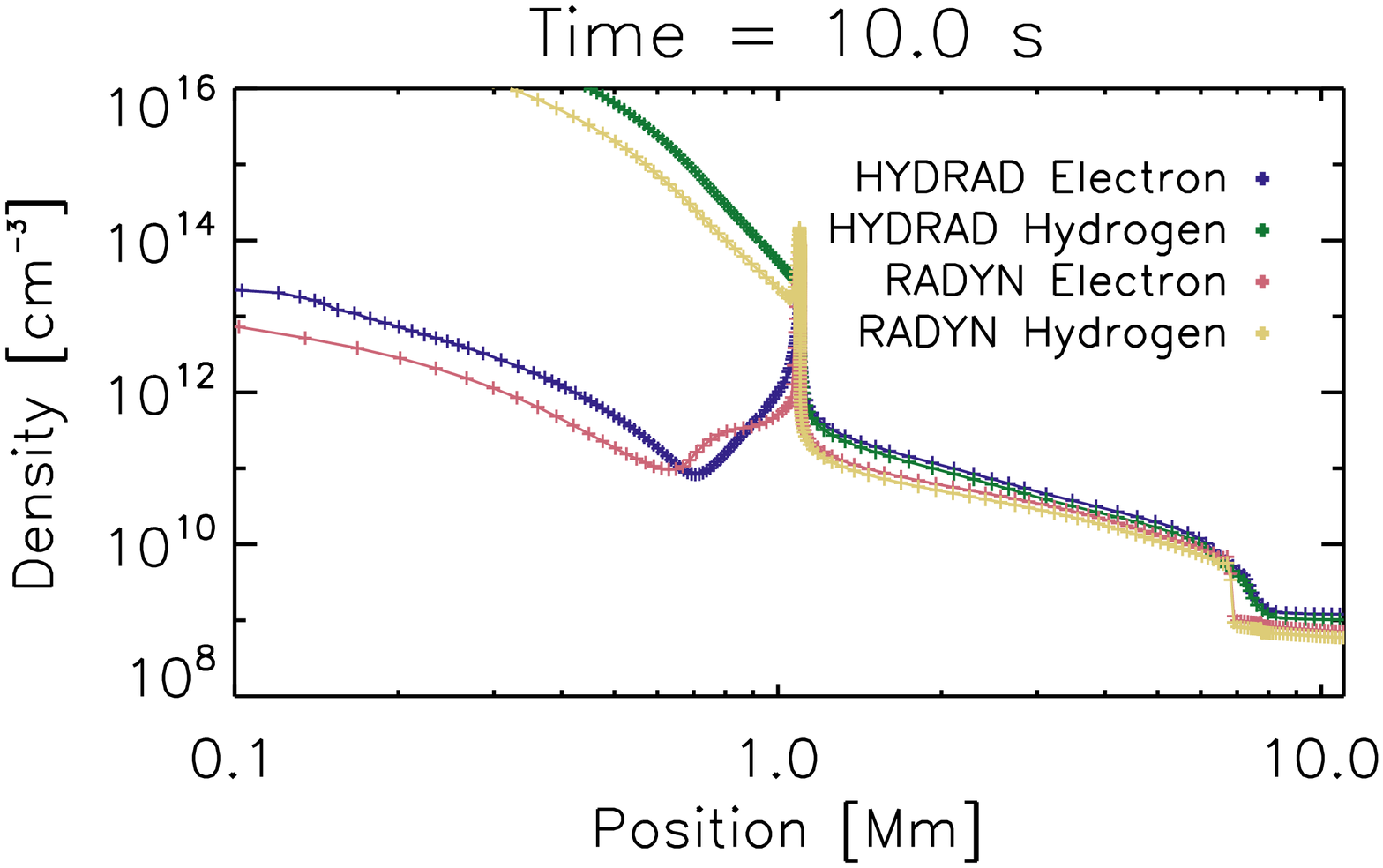}
\end{minipage}
\begin{minipage}[b]{0.32\linewidth}
\centering
\includegraphics[width=\textwidth]{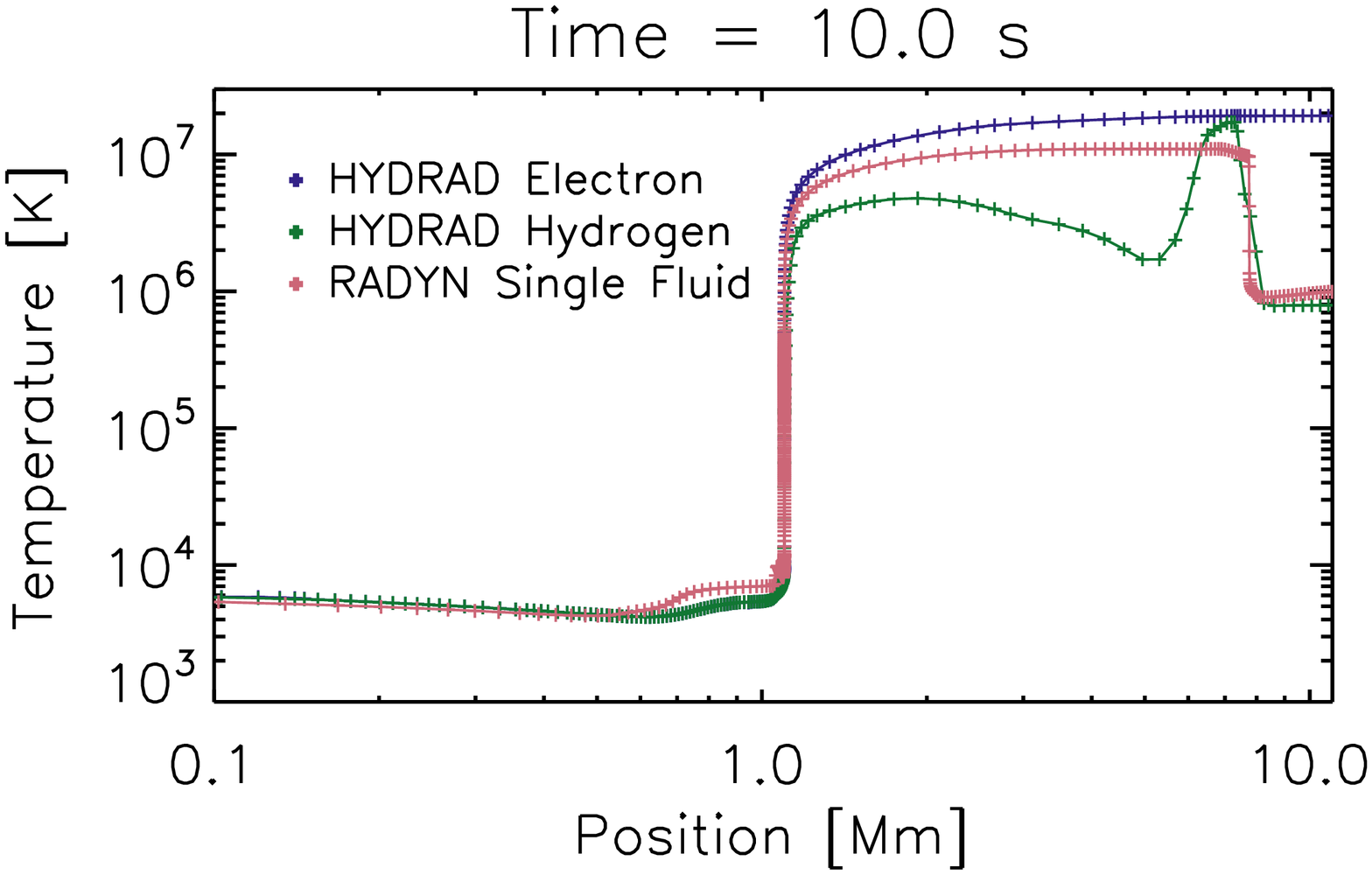}
\end{minipage}
\begin{minipage}[b]{0.32\linewidth}
\centering
\includegraphics[width=\textwidth]{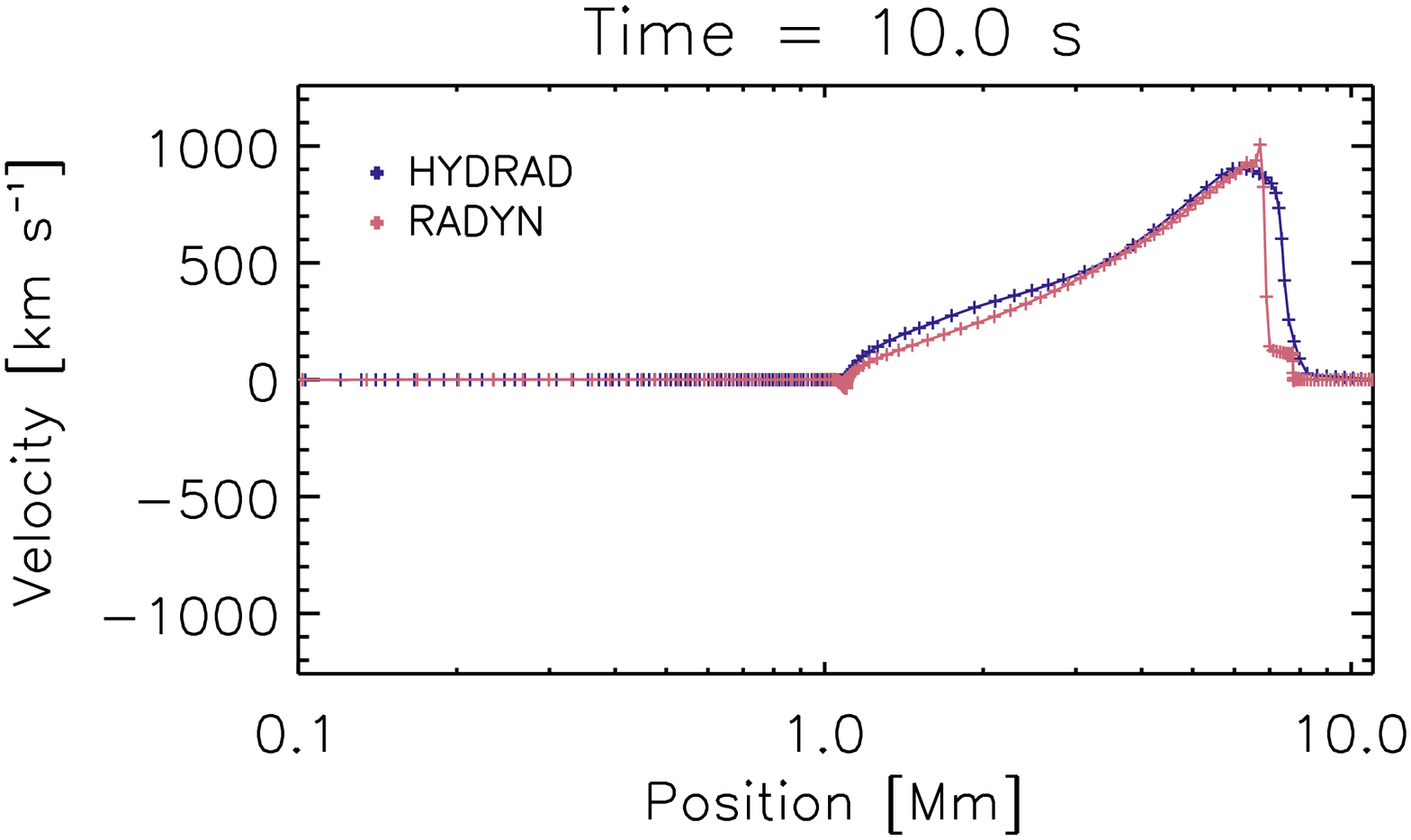}
\end{minipage}
\begin{minipage}[b]{0.32\linewidth}
\centering
\includegraphics[width=\textwidth]{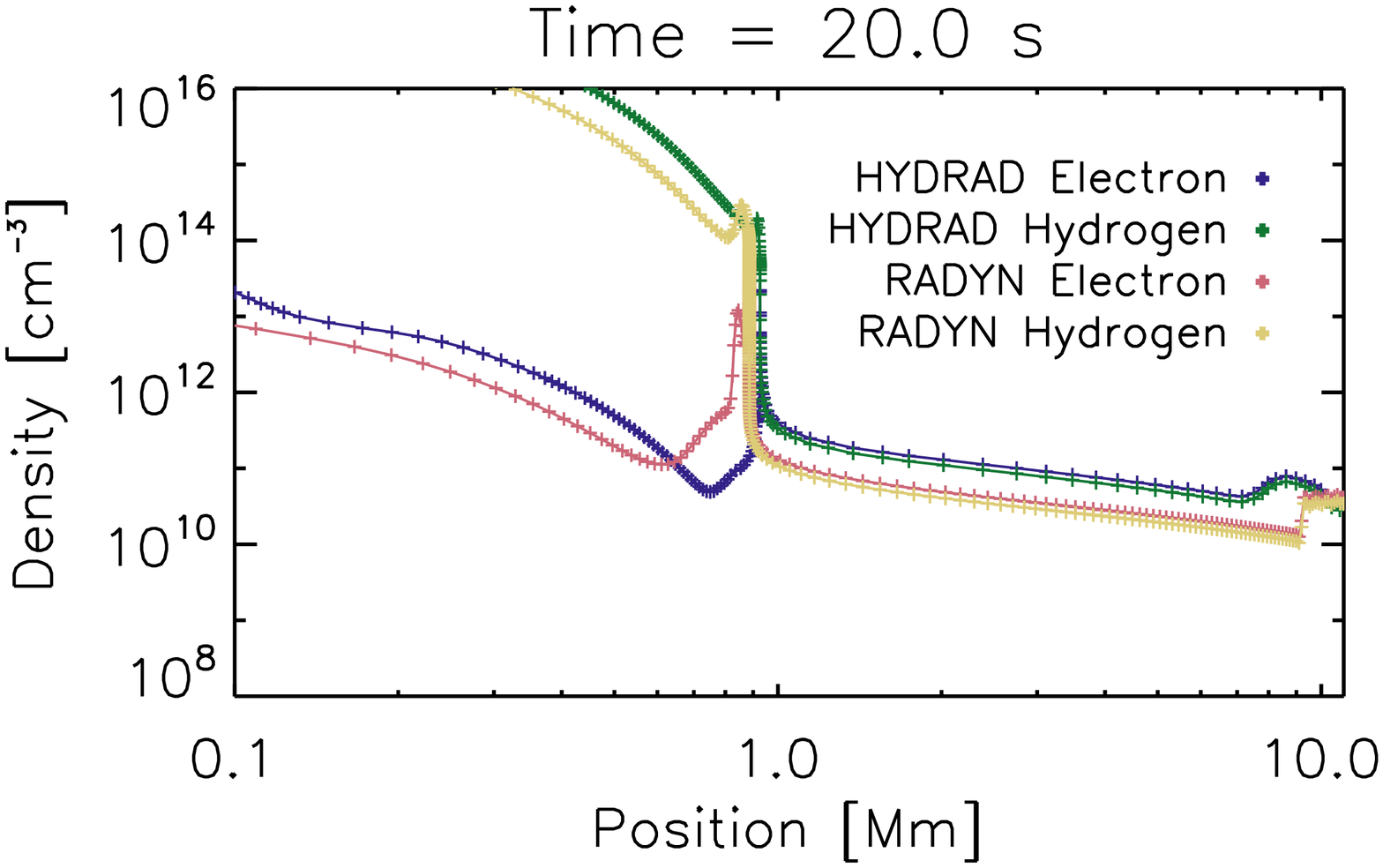}
\end{minipage}
\begin{minipage}[b]{0.32\linewidth}
\centering
\includegraphics[width=\textwidth]{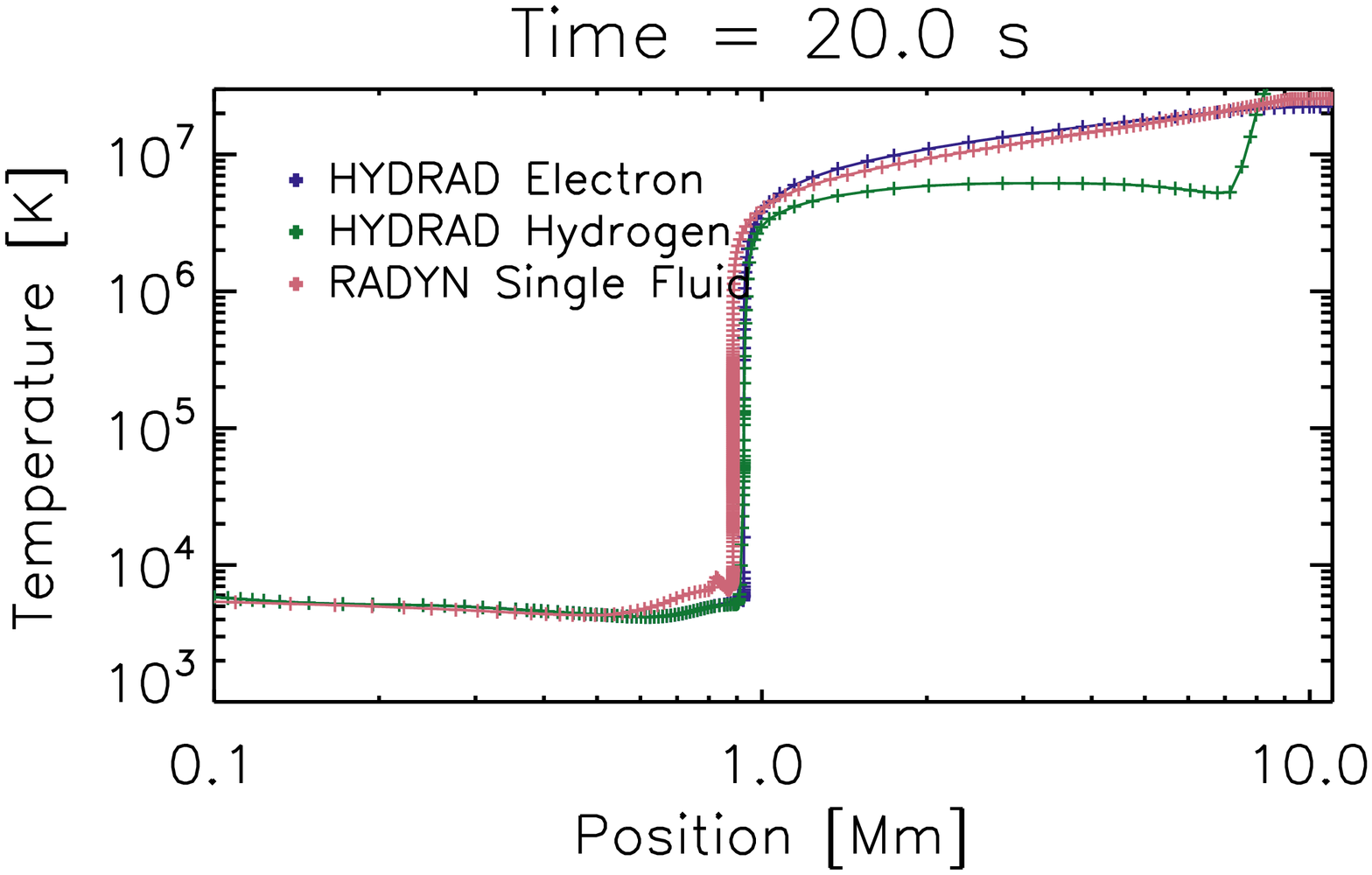}
\end{minipage}
\begin{minipage}[b]{0.32\linewidth}
\centering
\includegraphics[width=\textwidth]{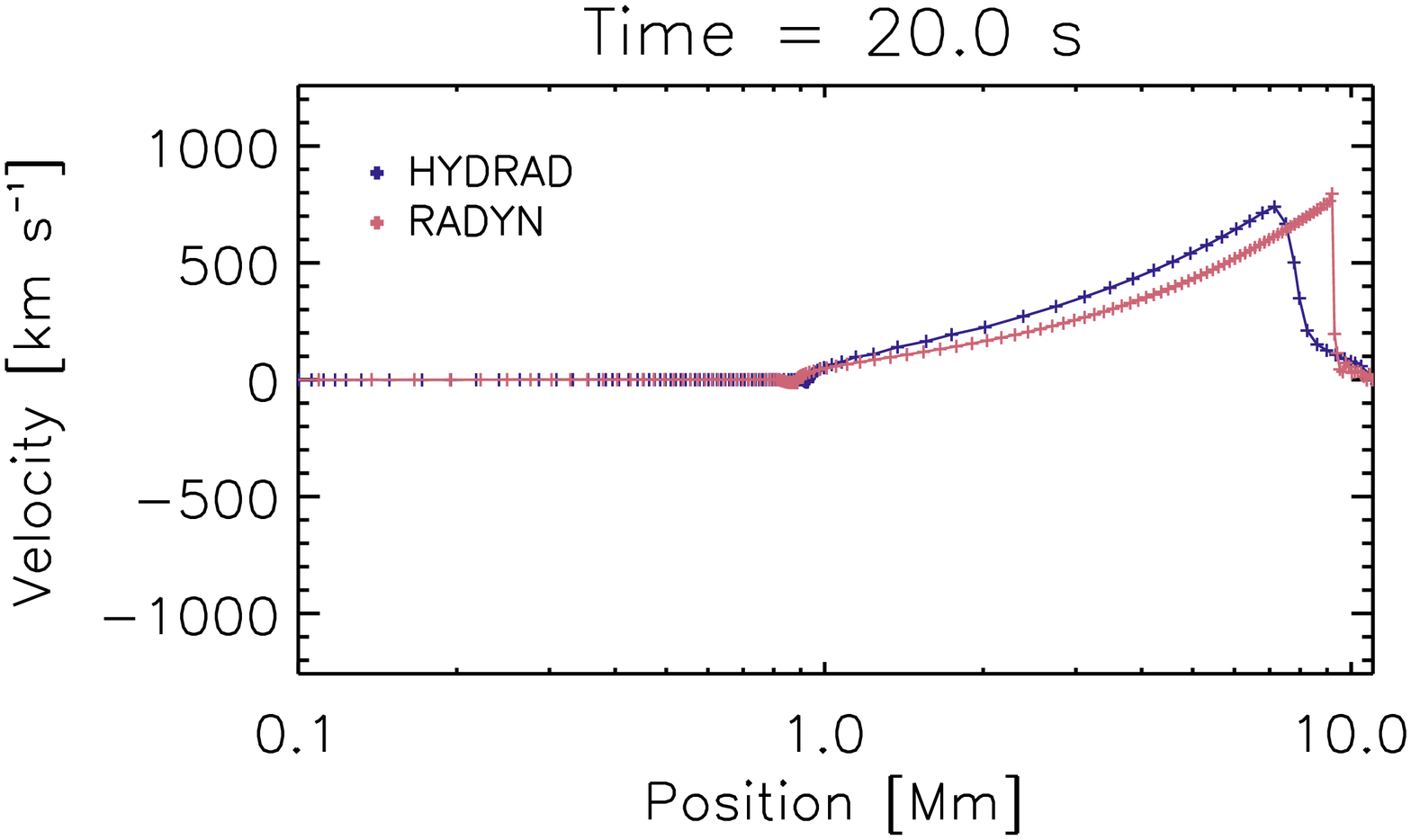}
\end{minipage}
\begin{minipage}[b]{0.32\linewidth}
\centering
\includegraphics[width=\textwidth]{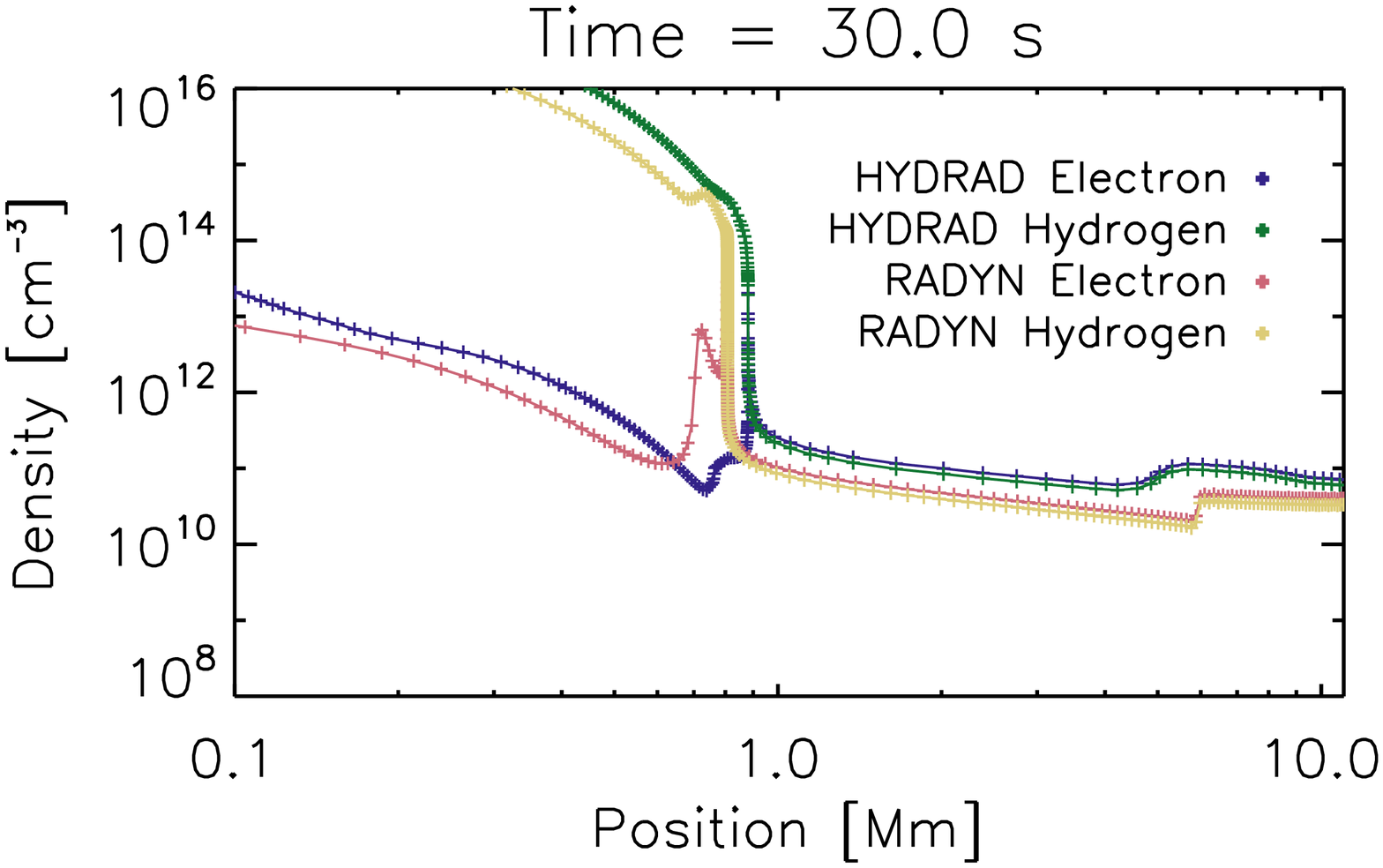}
\end{minipage}
\begin{minipage}[b]{0.32\linewidth}
\centering
\includegraphics[width=\textwidth]{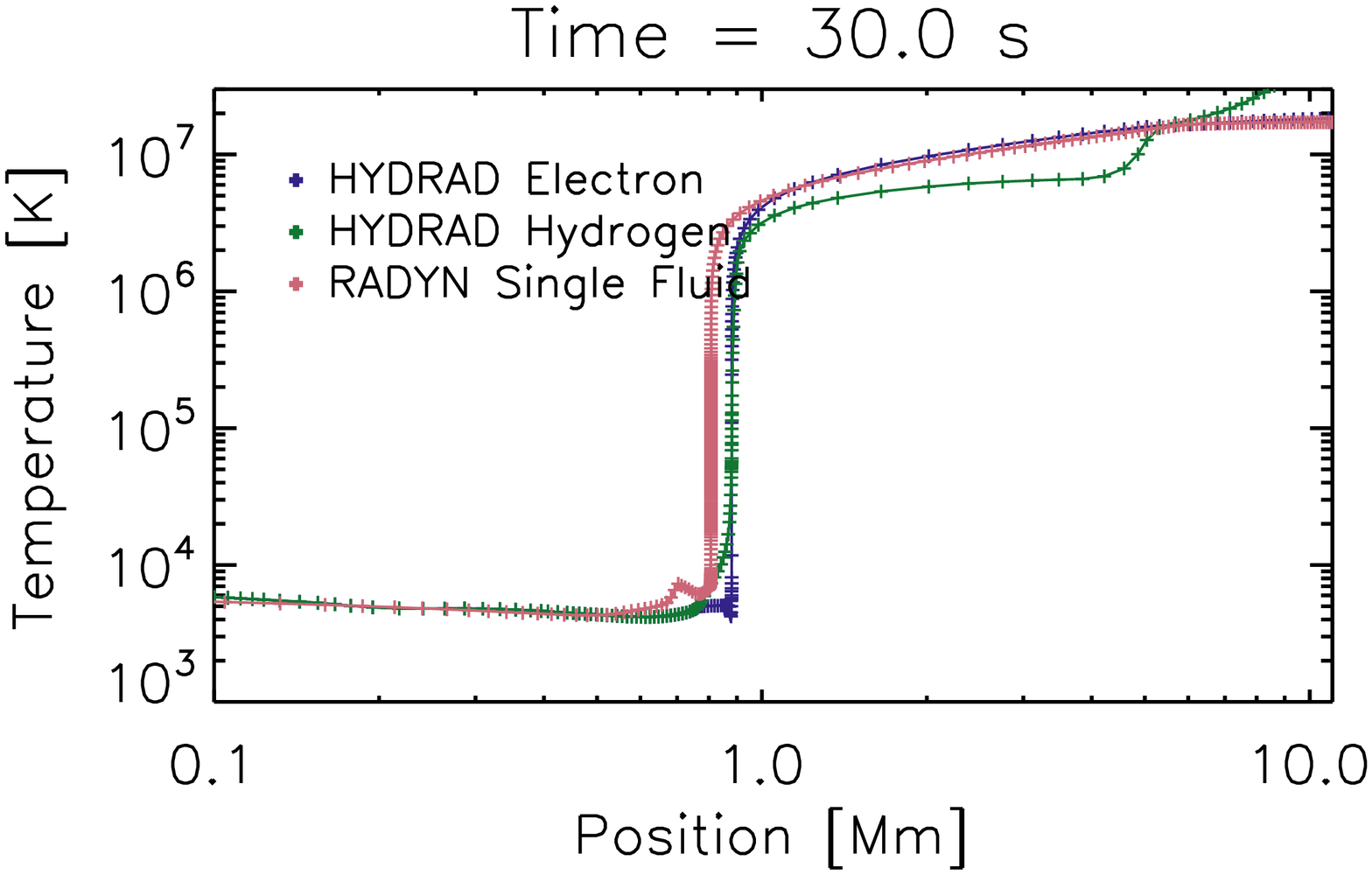}
\end{minipage}
\begin{minipage}[b]{0.32\linewidth}
\centering
\includegraphics[width=\textwidth]{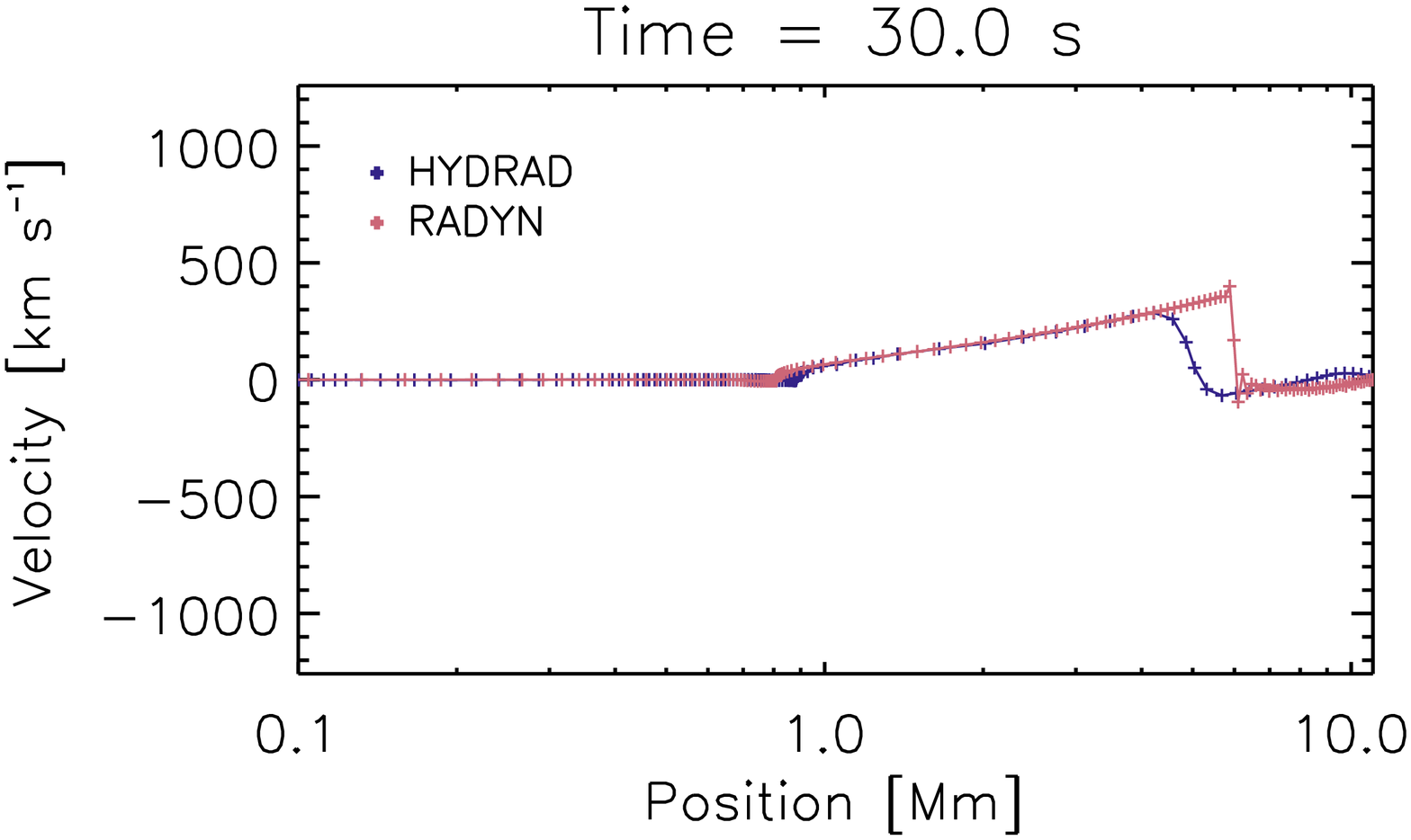}
\end{minipage}
\caption{Similar to the previous figure, using model 012 from the F-CHROMA website, which has a beam heating rate 10 times stronger than model 078.  Once again, despite significant differences in numerics and physics, the two compare well in general.  Movies of these figures are available in the electronic version of this paper with the full duration of the simulation at 0.1\,s cadence, along with a movie showing the evolution of the hydrogen level populations.  }
\label{fig:012}
\end{figure*}

In Figure \ref{fig:012}, we show a comparison to model 012 from the F-CHROMA website, which has an electron beam with the same cut-off and spectral index, but an energy flux 10 times higher than the previous case.  We find similar agreement: the overall evolution is comparable, but the details differ because of differences in both the physics and numerics.  In particular, following the cessation of heating, we once again find that the chromospheric electron density falls more rapidly in HYDRAD than RADYN, but is otherwise comparable. 

\leavevmode \newline

\acknowledgments  
JWR was supported by a Jerome and Isabella Karle Fellowship for this work.  This work was supported by NASA's \textit{Hinode} project. \textit{Hinode} is a Japanese mission developed and launched by ISAS/JAXA, with NAOJ as domestic partner and NASA and STFC (UK) as international partners. It is operated by these agencies in co-operation with ESA and NSC (Norway).  The numerical simulations were performed under a grant of computer time from the Department of Defense
High Performance Computing Program.  The authors would like to thank the anonymous referee for providing many helpful comments that significantly improved the model.  The authors would like to thank Drs.\ Tiago Pereira, Han Uitenbroek, Mats Carlsson, Jorrit Leenaarts, Joel Allred, Graham Kerr, and Jaroslav Dud\'ik for addressing many questions that we had while working on various stages of this paper.  We would like to thank Sam Cable for his efforts in parallelizing HYDRAD.  This research has made use of NASA's Astrophysics Data System.  IRIS is a NASA small explorer mission developed and operated by LMSAL with mission operations executed at NASA Ames Research center and major contributions to downlink communications funded by ESA and the Norwegian Space Centre.  We have made use of color-blind safe color tables where possible, using the IDL routine ``distinct\_colors'' by Paul Tol (\url{https://personal.sron.nl/~pault/}).   The research leading to these results has received funding from the European Community's Seventh Framework Programme (FP7/2007-2013) under grant agreement no. 606862 (F-CHROMA), and from the Research Council of Norway through the Programme for Supercomputing.

\bibliography{apj}
\bibliographystyle{aasjournal}

\end{document}